\documentclass[prd,showpacs,preprintnumbers,amsmath,amssymb]{revtex4}
\usepackage{graphics,color}
\usepackage{epsfig}
\usepackage{dcolumn}
\usepackage{bm}

\begin{document}
\title{$D\bar{D}$ production and their interactions}
\author{Yan-Rui Liu}\email{yrliu@th.phys.titech.ac.jp}
\affiliation{Department of Physics, H-27, Tokyo Institute of Technology, Meguro,
Tokyo 152-8551, Japan\\ Institute of High Energy Physics, P.O.
Box 918-4, Beijing 100049, People's Republic of China}
\author{Makoto Oka}\email{oka@th.phys.titech.ac.jp}
\affiliation{Department of Physics, H-27, Tokyo Institute of Technology, Meguro,
Tokyo 152-8551, Japan}
\author{Makoto Takizawa}\email{takizawa@ac.shoyaku.ac.jp}
\affiliation{Showa Pharmaceutical University,
Machida, Tokyo 194-8543, Japan}
\author{Xiang Liu}
\affiliation{School of Physical Science and Technology, Lanzhou
University, Lanzhou 730000, People's Republic of China}
\author{Wei-Zhen Deng}
\author{Shi-Lin Zhu}\email{zhusl@pku.edu.cn}
\affiliation{Department of Physics, Peking University, Beijing
100871, People's Republic of China}

\date{\today}

\begin{abstract}

S- and P- wave $D\bar{D}$ scatterings are studied in a meson
exchange model with the coupling constants obtained in the heavy quark effective theory. With the extracted P- wave
phase shifts and the separable potential approximation, we include
the $D\bar{D}$ rescattering effect and investigate the production
process $e^+e^-\to D\bar{D}$. We find that it is difficult to explain the
anomalous line shape observed by the BES Collaboration with this
mechanism. Combining our model calculation and the experimental
measurement, we estimate the upper limit of the nearly
universal cutoff parameter to be around 2 GeV.
With this number, the upper limits of the binding energies of the S- wave $D\bar{D}$ and $B\bar{B}$ bound states are obtained. Assuming that the S- wave and P- wave interactions rely on the same cutoff, our study provides a way of extracting the information about S- wave molecular bound states from the P- wave meson pair production.
\end{abstract}


\maketitle

\section{Introduction}\label{sec1}

Recently observed charmoniumlike states, called $X$, $Y$ or $Z$, have
motivated heated discussions on their properties (see Refs.
\cite{Hmesons,charmonium,XYZ,newHadron,rev-new} for the detailed
review). All of them are above the $D\bar{D}$
threshold and most states are near thresholds of two mesons. Various
interpretations have been proposed, such as tetraquark or S-wave
molecular states, hybrid states, dynamically generated states, or
mixing states of $c\bar{c}$ state and exotic components, while the possibility
that they are just $c\bar{c}$ states has not been excluded yet
\cite{xyz-screen,Liu:2009fe}. Among these possibilities, the
molecular interpretation is worth consideration, because the molecular states are expected to appear near the threshold.

To understand whether the proposed molecules exist or not, one may
study the bound state solution of the heavy meson-antimeson
systems. The meson exchange models are widely used in describing interactions of two hadrons
\cite{Tornqvist,
meson-meson-ps,3872-omep,4430-omep,Liu:2007bf,3872-quark,mole-pos,Liu:2009ei,mole-quark,mole-more}.
Other approaches include the gluon exchange models
\cite{wong,ding-4430} the unitarized model \cite{mole-dynamical},
lattice QCD \cite{mole-lat}, and QCD sum rule formalism
\cite{mole-qsr}. According to these theoretical calculations, it
seems that the existence of heavy quark molecules is inevitable,
especially for the isoscalar hidden bottom states, which awaits
the future experimental confirmation.

Among the hidden-charm meson-meson systems, the $D\bar{D}$ system
is the simplest one. Since $D$ is a pseudoscalar meson, the
total angular momentum of the system is equal to the orbital
momentum. The possible quantum numbers of the system are
$J^{PC}=0^{++}$, $1^{--}$, $2^{++}$, etc. There is no mixing
between different partial waves, unlike in the case of the
deuteron or $D\bar{D}^*$, where for instance the $S-D$- wave mixings are
important. For the other charm meson and anticharm meson system,
there may exist open charm decay channels which renders the
analysis more complicated. So we focus on the $D\bar{D}$ system
and discuss the isoscalar case within one meson exchange model
intensively.

The investigations in the literatures indicate that such a scalar
$D\bar{D}$ bound/resonance state may exist. In fact, a $D\bar{D}$ bound
state was obtained around 3.1 MeV in a quark-based model
\cite{wong}. A quasibound state was also found with the
unitarized method \cite{mole-dynamical,DDbar-dynamical}. The
vector meson exchange results in a possible binding solution
\cite{meson-meson-ps}. Our previous results cannot exclude its
existence within the meson exchange framework \cite{mole-pos} and
chiral quark model \cite{mole-quark}, either. However, a very
recent analysis indicates that the existence of a $D\bar{D}$ bound
state is difficult to understand \cite{ddbar-matter}. In this
paper, we will reanalyze this issue.

Despite these efforts, the interaction in the heavy meson
systems is still poorly known. It is expected that the study of the
scattering provides us with additional information besides solving the
bound state problems. For example, the S- wave $DK$ scattering
lengths and phase shifts give us additional information about whether
the molecular interpretation for $D_{sJ}(2317)$ is reasonable or
not \cite{DK-ph}. The scattering of $D$ and $D^*$ off the X(3872)
may reflect the $D\bar{D}^*$ interaction if X(3872) is a molecular
state \cite{DXscattering}. To better understand whether the
$D\bar{D}$ system may form a bound state, we will calculate the
partial wave scattering phase shifts and the relevant cross
sections. This is the first major part of the present study.

In addition to the observation of these unexpected hidden-charm X,
Y, Z mesons, the BES Collaboration recently announced an anomalous
line shape of the $e^+e^-\to$ hadrons total cross sections in Ref.
\cite{bes-anom-line}. The structure is slightly above the
$D\bar{D}$ threshold and slightly lower than $\psi(3770)$. A
similar anomalous line shape was also observed in the $D\bar{D}$
production \cite{bes-DDbar-line}. The di-resonance assumption is
one possible choice to understand such a structure
\cite{diresonance}. However, it is difficult to identify just from
the cross section whether this structure is due to a bound state,
a resonance or the final state interactions (FSI). More studies
are required.

The dominant decay mode of $\psi(3770)$ is the P- wave $D\bar{D}$.
The FSI effects \cite{3770-FSI-xiang,3770-FSI-zhao} have been
considered in understanding its large non-$D\bar{D}$ decay
observed by BES \cite{3770-nonddbar-bes1,3770-nonddbar-bes2}.
Their results indicate that FSI has non-negligible contributions.
If the $D\bar{D}$ interaction were really strong,  the
rescattering effect would also lead to the anomalous line shape in
their production. In Ref. \cite{3770lineshape}, part of the FSI
effects in the $D\bar{D}$ production has been included. No
anomalous line shape appears. Here, we will study the rescattering
effects in the process $e^+e^-\to D\bar{D}$ based on the
calculated phase shifts and the Yamaguchi separable potential approximation
\cite{separable}. Therefore, the multiple scattering effects are
included. This is the other major part of the present study.

With a heavier meson mass, a smaller kinetic energy and nearly
the same potentials according to the heavy quark symmetry, the
bottom analogous systems are more interesting. We will extend
our study to the isoscalar $B\bar{B}$ cases.

We organize our paper as follows. In Sec. \ref{sec2}, we present
the relevant Lagrangian, the derived potentials, and the
definition of the threshold parameters. The numerical results for
the S- and P- wave $D\bar{D}$ systems are given in Secs. \ref{sec3}
and \ref{sec4}, respectively. In Sec. \ref{sec5}, we consider the
rescattering effect in $e^+e^-\to D\bar{D}$. In Sec. \ref{sec6},
the results for the $B\bar{B}$ cases are presented. The last
section is our discussion.

\section{The effective potential and threshold parameters}\label{sec2}

One pion exchange between $D$ and $\bar{D}$ is forbidden because of the
parity conservation. What we need are the couplings of the $D$ meson
with the light scalar and vector mesons. The relevant effective
Lagrangian in the heavy quark limit reads
\cite{Casalbuoni,bardEH,3872-omep}
\begin{eqnarray}
{\cal L}=g_\sigma Tr[H\sigma \bar {H}]-i\beta_V Tr[H v^\mu
\rho_\mu\bar{H}],
\end{eqnarray}
where the field $H$ denotes the degenerate $(0^-, 1^-)$ doublet
\begin{eqnarray}
H&=&\frac{1+\not v}{2 }[P^{*\mu}\gamma_\mu+iP \gamma_5],
\end{eqnarray}
and $v^\mu=(1,0,0,0)$ denotes the velocity of the heavy mesons.
The vector meson field has the form
\begin{eqnarray}
\rho_\mu=i\frac{g_V}{\sqrt2}\hat{\rho}_\mu,\qquad \hat{\rho}_\mu=
\left(\begin{array}{ccc}
\frac{\rho^{0}}{\sqrt{2}}+\frac{\omega}{\sqrt{2}}&\rho^{+}&K^{*+}\\
\rho^{-}&-\frac{\rho^{0}}{\sqrt{2}}+\frac{\omega}{\sqrt{2}}&
K^{*0}\\
K^{*-} &\bar{K}^{*0}&\phi
\end{array}\right)_\mu.\label{vector}
\end{eqnarray}
For the coupling constant $g_\sigma$, we use the value derived
according to the chiral multiplets assumption \cite{bardEH},
$g_\sigma=0.76$. The values $g_V=m_\rho/f_\pi=5.8$ and $\beta_V=0.9$
are obtained from the vector meson dominance
\cite{Casalbuoni,vmdpara}.

From the Lagrangian, one may derive the $D\bar{D}$ scattering
amplitude and then the effective potentials from the $\sigma$,
$\rho$ and $\omega$ exchange, which are all Yukawa type
\begin{eqnarray}\label{pot-org}
{\cal V}_\sigma(r)&=& -\frac{g_\sigma^2}{4\pi r}e^{-m_\sigma r},\nonumber\\
{\cal V}_\rho(r)&=& -3\frac{(\beta_V g_V)^2}{16\pi r}e^{-m_\rho
r},\nonumber\\
{\cal V}_\omega(r)&=&-\frac{(\beta_V g_V)^2}{16\pi r}e^{-m_\omega r}.
\end{eqnarray}
Here we consider only potentials for the isoscalar $D\bar{D}$
system. For the isovector case, one changes the factor (-3) in
$V_\rho(r)$ to (+1). So the interaction for the $I=0$ system is
more attractive.

In principle, one may solve the bound state and the scattering
problem by inserting these potentials into the Schr\"odinger
equation directly. However, we will see unphysical results appear
because all the mesons are assumed to be pointlike particles. The
situation is very similar to the $N\bar{N}$ case where the very
short range interaction is unclear \cite{NNbar}. For the realistic
system of composite particles, a form factor at each interacting
vertex is necessary. We will use the monopole type form factor
\begin{eqnarray}
F(q)=\frac{\Lambda^2-m^2}{\Lambda^2-q^2},
\end{eqnarray}
where $\Lambda\sim$1 GeV is the cutoff, $m$ is the exchanged meson
mass, and $q$ is its four-momentum. The improved potentials are
\cite{mole-pos}
\begin{eqnarray}\label{pot-mod}
V_\sigma=-\frac{g_\sigma^2}{4\pi}[\frac{1}{r}(e^{-m_\sigma
r}-e^{-\Lambda r} )-
\frac{\Lambda^2-m_\sigma^2}{2\Lambda}e^{-\Lambda r}
],\nonumber\\
V_\rho=-3\frac{(\beta_V g_V)^2}{16\pi}[\frac{1}{r}(e^{-m_\rho
r}-e^{-\Lambda r})-
\frac{\Lambda^2-m_\rho^2}{2\Lambda}e^{-\Lambda r} ],\nonumber\\
V_\omega=-\frac{(\beta_V g_V)^2}{16\pi}[\frac{1}{r}(e^{-m_\omega
r}-e^{-\Lambda r})-
\frac{\Lambda^2-m_\omega^2}{2\Lambda}e^{-\Lambda r} ].
\end{eqnarray}
Here we use one cutoff to describe the system. The case
$\Lambda\to \infty$ gives the former potentials. We will call this
case the point particle limit.

After one gets the phase shifts with these potentials, the
threshold parameters can be derived through the definition
\begin{equation}\label{defthpar}
\lim_{k\to 0}k^{2L+1}\cot(\delta_L)\equiv\frac{1}{a_L},
\end{equation}
where $a_0$ ($a_1$) denotes the scattering length (volume) for the
S (P)- wave interaction. With this convention, $a_L$ is negative if
the interaction is repulsive. For a system with attractive
interactions, $a_L>0$ if there is no bound state and $a_L<0$ when
one bound state appears.

\section{The S wave $D\bar{D}$ system}\label{sec3}

We mainly explore whether there exists an S-wave bound state. From
the previous studies \cite{3872-omep,mole-pos,4430-omep}, we have
learned that the numerical results are very sensitive to the
cutoff $\Lambda$. Since 
the contact
interaction of the form $\sim \delta ({\vec r})$ does not exist, it is
instructive to study the case $\Lambda\to\infty$ first. We solve
the bound state with the potentials in Eq. (\ref{pot-org}). The
$\sigma$ meson is a broad scalar resonance with strong coupling to $I=0$ $\pi\pi$ S- wave scattering states. Its mass is not definite but around 400$\sim$600 MeV \cite{sigmas-ishi,sigmas-igi,sigmas-col}. We choose two limit cases, $m_\sigma=400$ MeV and 600 MeV. Other parameters are
$m_\rho=775.49$ MeV, $m_\omega=782.65$ MeV, and $m_D=1867.23$ MeV.
If one ignores the vector meson potentials ${\cal V}_\rho$ and
${\cal V}_\omega$ (we label this case NV), there are no
binding solutions. After the inclusion of the vector exchange potentials (we
label this case VC), we get a very deep bound state with the
binding energy around 980 MeV, which indicates this is not a
physical solution. A smaller cutoff suppresses the contributions
from the short range vector meson interactions. A number around 1
GeV should be reasonable. In the following calculation, we take
$\Lambda=$0.8, 1.0, 1.2, 1.5, and 2.0 GeV and compare the results (note the lower bound should satisfy
$\Lambda>m_\omega$).

\begin{figure}[htb]
\centering
\begin{tabular}{cc}
\scalebox{0.6}{\includegraphics{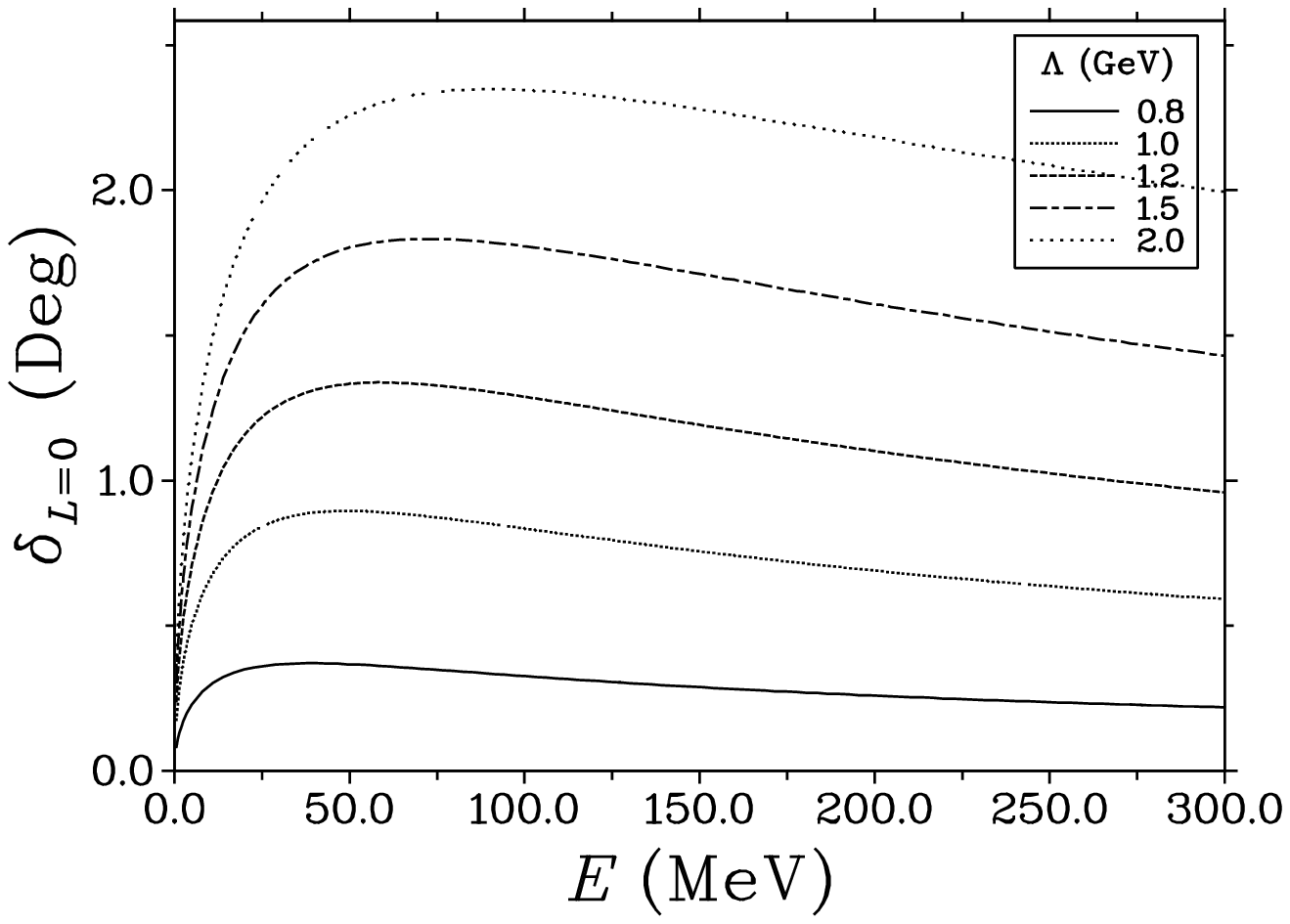}}&
\scalebox{0.6}{\includegraphics{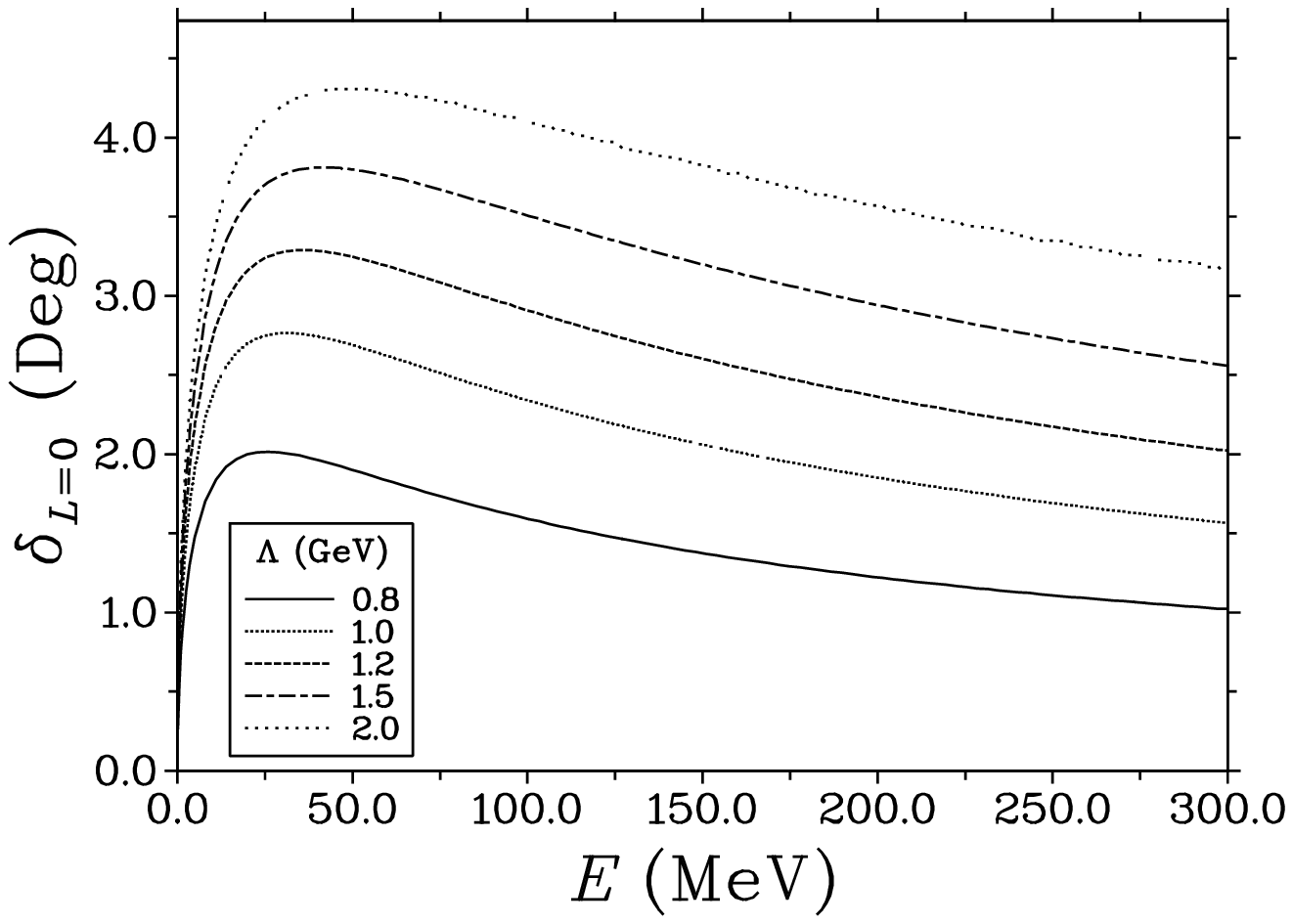}}\\
(a)&(b)\\
\scalebox{0.6}{\includegraphics{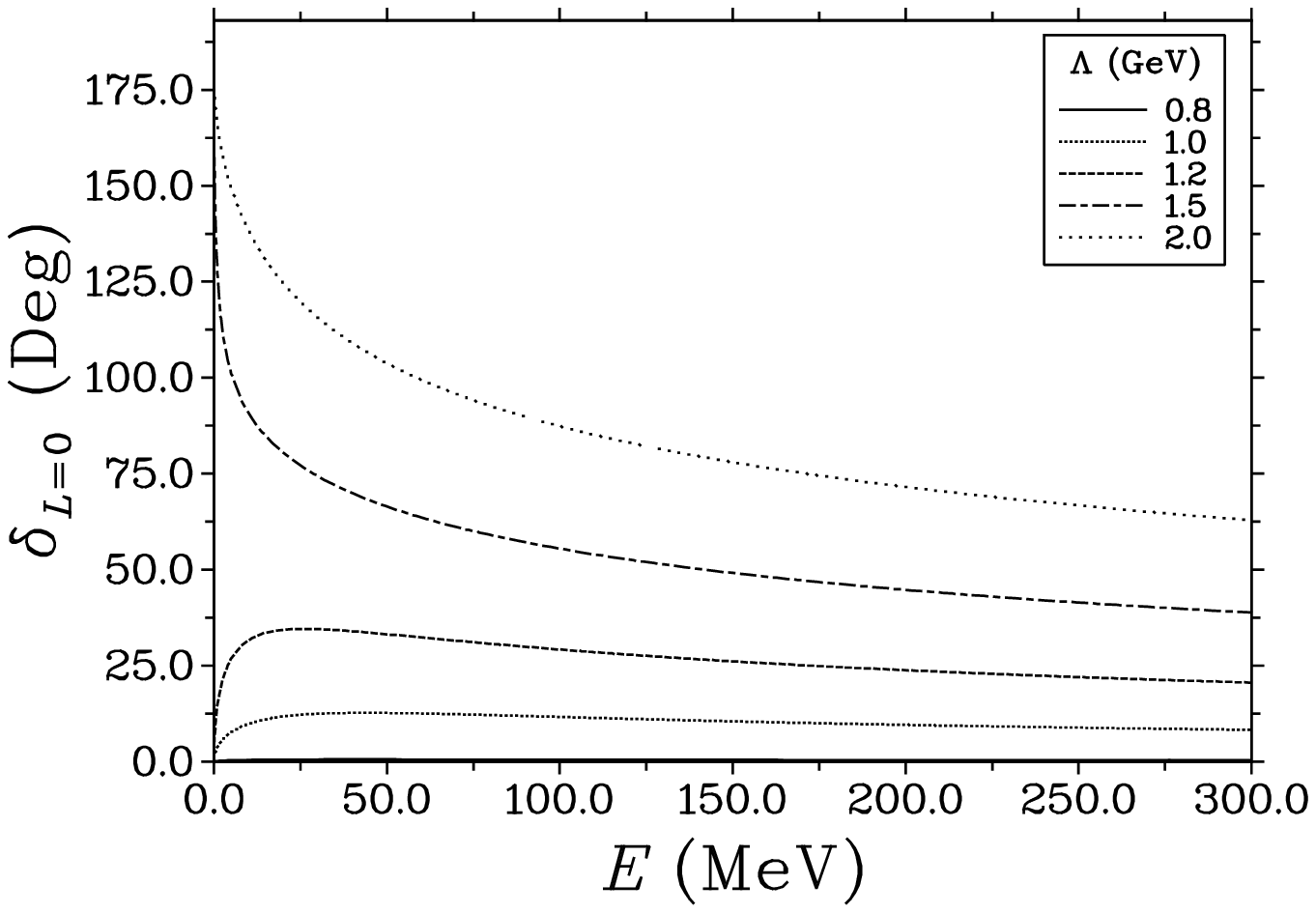}}&
\scalebox{0.6}{\includegraphics{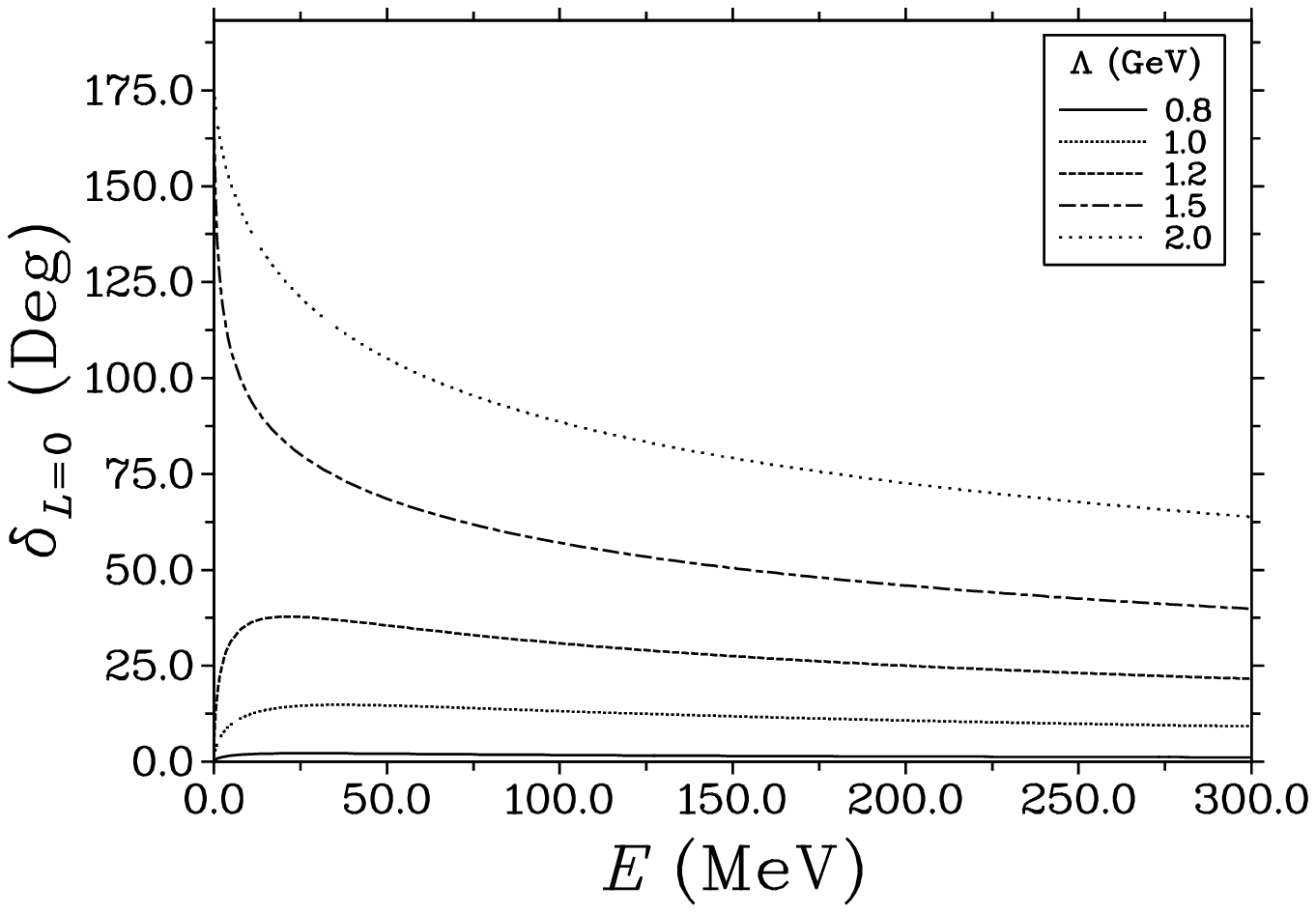}}\\
(c)&(d)
\end{tabular}
\caption{The phase shifts for the S wave $D\bar{D}$ scattering
with various parameters. The upper (lower) two diagrams correspond
to the cases without (with) vector meson exchange contributions.
The left (right) two diagrams are obtained with $m_\sigma$=600
(400) MeV. The cutoff $\Lambda$ is in units of GeV.}\label{ph-DDbar-S}
\end{figure}

\begin{table}[htb]
\centering \caption{S- wave $D\bar{D}$ scattering lengths in units
of fm. NV (VC) indicates the contributions from vector
mesons are omitted (included). The number of * in the table
indicates existence of a bound state. The binding energies are given in Table
\ref{EB}.}\label{SDDbar}
\tabcolsep=9pt
\begin{tabular}{c|c|cccccc}\hline
$m_\sigma$  & Vector meson  &\multicolumn{6}{c}{$\Lambda$ (GeV)}\\

(MeV)&exchange& 0.8 & 1.0 & 1.2 & 1.5 & 2.0 & $\infty$\\\hline
600  & NV & 0.0091 &0.020 &0.027 & 0.035 & 0.041 & 0.051\\
400  & NV &0.064&0.081&0.092& 0.10&0.11 &0.12\\
600  & VC & 0.014 &0.31 &1.23  &-5.46(*)& -1.14 (*)& 0.60(*)\\
400  & VC &0.068&0.40&1.56&-4.23(*)&-1.10(*)&0.74(*)\\
\hline
\end{tabular}
\end{table}

\begin{table}[htb]
\centering \caption{The binding energies for the different cases
in units of MeV. The three values for the S- wave $B\bar{B}$ in the
case of $\Lambda\to\infty$ correspond to the ground state, the
first and second radially excited states, respectively.}\label{EB}
\tabcolsep=9pt
\begin{tabular}{c|c|cccc}\hline
Systems&$m_\sigma$ &\multicolumn{4}{c}{$\Lambda$ (GeV)}\\
&(MeV) &  1.2 & 1.5 & 2.0 & $\infty$\\\hline
$D\bar{D}$(S wave)&600  &$\times$&-0.8  &-29.4&-974.5\\
&400  &$\times$ &-1.4&-31.8&-980.9\\\hline
$B\bar{B}$ (S wave)&600  & -8.3&-57.4&-186.0&-4916.7 (n=1)\\
&&&&&-444.4 (n=2)\\
&&&&&-1.3 (n=3)\\
&400  &  -10.1& -60.7& -190.5&-4924.8 (n=1)\\
&&&&&-449.7 (n=2)\\
&&&&&-2.0 (n=3)\\\hline
$B\bar{B}$ (P wave)&600&$\times$&$\times$&$\times$&-377.4\\
&400&$\times$&$\times$&-0.6&-383.1
\\\hline
\end{tabular}
\end{table}

We present the calculated S- wave phase shifts of the elastic
$D\bar{D}$ scattering in Fig. \ref{ph-DDbar-S}. In these diagrams, $E$ is the energy of the $D$ meson in the center of mass frame. According to the Levinson theorem, the phase shift approaches
$180^\circ$ when $E\to 0$ if a bound state exists. The results in
the figure indicate that a $D\bar{D}$ bound state is possible if
the short range attraction is strong, e.g. $\Lambda\geq
1.5$ GeV. If the cutoff is around 1.2 GeV or less, the S- wave $D\bar{D}$
bound state does not exist.

\begin{figure}[htb]
\centering
\begin{tabular}{cc}
\scalebox{0.6}{\includegraphics{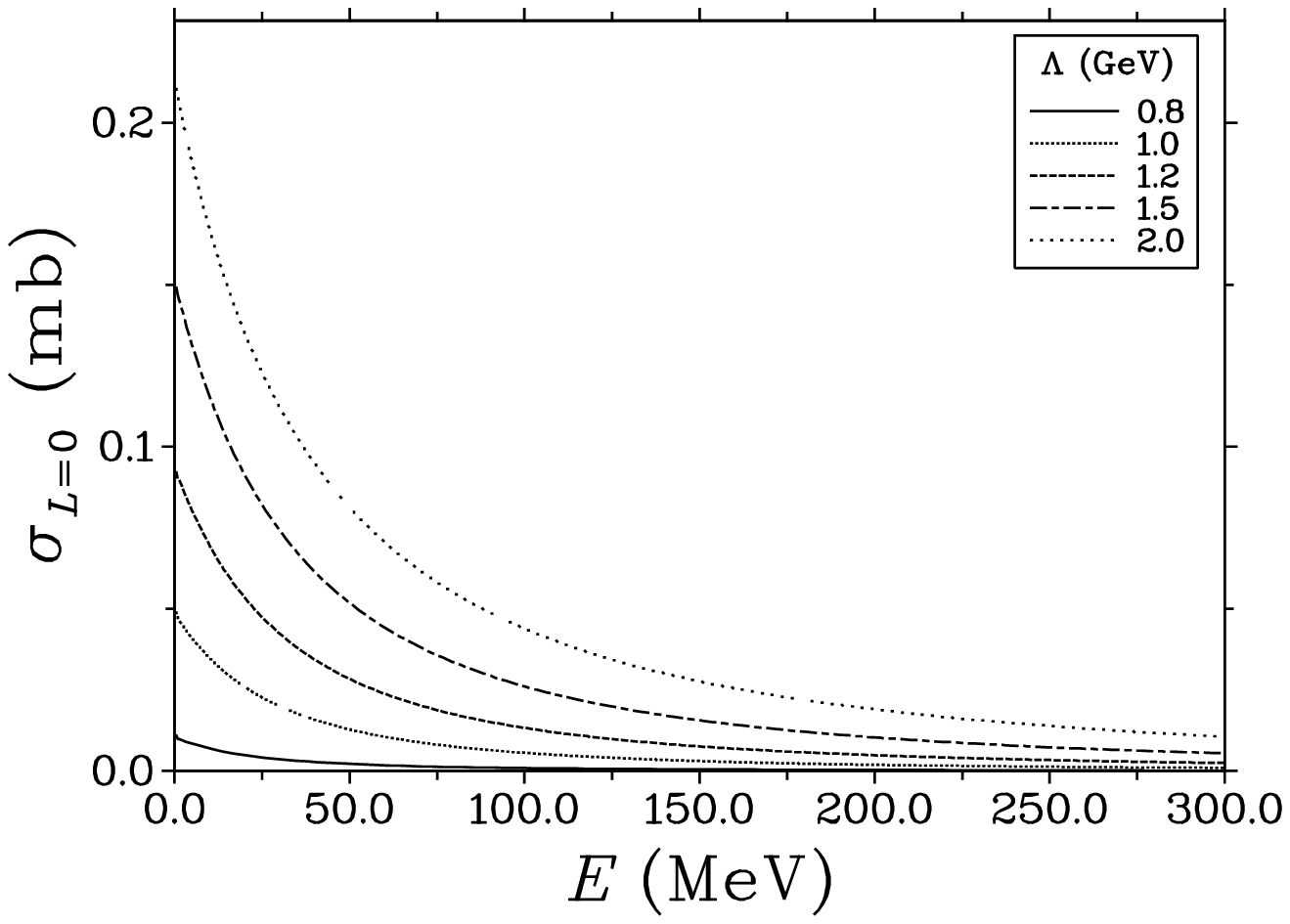}}&
\scalebox{0.6}{\includegraphics{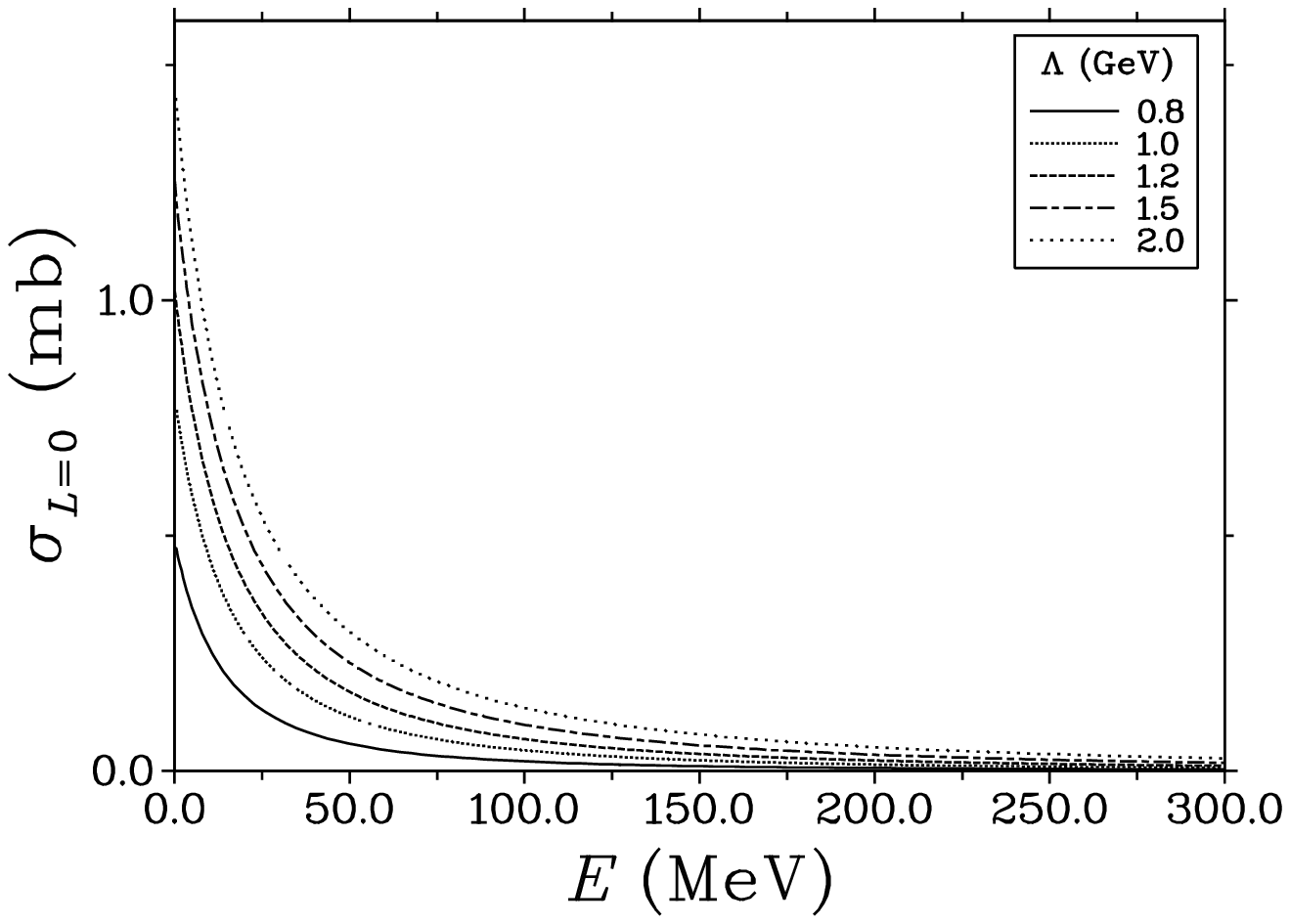}}\\
(a)&(b)\\
\scalebox{0.6}{\includegraphics{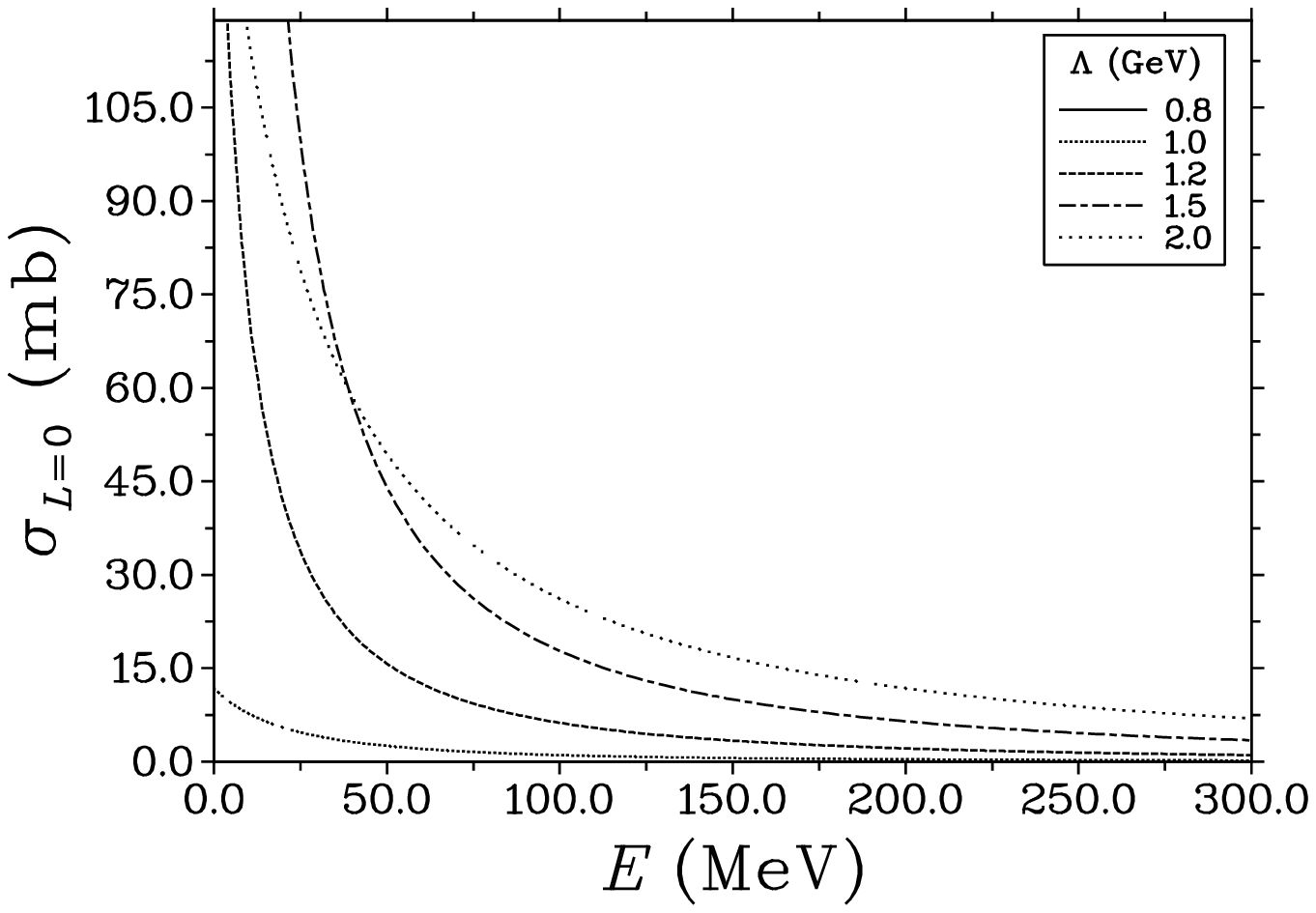}}&
\scalebox{0.6}{\includegraphics{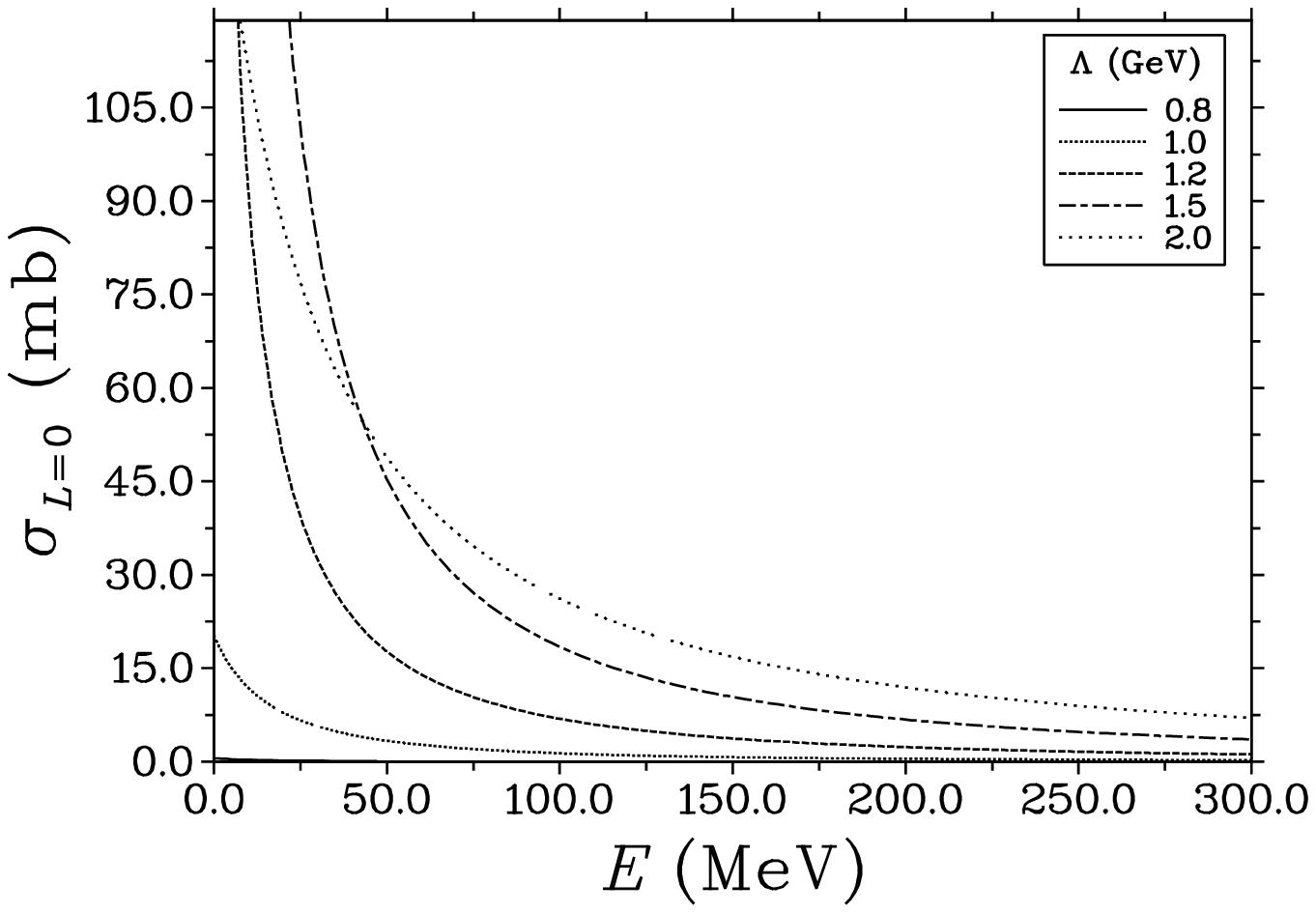}}\\
(c)&(d)
\end{tabular}
\caption{The S- wave total cross sections for the $D\bar{D}$ scattering
with various parameters. The upper (lower) two diagrams correspond
to the cases without (with) vector meson exchange contributions.
The left (right) two diagrams are obtained with $m_\sigma$=600
(400) MeV. The cutoff $\Lambda$ is in units of GeV. }\label{cs-DDbar-S}
\end{figure}

With the obtained phase shifts, it is not difficult to derive the
S- wave $D\bar{D}$ scattering lengths. We list them in Table
\ref{SDDbar}. The negative scattering lengths suggest that one
bound state exists, which can be understood from the diagrams (c)
and (d) in Fig. \ref{ph-DDbar-S}. The results for the case
$\Lambda\to\infty$ are also given in the table. The phase shift goes up from 180$^\circ$ in the case of VC, so the derived scattering length is positive.

We also revisit the bound state problem which was studied in Ref.
\cite{mole-pos}. The binding energies are summarized in Table
\ref{EB}.

It is straightforward to get the S- wave total cross sections from the
phase shifts. We present them in Fig. \ref{cs-DDbar-S}. One may
check the values at the threshold with the formula
$\sigma_{L=0}=4\pi a_0^2$. This explains why the cross section at
threshold with $\Lambda=2.0$ GeV is smaller than that with
$\Lambda=1.2$ (or 1.5) GeV in the case of VC.

From the above results, one concludes that an S- wave $D\bar{D}$
bound state may exist when the short range attraction is strong. Unfortunately the behavior of the short range
interaction is not completely understood. Whether the vector meson exchange
interaction is important, or equivalently whether $\Lambda$ is large, needs further study. The determination of a reasonable range for this parameter is one major task in this framework. $\Lambda\to\infty$ is not a realistic case because the binding energy around 1 GeV is too large to be explained by the $\rho$ and $\omega$ meson exchanges. In fact, if this is the case, the $D\bar{D}$ bound state is so compact that it should be represented by a 4-quark system, and it is not consistent with the $D\bar{D}$ molecular state. In the cases of a finite cutoff, a value larger than 2.0 GeV will lead to the binding energy more than 30 MeV while that around 1.2 GeV does not result in a binding solution. However, it is difficult to identify the reasonable range without further information. The extraction of the scattering length from lattice QCD simulations or experimental measurements will be helpful. One will see that the P- wave $B\bar{B}$ production is another observable to constrain the range of $\Lambda$. We will come back to this point later.

\section{The P-wave $D\bar{D}$ system}\label{sec4}

\begin{figure}[htb]
\centering
\begin{tabular}{cc}
\scalebox{0.6}{\includegraphics{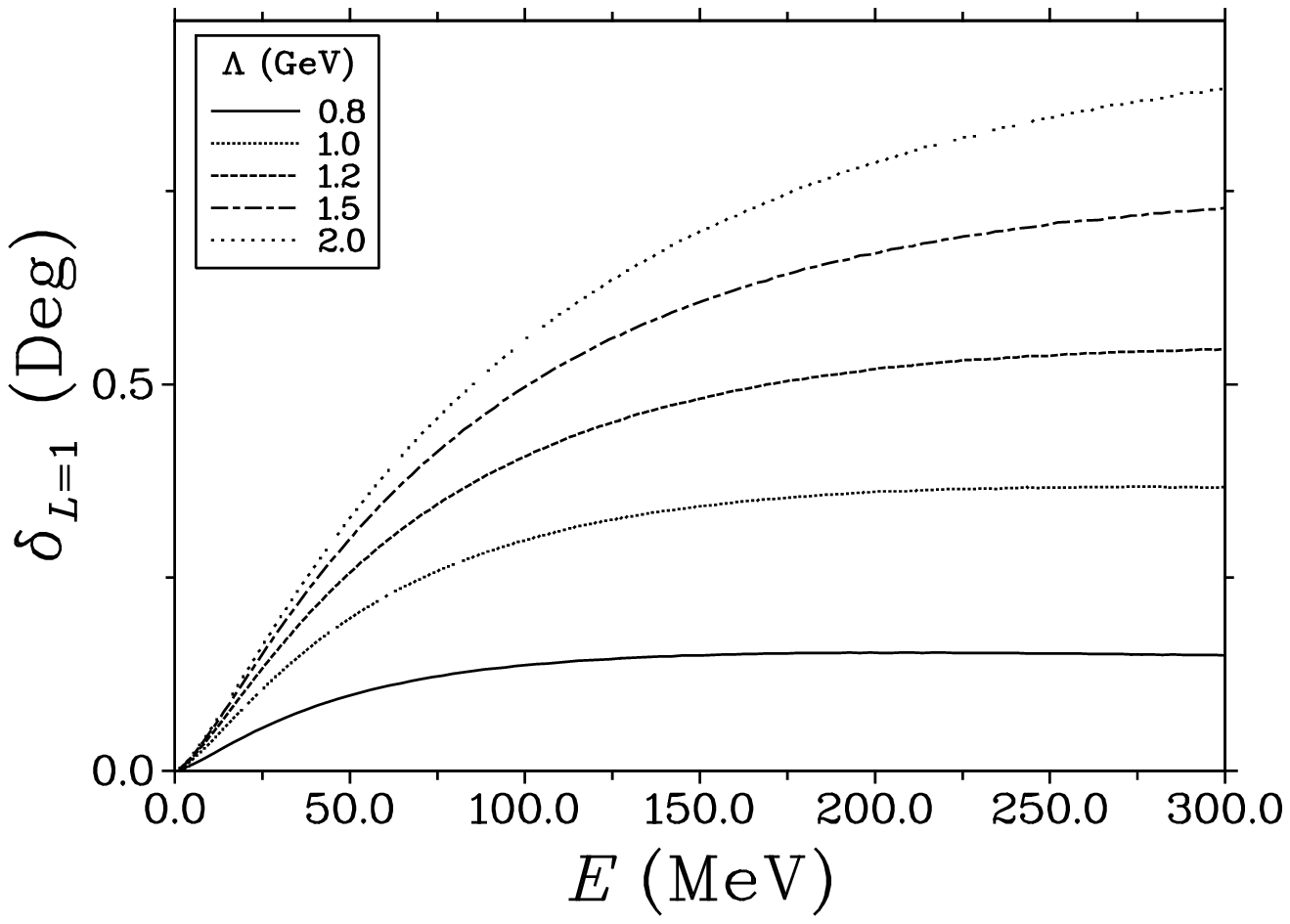}}&
\scalebox{0.6}{\includegraphics{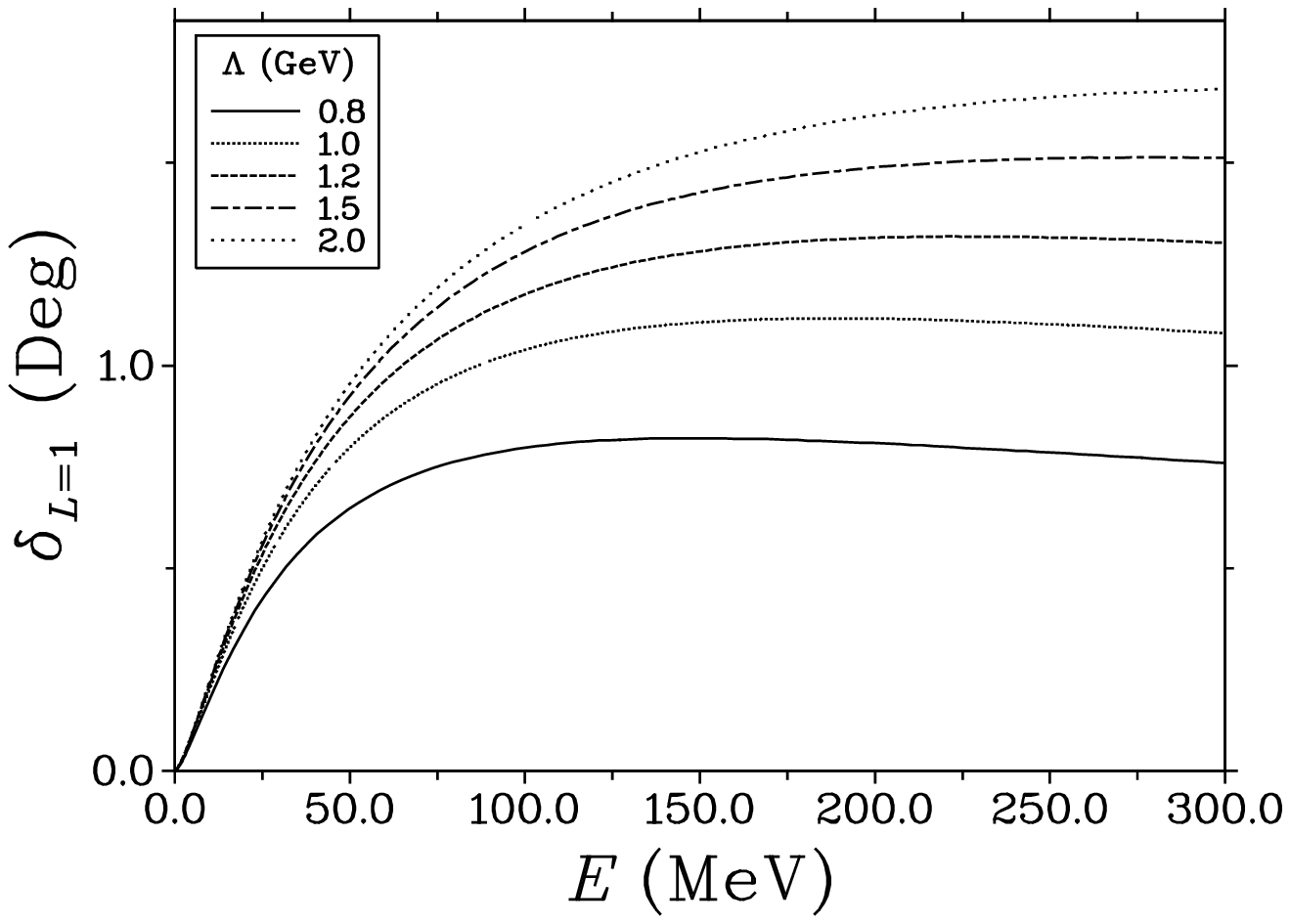}}\\
(a)&(b)\\
\scalebox{0.6}{\includegraphics{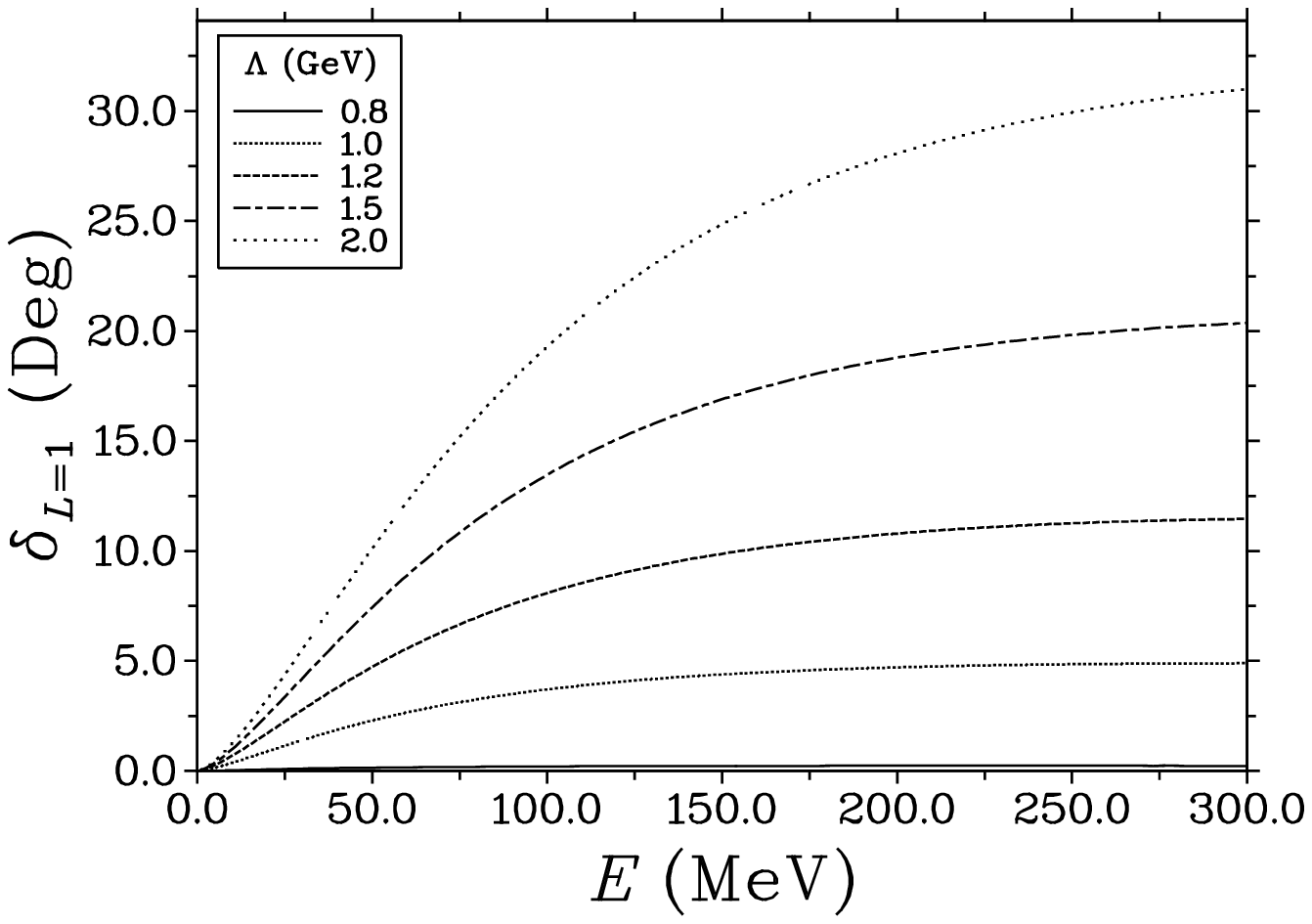}}&
\scalebox{0.6}{\includegraphics{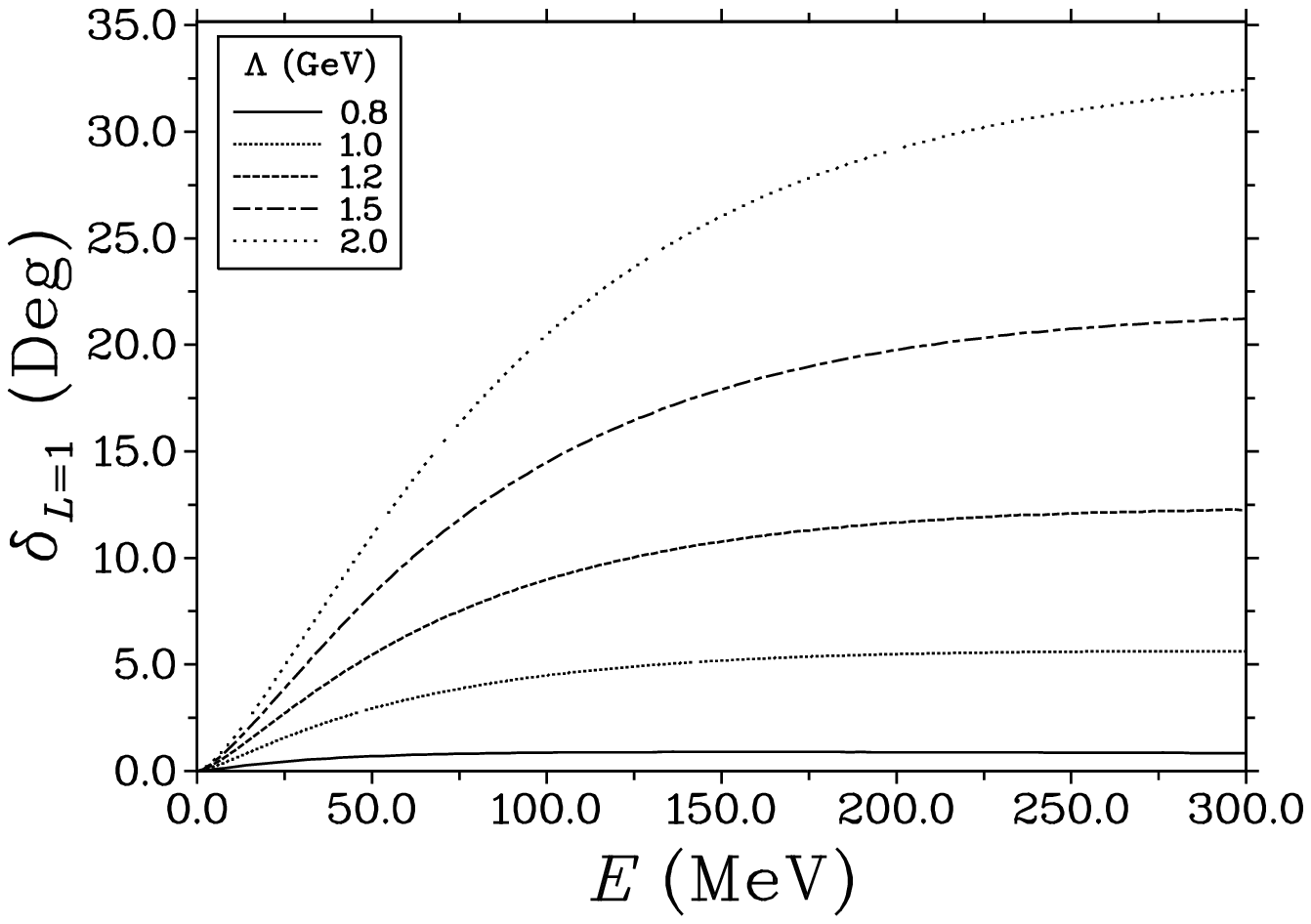}}\\
(c)&(d)
\end{tabular}
\caption{The phase shifts for the P- wave $D\bar{D}$ scattering
with various parameters. The upper (lower) two diagrams correspond
to the cases without (with) vector meson exchange contributions.
The left (right) two diagrams are obtained with $m_\sigma$=600
(400) MeV. The cutoff $\Lambda$ is in units of GeV.  }\label{ph-DDbar-P}
\end{figure}

\begin{table}[htb]
\centering \caption{The P- wave $D\bar{D}$ scattering volumes in
units of fm$^3$. NV (VC) indicates the contributions
from vector mesons are omitted (included).}\label{PDDbar}
\tabcolsep=9pt
\begin{tabular}{c|c|cccccc}\hline
$m_\sigma$  & Vector meson  &\multicolumn{6}{c}{$\Lambda$ (GeV)}\\
(MeV)&exchange& 0.8 & 1.0 & 1.2 & 1.5 & 2.0 & $\infty$\\\hline
600  & NV & 0.0014&0.0024&0.0029&0.0032&0.0033&0.0034\\
400  & NV &0.015&0.016&0.017&0.017&0.017 &0.017\\
600  & VC &0.0019 & 0.023&0.041&0.058&0.070&0.085\\
400  & VC &0.015&0.037&0.055&0.072&0.087&0.10\\
\hline
\end{tabular}
\end{table}

\begin{figure}
\centering
\begin{tabular}{cc}
\scalebox{0.6}{\includegraphics{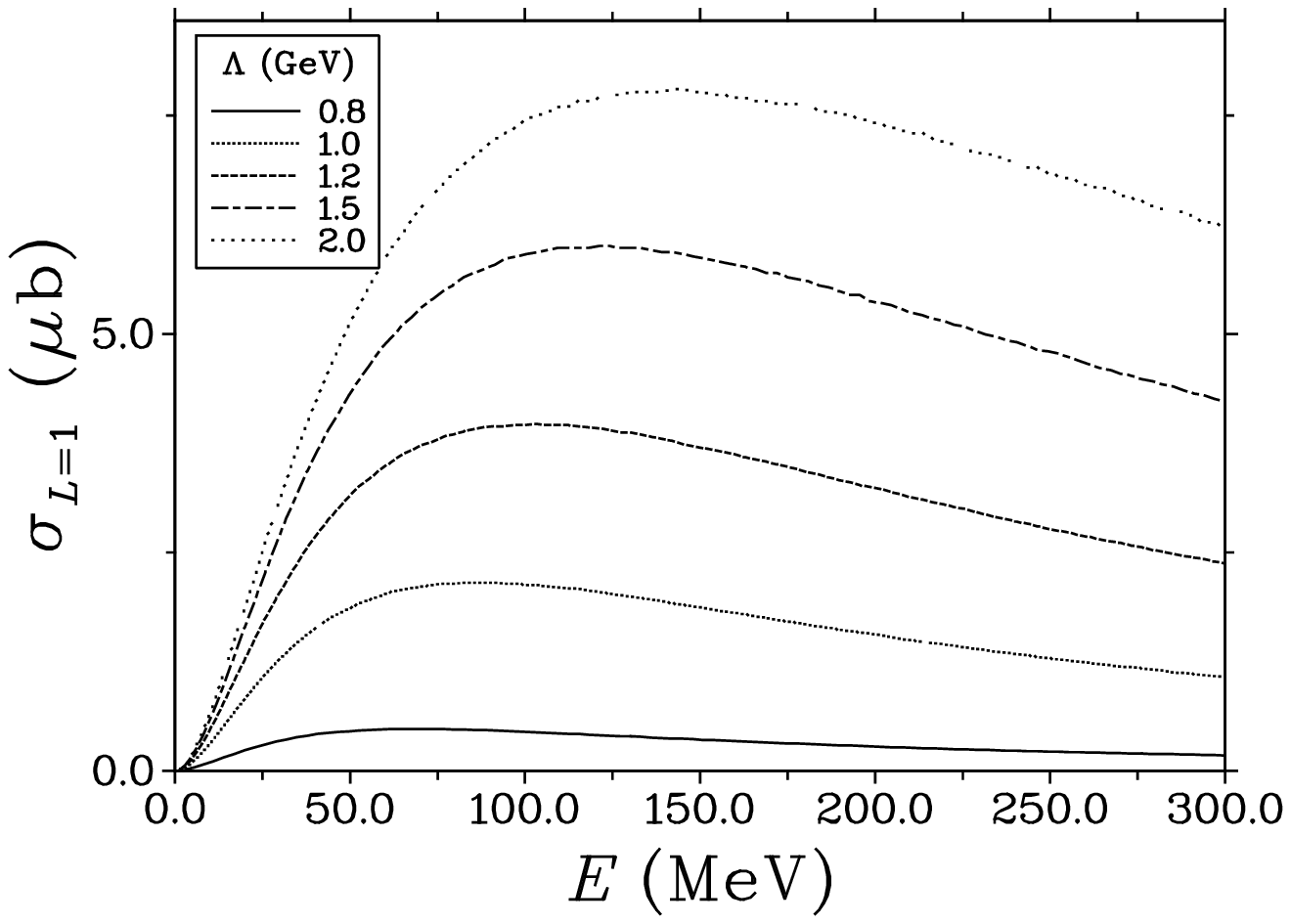}}&
\scalebox{0.6}{\includegraphics{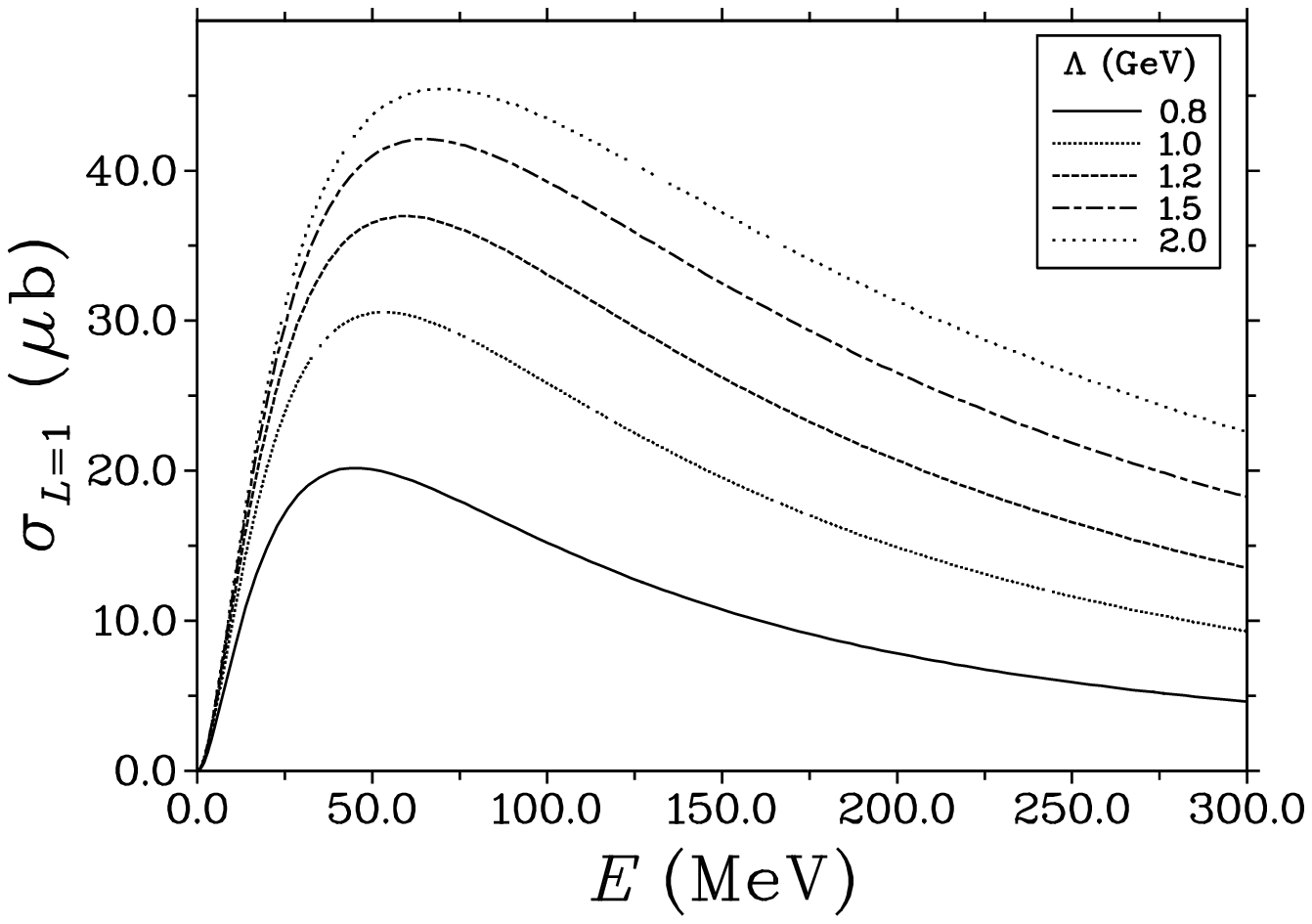}}\\
(a)&(b)\\
\scalebox{0.6}{\includegraphics{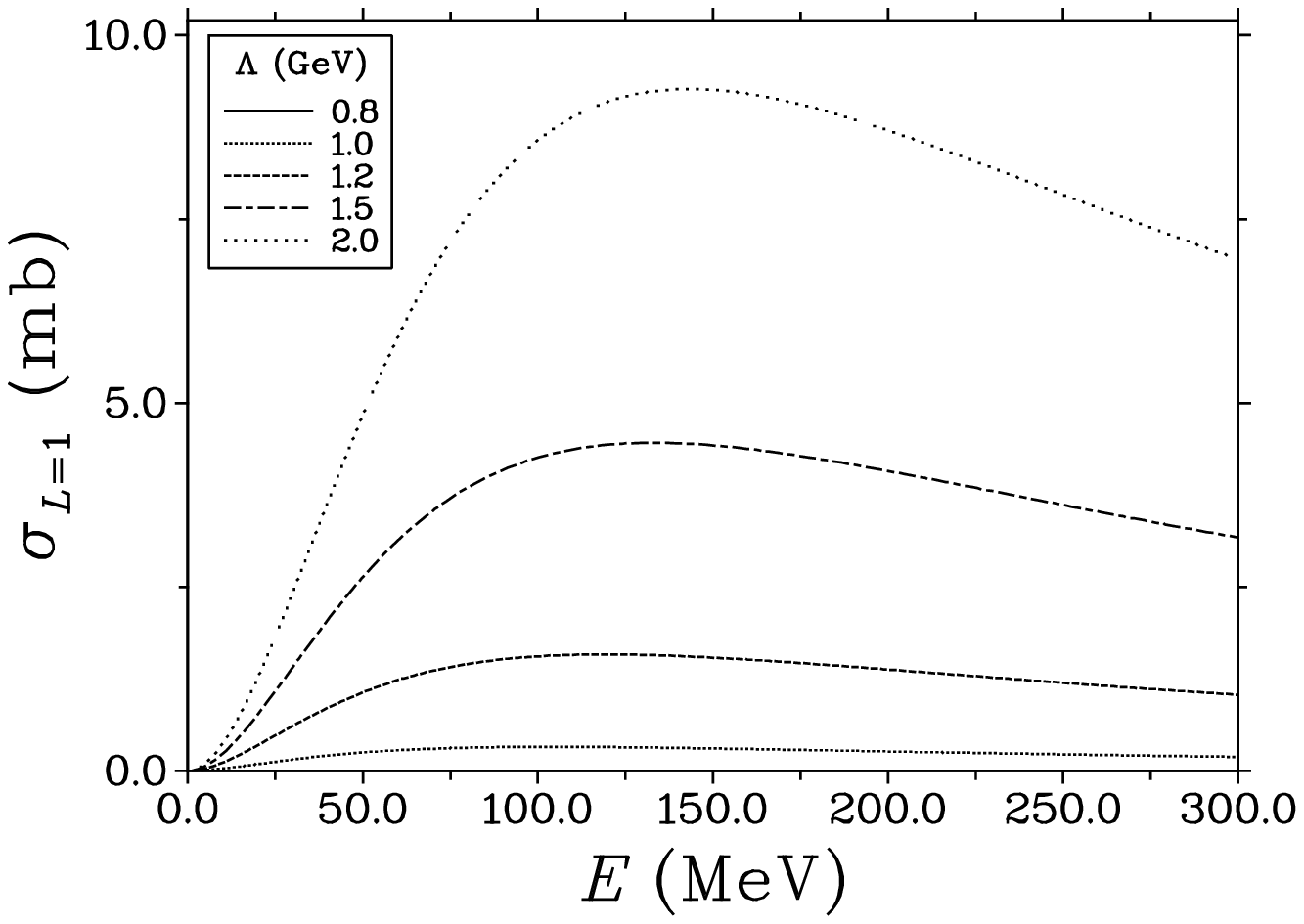}}&
\scalebox{0.6}{\includegraphics{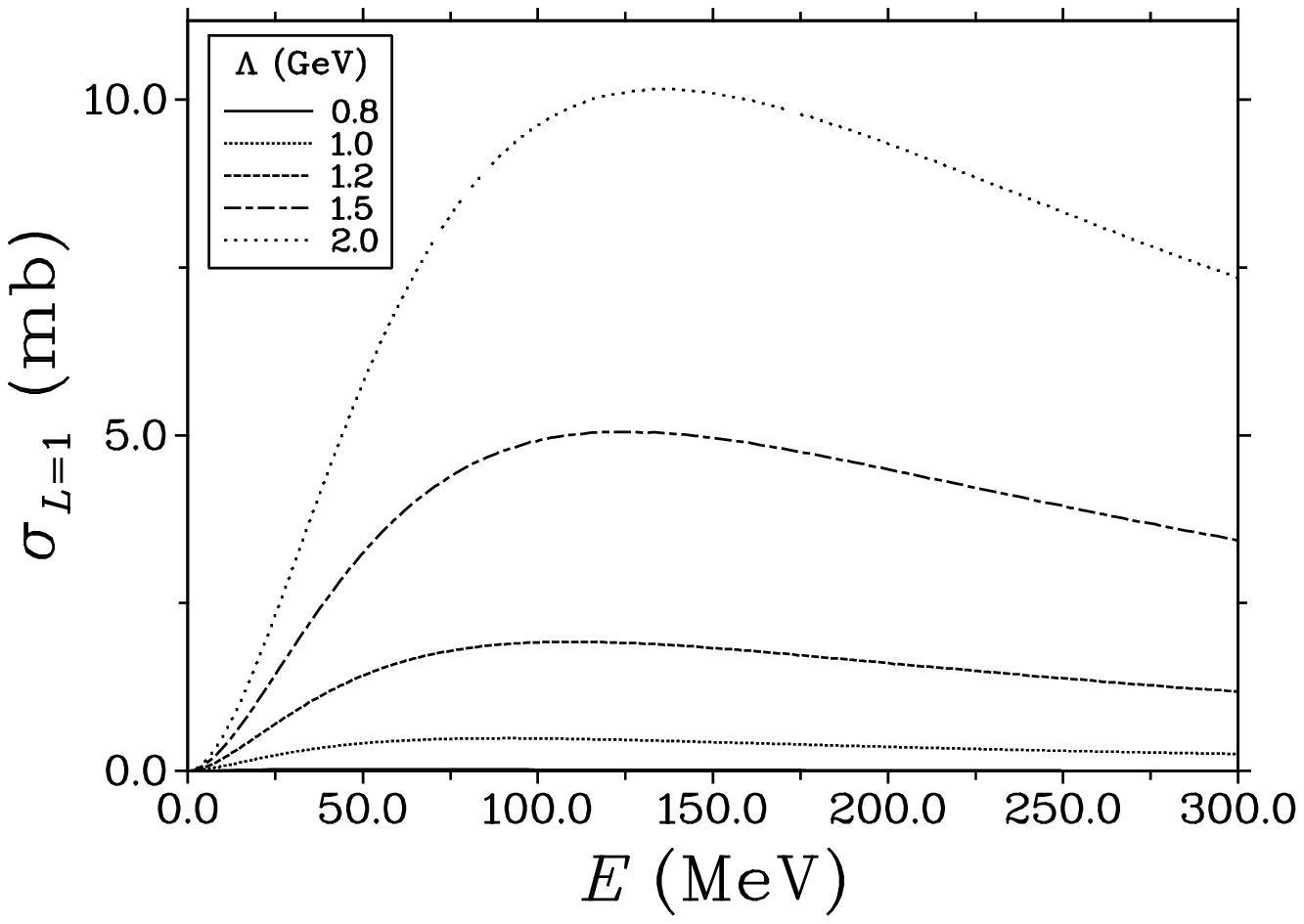}}\\
(c)&(d)
\end{tabular}
\caption{The P- wave total cross sections for the $D\bar{D}$ scattering
with various parameters. The upper (lower) two diagrams correspond
to the cases without (with) vector meson exchange contributions.
The left (right) two diagrams are obtained with $m_\sigma$=600
(400) MeV. The cutoff $\Lambda$ is in units of GeV.  }\label{cs-DDbar-P}
\end{figure}

The P-wave centrifugal barrier makes the interaction weaker than that in the
S-wave case. Actually, we do not find any binding solutions even
in the point particle limit ($\Lambda\to\infty$). In Fig. \ref{ph-DDbar-P}, we show the
phase shifts derived with different parameters. The diagrams
indicate that the attraction is not strong enough to form a bound
state nor a resonance. Similar to the S- wave case, one may get the
scattering volumes from the phase shifts according to the
definition Eq. (\ref{defthpar}). We present the numerical results
in Table \ref{PDDbar}. The stronger the attraction is, the larger
the scattering volume becomes. The obtained P- wave scattering
cross sections are shown in Fig. \ref{cs-DDbar-P}.

It is very interesting to note that there is a resonancelike
structure or bump in the cross section, although no bound state or
resonance pole (where the scattering phase shift crosses 90$^\circ$) exists in this channel. The structure appears when the
total energy of motion is around 40$\sim$150 MeV, depending on the
parameters. The P- wave interaction probably has effects on the
production of $\psi(3770)$ or $D\bar{D}$. We will explore this
issue in the following section.

To compare with the future experimental measurements of the
scattering cross sections, one has to sum up different partial
wave contributions. Here we have performed the calculation up to P
wave since the higher partial waves yield smaller contributions.
Because the S- wave cross section is much larger than the P- wave one, the total line shape of the cross sections $\sigma_{tot}=\sigma_{L=0}+\sigma_{L=1}$ is  similar to that of the S- wave case. That is, the resonancelike structure of
the P- wave interaction does not appear in the total cross section.

\section{The rescattering effect in $e^+e^-\to D\bar{D}$ production}\label{sec5}

\subsection{Formulation}

\begin{figure}[htb]
\includegraphics[scale=0.6]{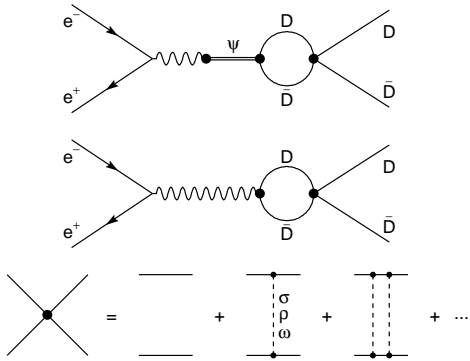}
\caption{The rescattering effects in the $D\bar{D}$ production
process.}\label{fig-fsi}
\end{figure}

The above P- wave structure motivates us to calculate the
$D\bar{D}$ production by including their rescattering effect,
which may be helpful to understand the anomalous line shapes
observed by the BES Collaboration
\cite{bes-anom-line,bes-DDbar-line}. The schematic diagram for
this effect in the process $e^+e^-\to D\bar{D}$ is plotted in Fig.
\ref{fig-fsi}. The $D\bar{D}$ pair comes mainly from $\psi(3770)$.
In a previous work \cite{3770lineshape}, the production near the
threshold has been studied by including the single loop
contributions of the intermediate $D\bar{D}$, $D\bar{D}^*+c.c.$,
and $D^*\bar{D}^*$. In this work, we would like to consider the
multiple rescattering effects using a nonrelativistic method.

One may find the detailed procedure to derive the formula in the Appendix A. Here, we only present the resultant production cross section $\sigma_{prod}=\sigma_1+\sigma_2$. The first part gives the cross section without FSI,
\begin{eqnarray}
\sigma_1&=&\frac{\pi}{3}\alpha_e^2 \frac{(s-4m_D^2)^{3/2}}{s^{5/2}}\times |{\rm f.f.}|^2,
\end{eqnarray}
where $\alpha_e=1/137$ and f.f. indicates the contributions from intermediate vector resonances and background fields
\cite{3770lineshape,mao-1,mao-2}. Since we here focus only on the
illustration of FSI effect in the following analysis, we consider
one resonance $\psi(3770)$ and take simply the form of f.f. from
Ref. \cite{mao-1},
\begin{eqnarray}\label{f.f.}
{\rm f.f.}=-F_{D\bar{D}}(s)+\frac{g_{\psi D\bar{D}}Q_c f_\psi
m_\psi}{s-m_\psi^2+i m_\psi\Gamma_T}e^{i\phi},
\end{eqnarray}
where the coupling $g_{\psi D\bar{D}}$ is defined through $\langle D(p_1)\bar{D}(p_2)|\psi(p,\lambda)\rangle=-i g_{\psi D\bar{D}}\epsilon^{(\lambda)}\cdot(p_1-p_2)(2\pi)^4\delta^4(p-p_1-p_2)$, $Q_c$ is the electric charge of the
charm quark, the decay constant $f_\psi$ of $\psi(3770)$ is given by $\langle 0|\bar{c}\gamma_\mu c|\psi(\lambda)\rangle=f_\psi m_\psi \epsilon_\mu^{(\lambda)}$, $m_\psi$ ($\Gamma_T$) is the mass (width) of $\psi(3770)$, $\phi$ is a relative phase, and
$F_{D\bar{D}}(s)$ is an effective form factor describing the
coupling of the virtual photon with the $D\bar{D}$ pair. Here, we
assume
\begin{eqnarray}
F_{D\bar{D}}(s)=\frac{m_\psi^2 F_0}{s}
\end{eqnarray}
with $F_0$ as an adjustable constant \cite{mao-1}. One derives $g_{\psi D\bar{D}}$ from the
branching ratio of the strong decay and $f_\psi$ from the leptonic
decay width $\Gamma_{ee}$. Note one should consistently consider the rescattering effect in determining $g_{\psi D\bar{D}}$ from the $\psi(3770)$ decay. Now the decay width also has two parts $\Gamma=\Gamma_1+\Gamma_2$. We have $\Gamma_2/\Gamma_1=\sigma_2/\sigma_1$ at the peak $\sqrt{s}=m_\psi$.

\begin{figure}
\centering
\begin{tabular}{cc}
\scalebox{0.5}{\includegraphics{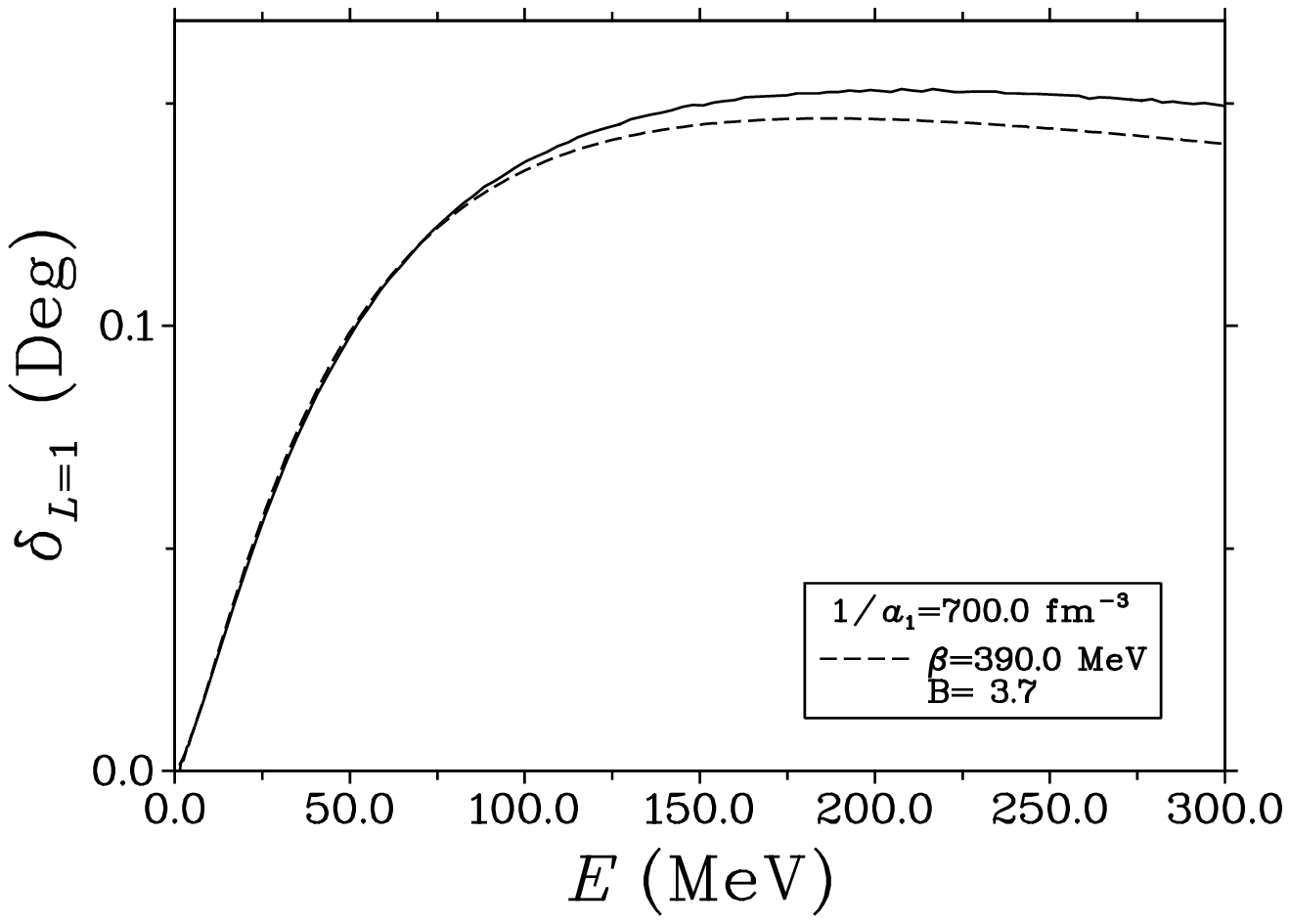}}&\scalebox{0.5}{\includegraphics{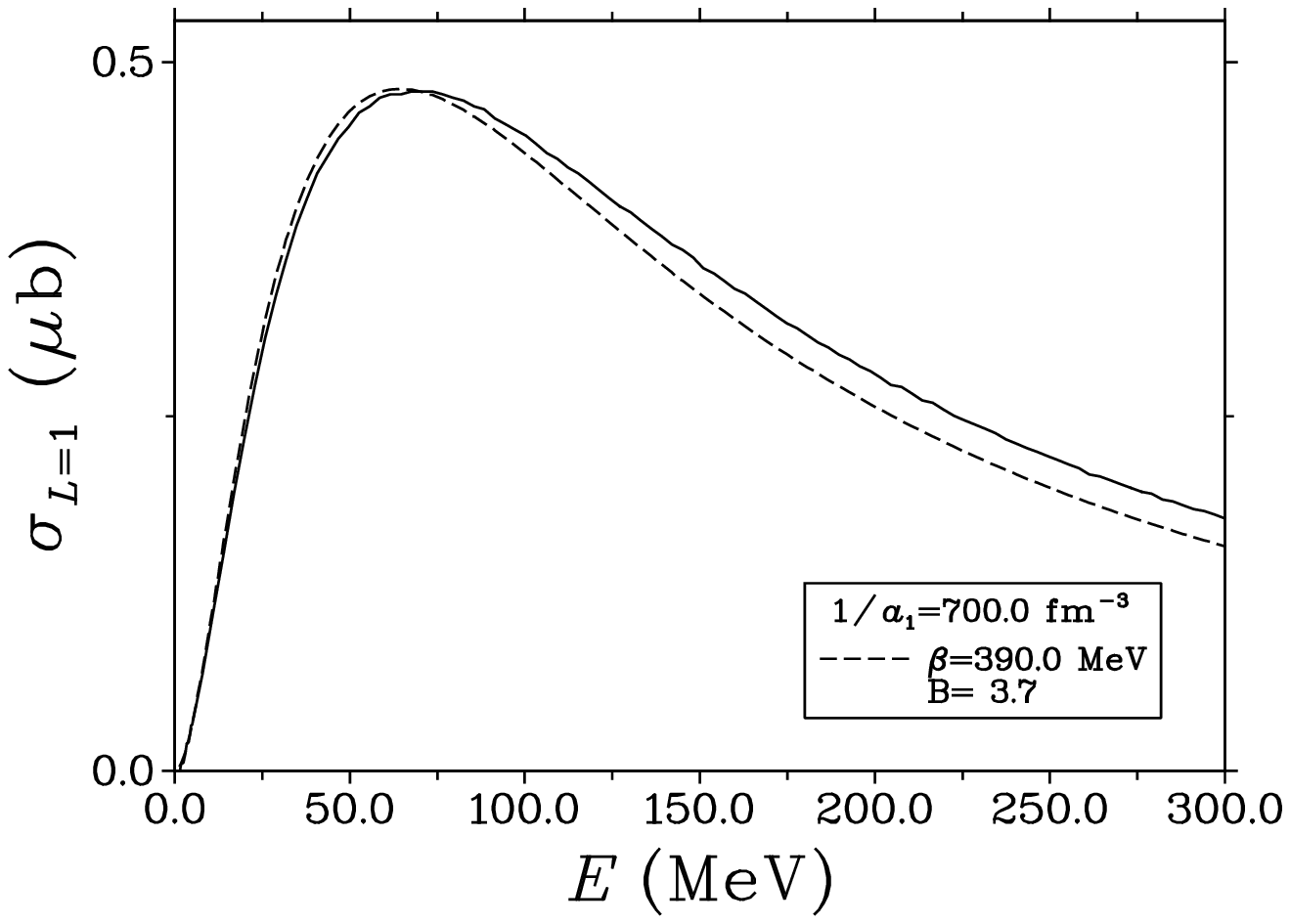}}\\
(a-1)&(a-2)\\
\scalebox{0.5}{\includegraphics{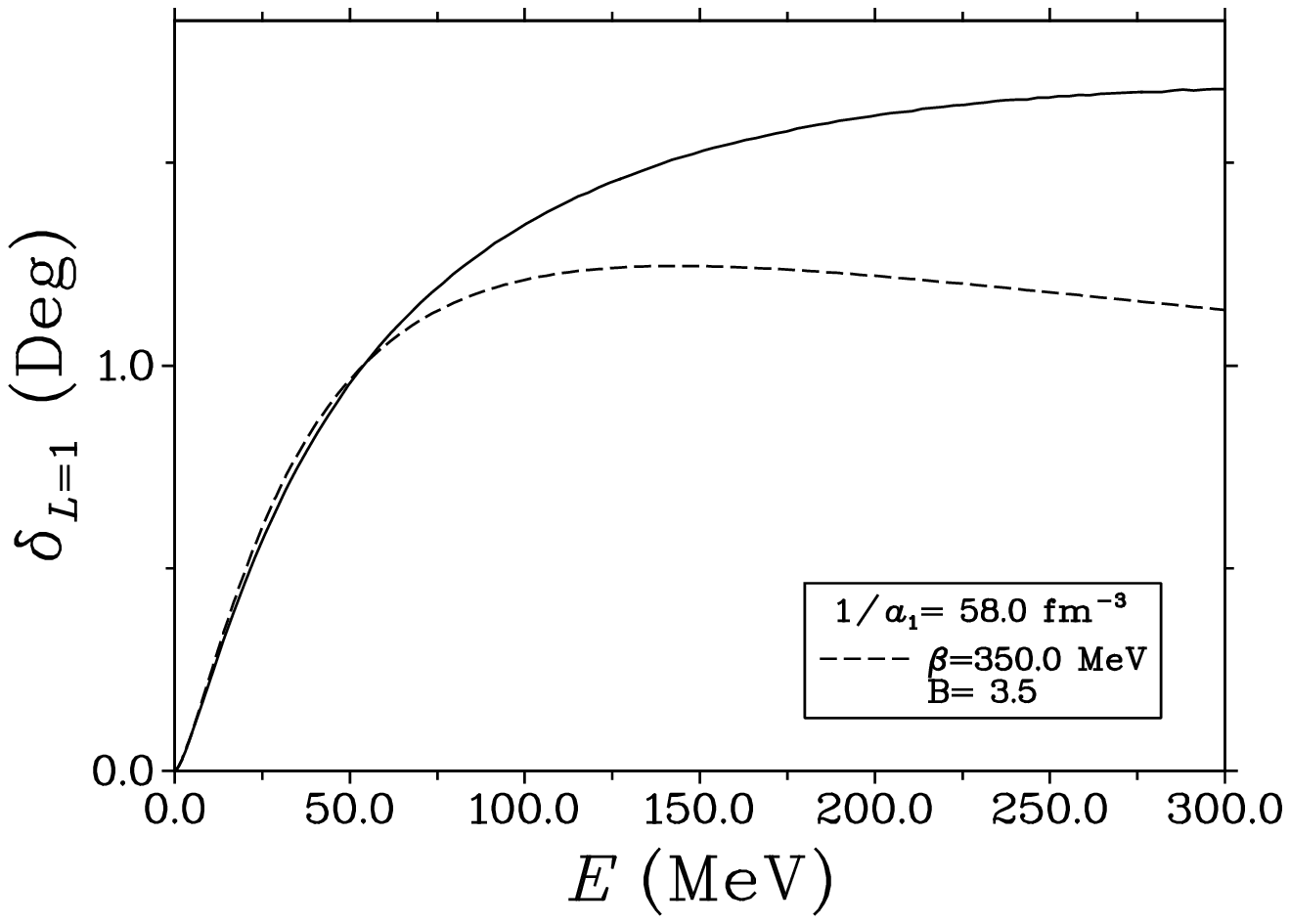}}&\scalebox{0.5}{\includegraphics{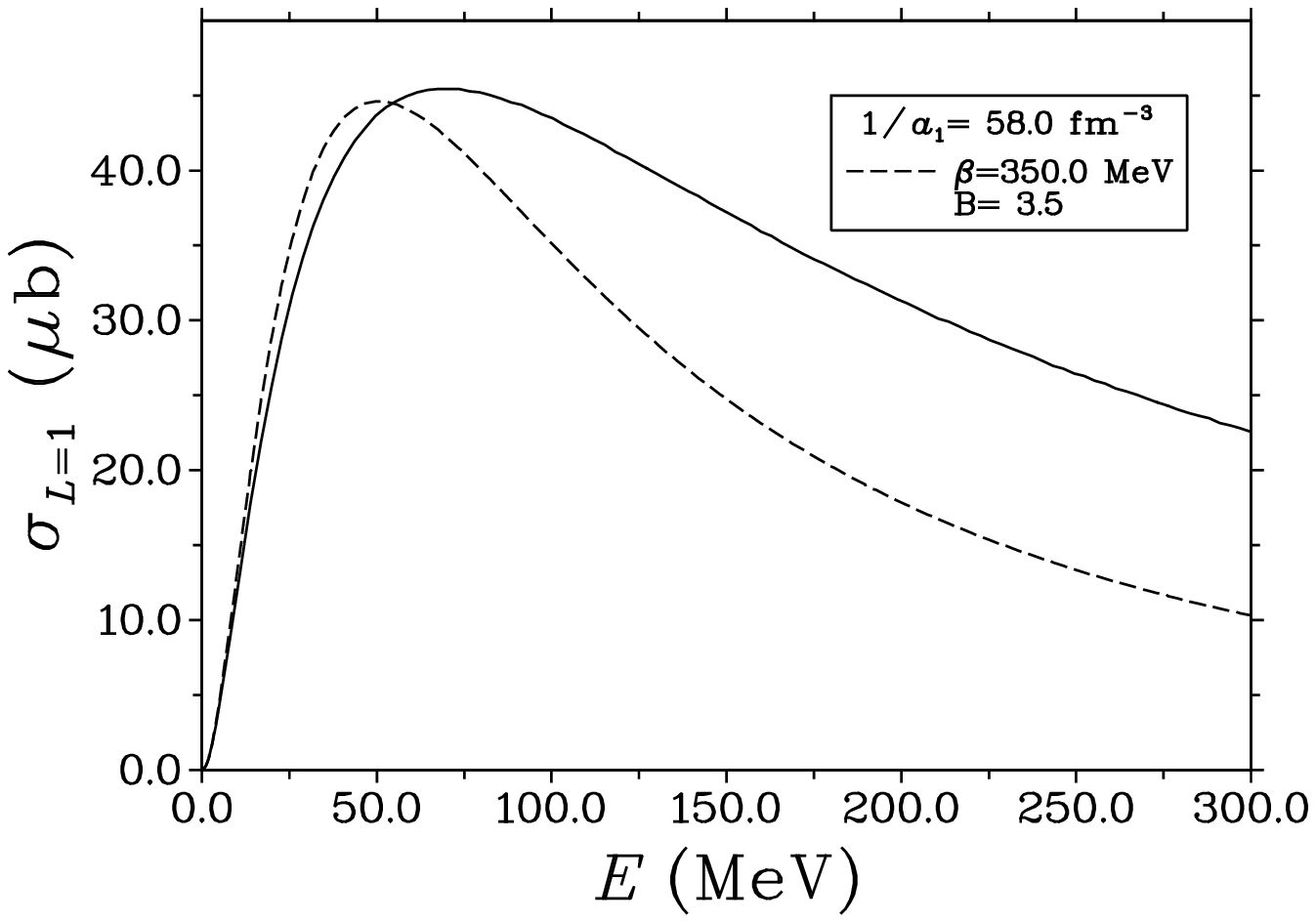}}\\
(b-1)&(b-2)\\
\scalebox{0.5}{\includegraphics{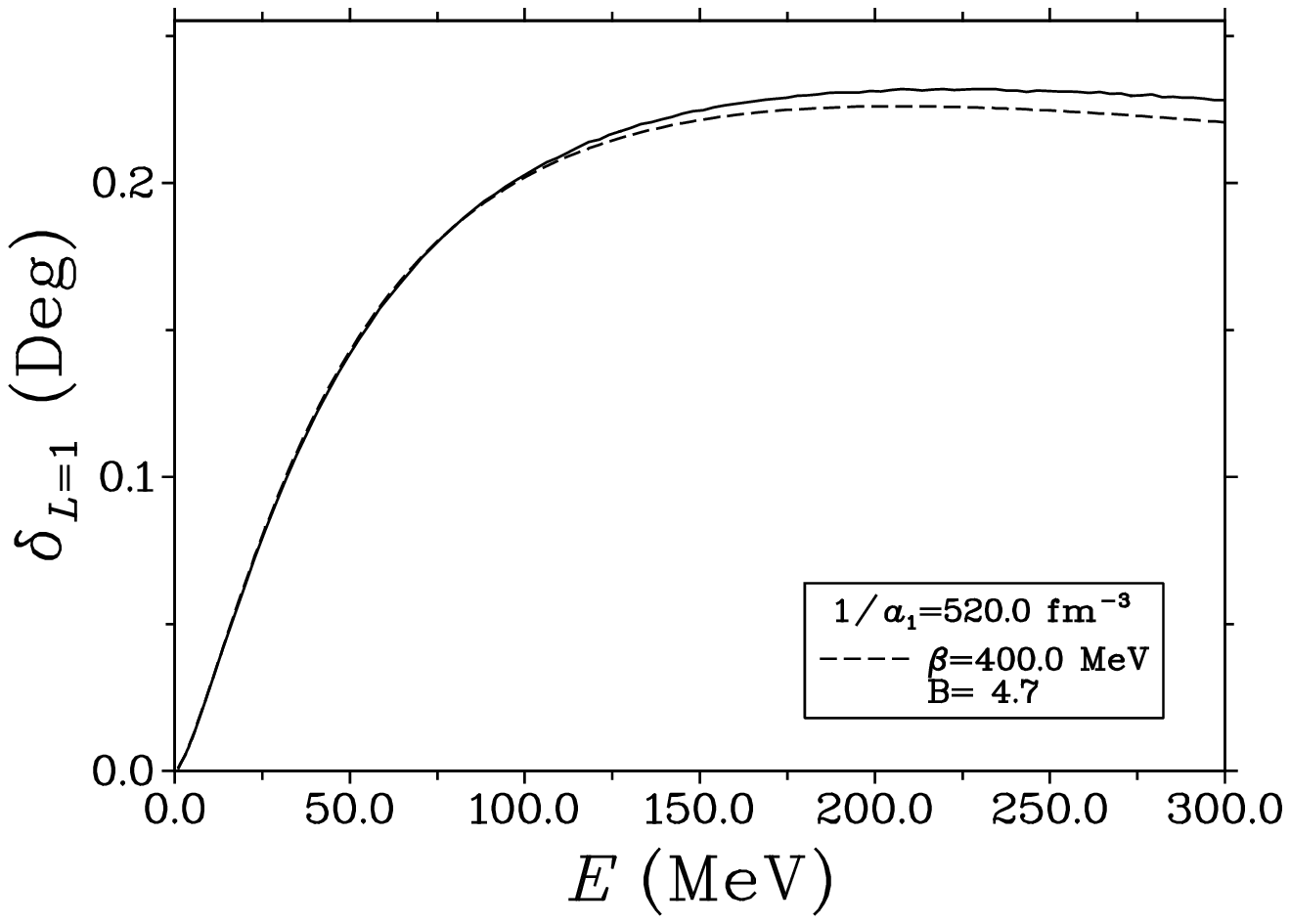}}&\scalebox{0.5}{\includegraphics{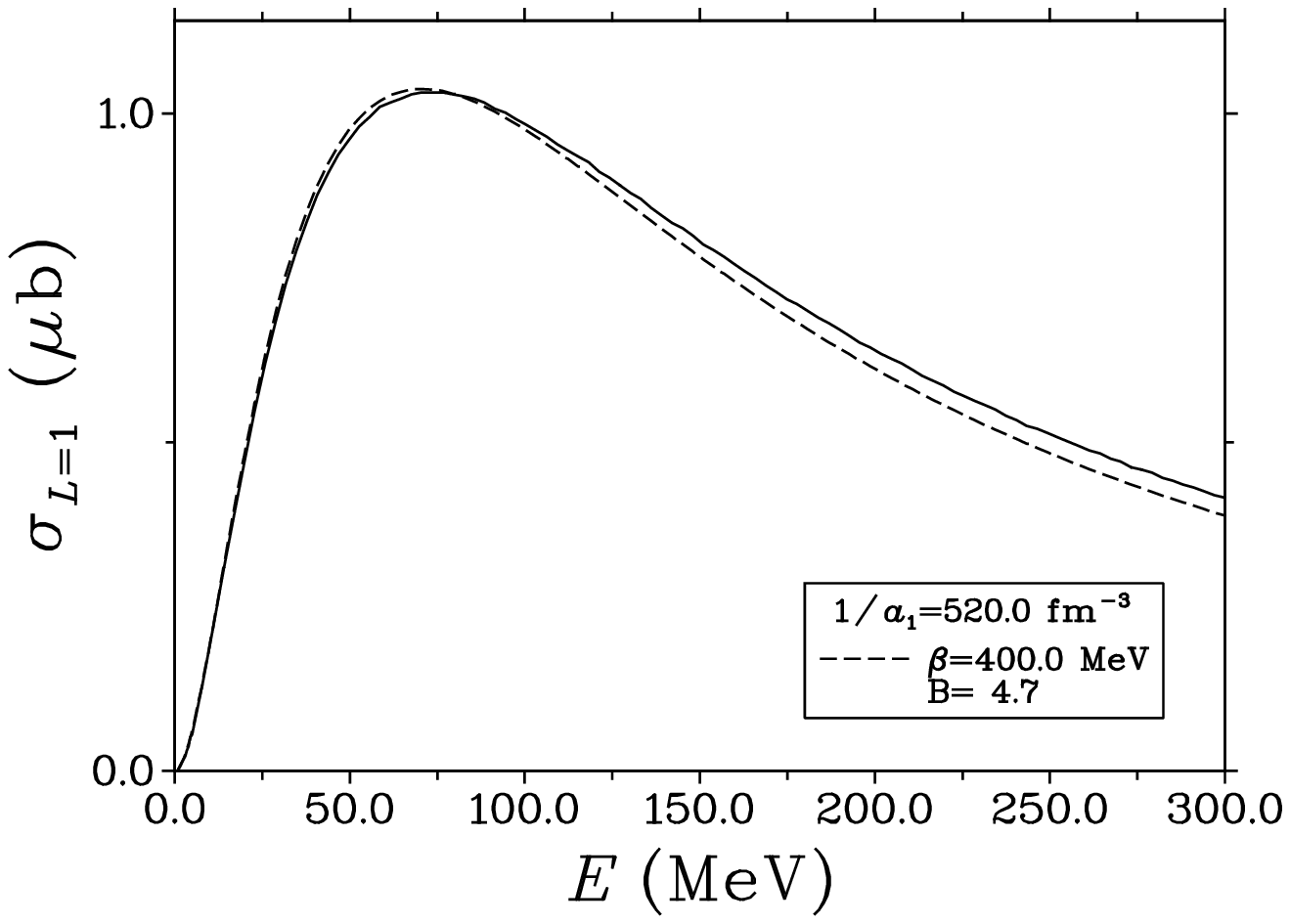}}\\
(c-1)&(c-2)\\
\scalebox{0.5}{\includegraphics{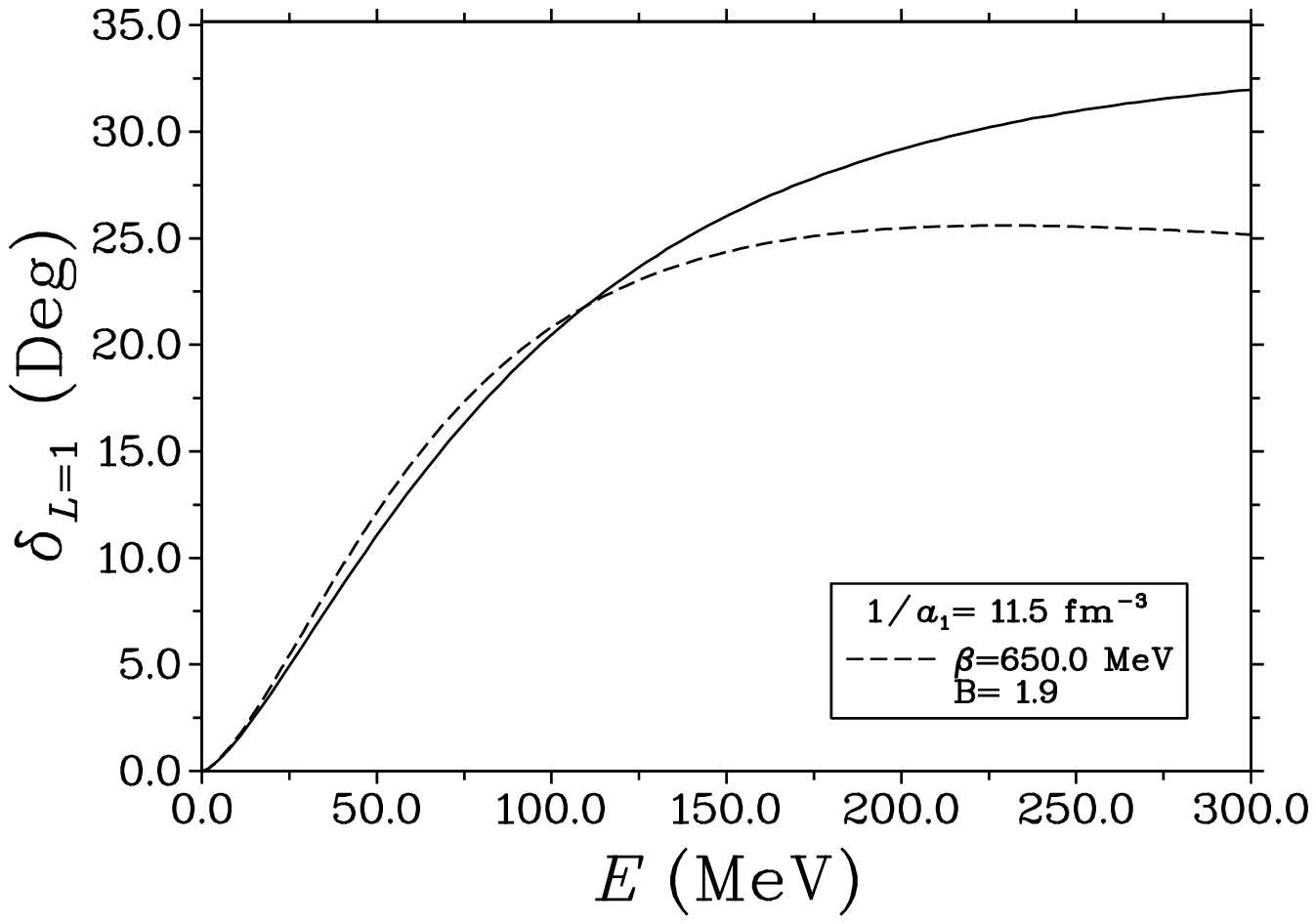}}&\scalebox{0.5}{\includegraphics{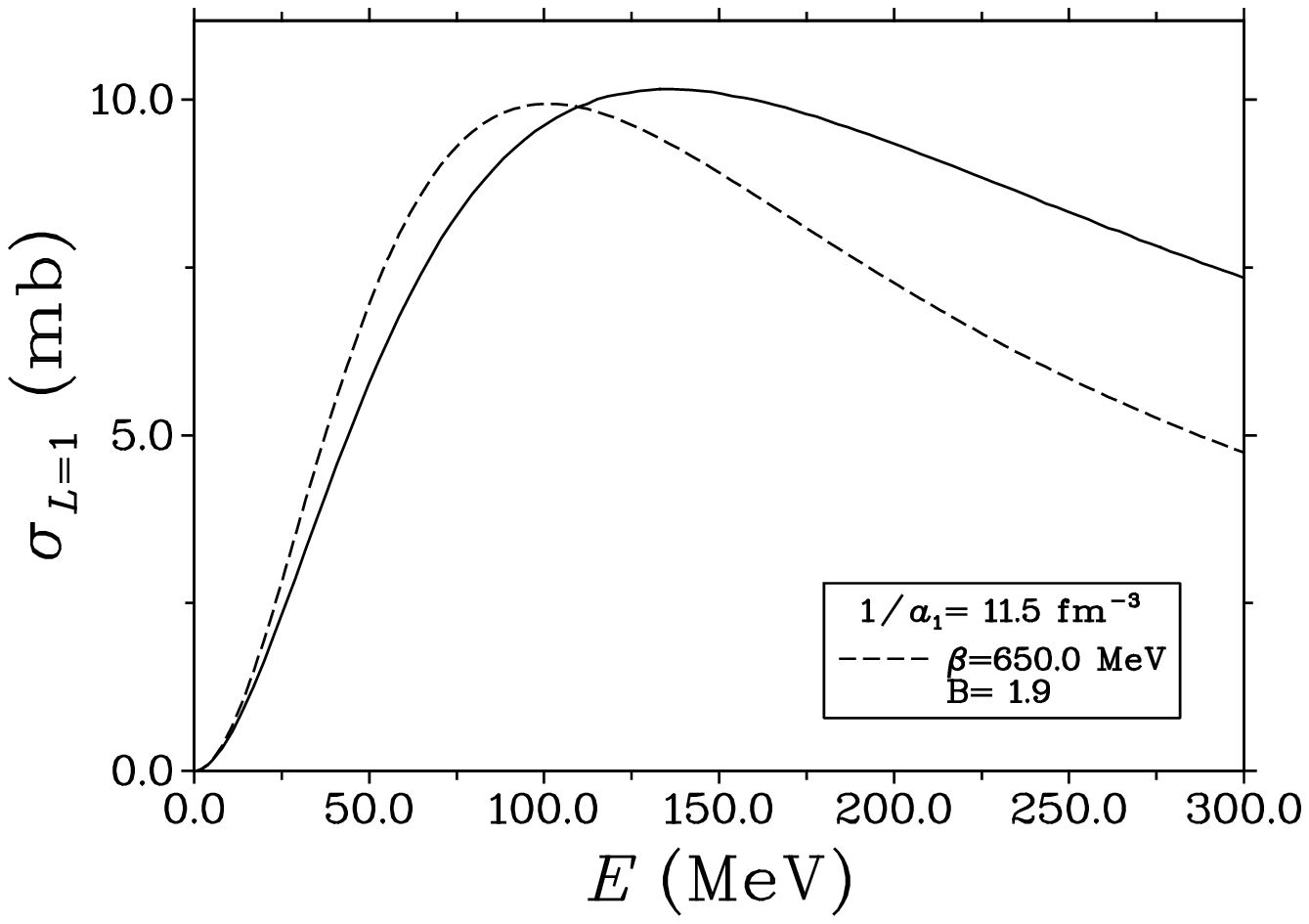}}\\
(d-1)&(d-2)
\end{tabular}
\caption{The reproduced P- wave $D\bar{D}$ phase shifts $\delta_{L=1}$
and cross sections $\sigma_{L=1}$ in the four cases, (a):
$1/a_1$=700.0 fm$^{-3}$, (b): $1/a_1$=58.0 fm$^{-3}$, (c):
$1/a_1$=520.0 fm$^{-3}$, and (d): $1/a_1$=11.5 fm$^{-3}$,
respectively. The solid line denotes the results calculated in the
former section.}\label{fit-ph-cs}
\end{figure}

The second part reflects the rescattering effect
\begin{eqnarray}
\sigma_2&=&\frac{64}{9}\frac{\pi^2\alpha_e^2\lambda
}{s^2m_D}\left\{\frac{2(ReU\times
ImU)W^r-[(ReU)^2-(ImU)^2]W^i}{{(W^r)}^2+ {(W^i)}^2}\right\}\times
|{\rm f.f.}|^2.
\end{eqnarray}
We have adopted
the Yamaguchi separable potential approximation \cite{separable} in deriving this formula and $\lambda$ is the coupling constant. This
approximation has recently been used to study the structure of
X(3872) \cite{taki}. In the above formula,
\begin{eqnarray}
ReU&=&-m_D^2\frac{\beta}{2\pi^2} {\cal P}\int_0^\infty d k\frac{k^4}{(k^2+2M^2)(k^2-\alpha^2)}t(k)F_p(k),\\
ImU&=&- m_D\frac{\beta\alpha^3}{4\pi\sqrt{s}}
t(\alpha)F_p(\alpha),\\
W^r&=&1-\frac{\lambda \beta^2}{6\pi^2} {\cal P}\int_0^\infty dk \frac{k^4}{k^2-\alpha^2}[t(k)]^2,\\
W^i&=&-\frac{\lambda \beta^2 \alpha^3}{12\pi}[t(\alpha)]^2,\\
t(k)&=&\frac{1}{(k^2+\beta^2)^2}+\frac{B}{\beta^2(k^2+\beta^2)},
\end{eqnarray}
where $\alpha^2\equiv m_D(\sqrt{s}-2m_D)$ and ${\cal P}$ means the
principal value integration. We will determine $\lambda$, $\beta$, and $B$ by reproducing the phase shifts and the scattering cross section
calculated in the previous section (see Appendix B). We have introduced a phenomenological form factor in the $D\bar{D}$ production vertex
\begin{eqnarray}
F_p(k)=\frac{\Lambda_p^2}{\Lambda_p^2+k^2},
\end{eqnarray}
where the cutoff $\Lambda_p$ around 1 GeV reflects the loop
contributions in considering rescattering effects. Here, $k$ means the virtuality and the coupling between $\psi(3770)$ and $D\bar{D}$ becomes small when the virtuality of the $D$ meson is large. This form factor helps to derive a physically reasonable $g_{\psi D\bar{D}}$, which will be obtained with the following expression:
\begin{eqnarray}
Br(\psi(3770)\to D\bar{D})\cdot\Gamma_T&=&{g_{\psi D\bar{D}}}^2\left\{\frac{(m_\psi^2-4m_D^2)^{3/2}}{24\pi m_\psi^2}\right.\nonumber\\
&&\left.+\frac{8}{9}\frac{\lambda }{m_Dm_\psi}\left[\frac{2(ReU\times
ImU)W^r-[(ReU)^2-(ImU)^2]W^i}{{(W^r)}^2+ {(W^i)}^2}\right]_{\sqrt{s}=m_\psi}\right\}.
\end{eqnarray}

\subsection{Results}

\begin{figure}
\centering
\begin{tabular}{cc}
\scalebox{0.6}{\includegraphics{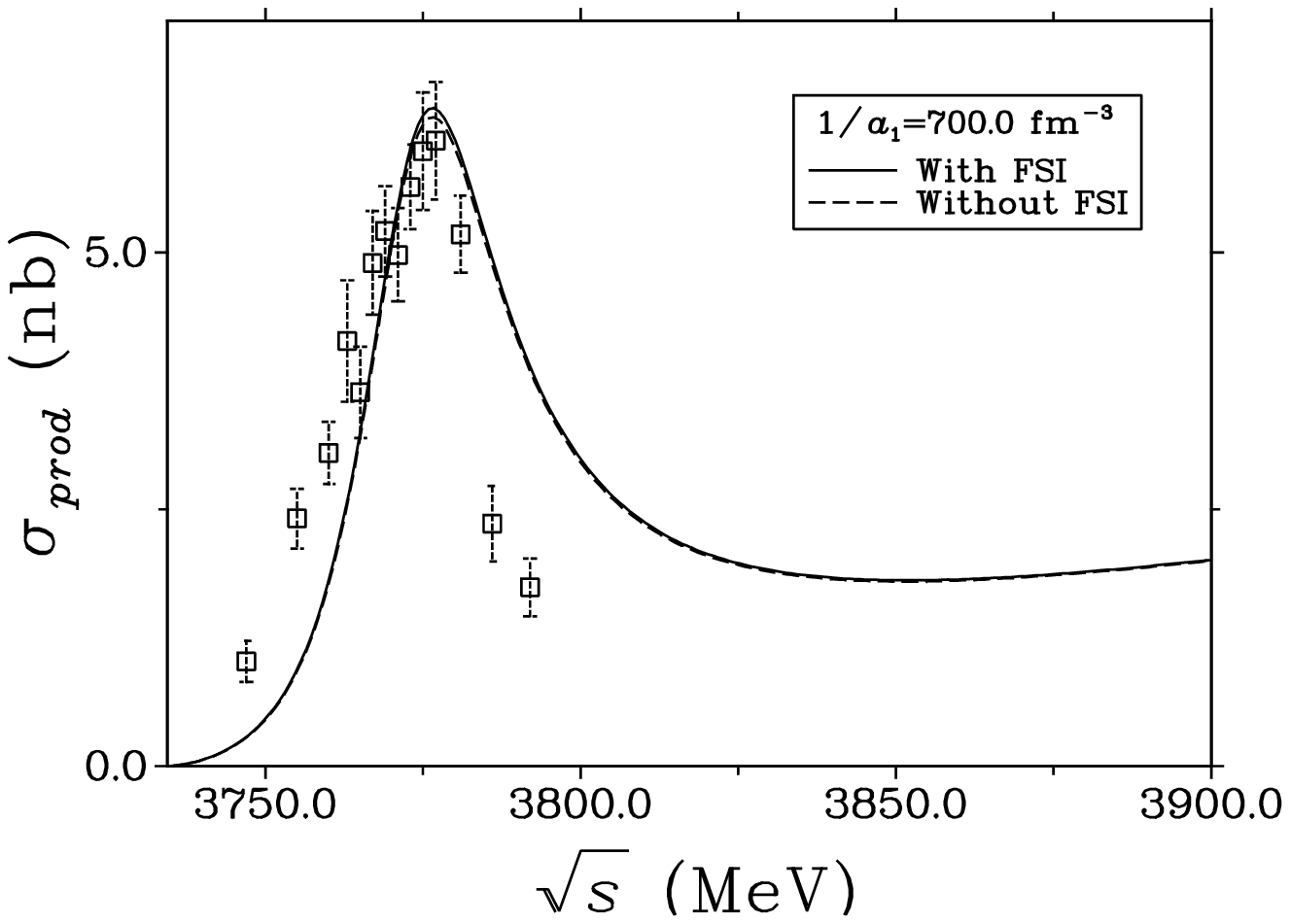}}&\scalebox{0.6}{\includegraphics{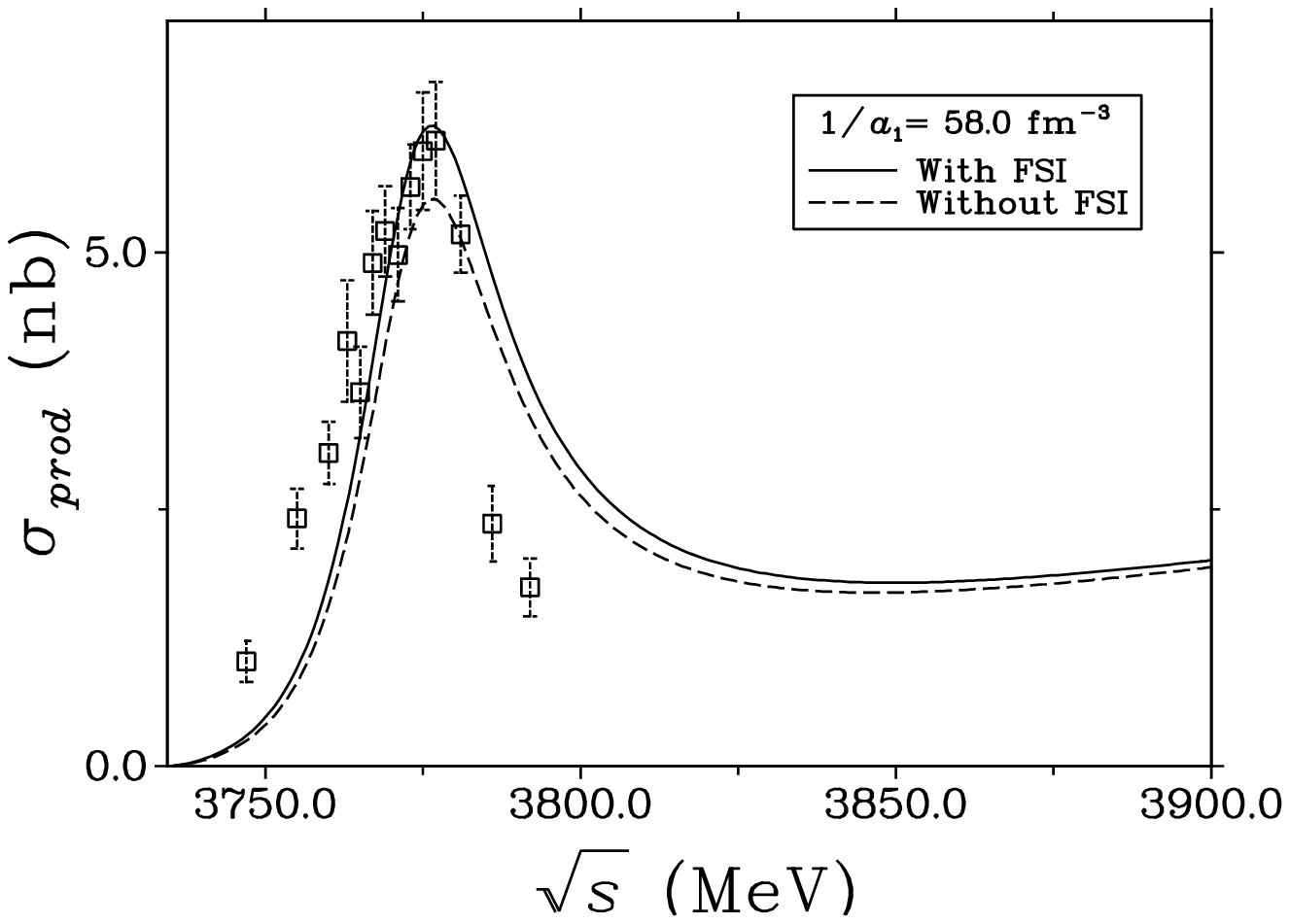}}\\
(a)&(b)\\
\scalebox{0.6}{\includegraphics{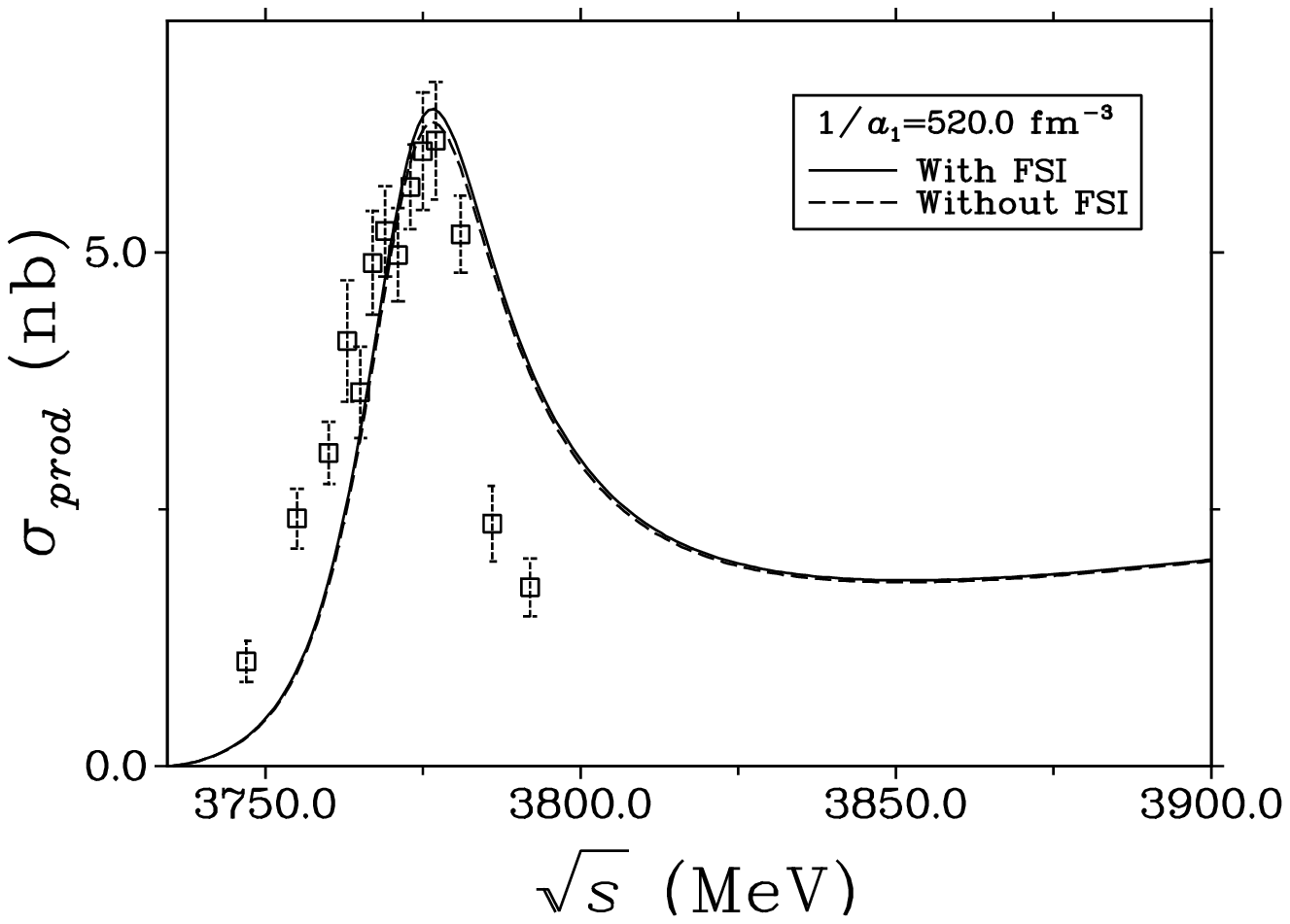}}&\scalebox{0.6}{\includegraphics{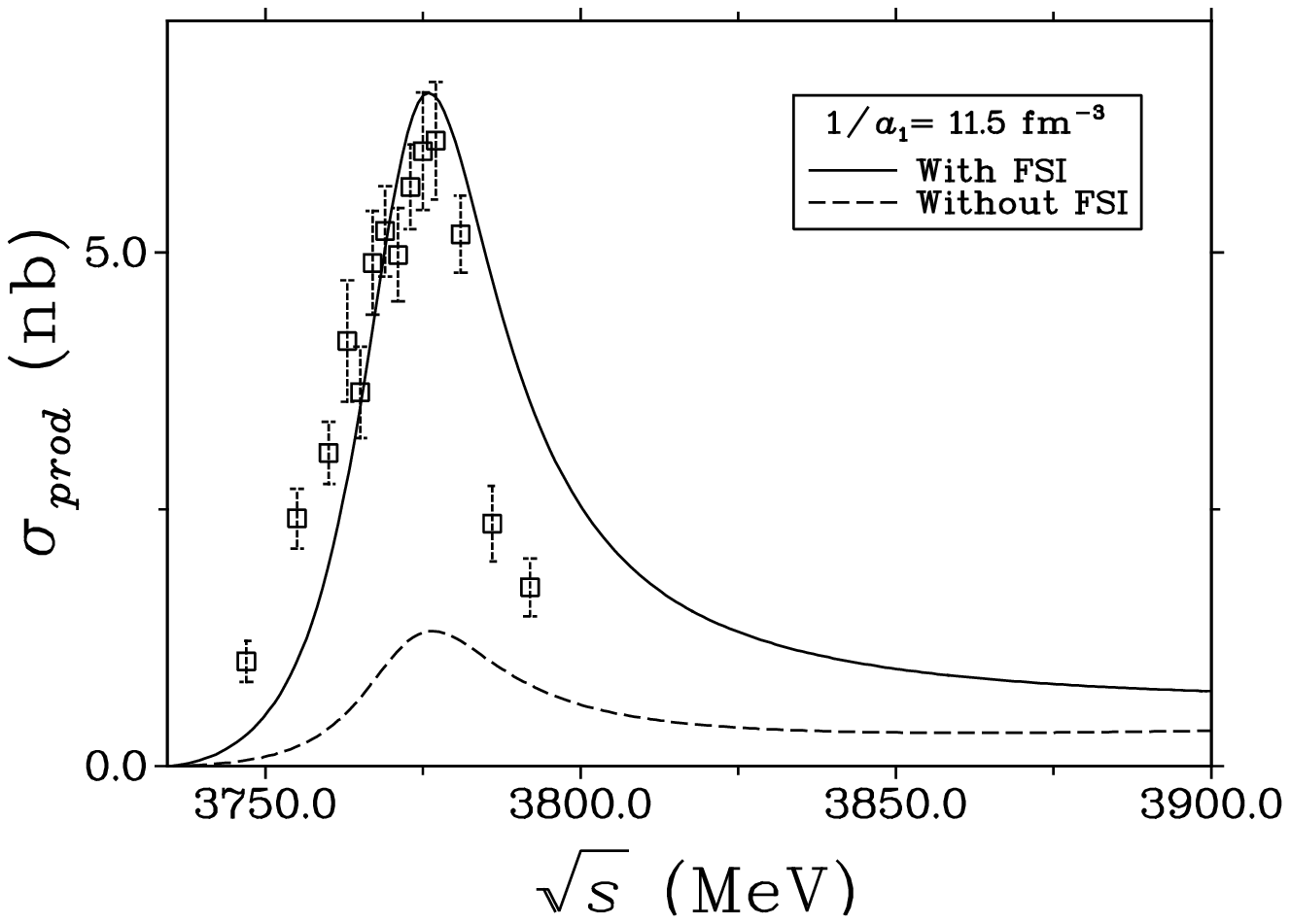}}\\
(c)&(d)
\end{tabular}
\caption{The obtained $D\bar{D}$ production cross sections
correspond to the four cases (a): $1/a_1$=700.0 fm$^{-3}$, (b):
$1/a_1$=58.0 fm$^{-3}$, (c): $1/a_1$=520.0 fm$^{-3}$, and (d):
$1/a_1$=11.5 fm$^{-3}$, respectively. We get the dash lines by ignoring the rescattering part $\sigma_2$ of the production cross section $\sigma_{prod}=\sigma_1+\sigma_2$. The experimental data are taken from Ref. \cite{bes-DDbar-line}.}\label{pr-cs}
\end{figure}

We choose four cases of parameters to compare with each other:
 (a) $m_\sigma=600$ MeV, $\Lambda=0.8$ GeV, and NV; (b) $m_\sigma=400$ MeV,
 $\Lambda=2.0$ GeV, and NV; (c) $m_\sigma=600$ MeV, $\Lambda=0.8$ GeV, and VC;
 and (d) $m_\sigma=400$ MeV, $\Lambda=2.0$ GeV, and VC. The first case has
 the weakest attraction, while the fourth one has the strongest attraction.

For the case (a), we have $1/a_1=700.0$ fm$^{-3}$. We found
($\beta$,$B$)=(370,8.5), (380,5.2), (390,3.7), (400,2.7), and
(410,2.2) [the unit of $\beta$ is MeV] can all roughly reproduce
the cross sections and the phase shifts. We take $\beta$=390 MeV,
$B$=3.7 as an example to illustrate the result. The reproduced
phase shifts and cross sections are presented in Fig.
\ref{fit-ph-cs}(a) and the calculated production cross sections
are given in Fig. \ref{pr-cs}(a). We also put the BES data \cite{bes-DDbar-line} in the diagram. When plotting
the latter diagram, we have used: $\Lambda_p$=1.0 GeV,
$m_\psi=3772.92$ MeV \cite{PDG}, $\Gamma_T=27.3$ MeV \cite{PDG}, $f_\psi=100.4$ MeV obtained from $\Gamma_{ee}=0.265$ keV \cite{PDG}, $g_{\psi D\bar{D}}=12.6$ from $B_r(\psi(3770)\to D\bar{D})=0.853$ \cite{PDG}, $F_0=5.0$ and $\phi=\pi/2$. One finds that the FSI contribution is small in this
case, and it certainly does not lead to an anomalous line shape.

For the case (b), we have $1/a_1=58.0$ fm$^{-3}$. The parameters
may be ($\beta$ (MeV),B)=(320,30), (330,10), (340,5.5), (350,3.5),
(360,2.7), (370,2.0), or (380,1.6).
As an example, we present the phase shifts and cross sections
corresponding to $\beta$=350 MeV, $B$=3.5 in Fig.
\ref{fit-ph-cs}(b) and the calculated production cross sections in
Fig. \ref{pr-cs}(b). Now the coupling constant becomes $g_{\psi D\bar{D}}=11.9$, while the other parameters are unchanged. The attraction is
stronger, but the line shape is still normal.

For the case (c), $1/a_1=520.0$ fm$^{-3}$. The values
($\beta$,B)=(370,50), (380,13), (390,7.0), (400,4.7), (410,3.5),
(420,2.6), (430,2.1), or (450,1.4) are acceptable where the unit
of $\beta$ is MeV. With $\beta$=400 MeV, $B$=4.7, the fitted phase
shifts and cross sections are plotted in Fig. \ref{fit-ph-cs}(c).
We present the calculated production cross sections in Fig. \ref{pr-cs}(c). In this case, we have the same coupling constant $g_{\psi D\bar{D}}=12.6$ as case (a). Because the contributions from the
vector meson exchange interactions are small,  the line shape is
also similar.

For the case (d), with $1/a_1=11.5$ fm$^{-3}$, one may use
($\beta$,B)=(550,60), (570,12), (590,5.8), (600,4.5), (650,1.9),
(700,1.0), or (800,0.2) to reproduce the phase shifts and the
cross sections. Figure \ref{fit-ph-cs}(d) shows an illustration with
$\beta=650$ MeV and B=1.9. We plot the corresponding production
cross section in the last diagram of Fig. \ref{pr-cs}. Now two parameters have been changed: $g_{\psi D\bar{D}}=5.6$ and $F_0=2.0$. For the
other combinations of $\beta$ and $B$, the magnitude changes
a little but it can be adjusted to the experimental data by varying $F_0$ and the phase angle $\phi$. Although the final state attraction is strong, the anomalous line shape does not appear even in this case.

\begin{figure}
\begin{tabular}{cc}
\scalebox{0.6}{\includegraphics{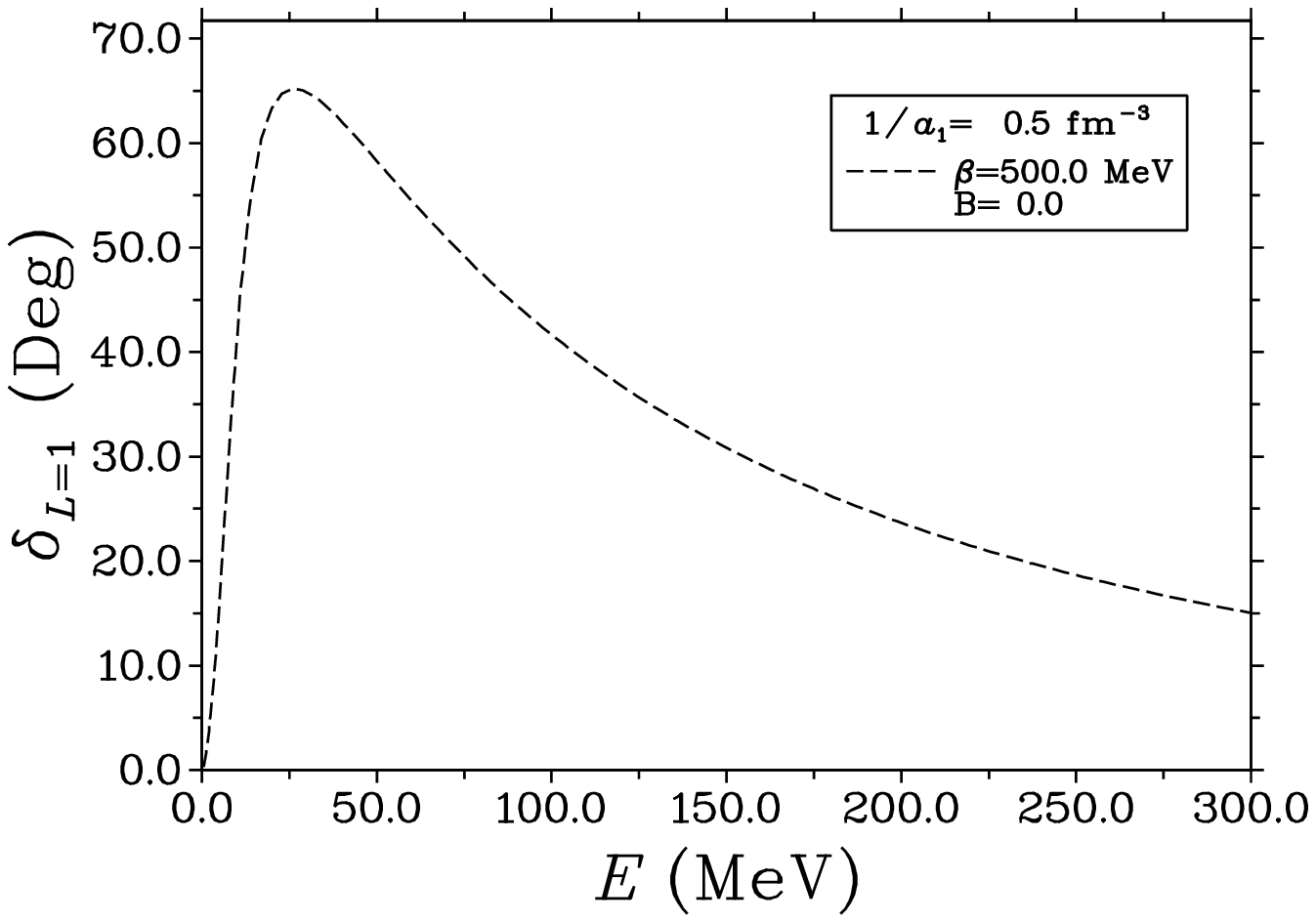}}&
\scalebox{0.6}{\includegraphics{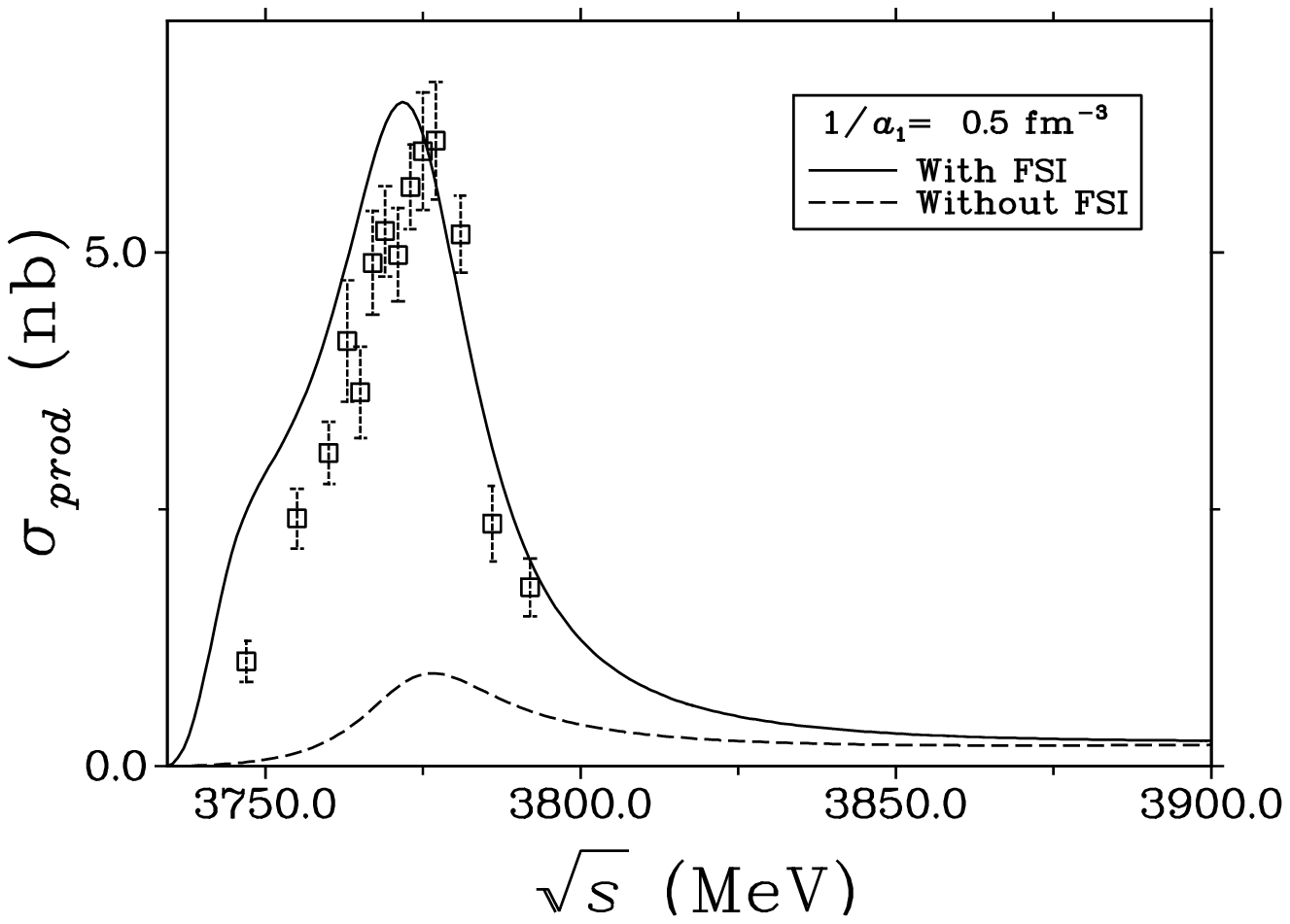}}\\
(a)&(b)
\end{tabular}
\caption{The P- wave $D\bar{D}$ phase shifts (a) and their
production cross section (b) in an extreme case with $1/a_1$=0.5
fm$^{-3}$, $\beta$=500 MeV and B=0. We get the dash line in the right diagram by ignoring the rescattering part $\sigma_2$ of the production cross section $\sigma_{prod}=\sigma_1+\sigma_2$. The experimental data are taken from Ref. \cite{bes-DDbar-line}.}\label{assumed}
\end{figure}

The results of these four cases (Fig. \ref{pr-cs}) tell us that the $D\bar{D}$ rescattering effects cannot change the
line shape of the production cross section. The reason is that the P- wave interaction is still not attractive enough. To see a cross section of a stronger attraction, let us arbitrarily consider an extreme case: $1/a_1=0.5$ fm$^{-3}$,  $\beta$=500.0 MeV, and B=0.0, where no bound state or resonance pole is formed, though a sharp rise of the P-wave phase shift is observed just above the threshold. The corresponding phase shifts and the resulting $D\bar{D}$ production cross sections are plotted in Fig. \ref{assumed}. We have used $g_{\psi D\bar{D}}=4.6$ and $F_0=1.5$. One finds that the anomalous line shape appears now. Also, the peak around 3770 MeV is shifted to a little lower position. The fact that the FSI may lower the mass of a bound state or a resonance reflects the couple channel effects. Because the scattering volume is much larger than the maximum number 0.1 fm$^3$ in Table \ref{PDDbar}, this case may not be realistic.

From the above results, we conclude that although the strong FSI
may lead to anomalous line shapes in the production processes in the unrealistic case, one
cannot interpret those observed by the BES Collaboration with this mechanism. It is worthwhile to study  further
whether the anomalous line shapes are due to other nearby resonances, new resonances, or
channel coupling effects.

\section{The $B\bar{B}$ system}\label{sec6}

All the previous studies may be extended to the $B\bar{B}$ case
naturally. Now $\Upsilon(4S)$ is slightly above the threshold. The
features, in principle, should be similar to those of the
$D\bar{D}$ case. But we will see more interesting results. We will
present the figures only if the line shape is strange. When
performing the numerical evaluation, the new parameters we have to
use are the meson masses \cite{PDG} $m_B=5279.34$ MeV,
$m_\Upsilon=10579.4$ MeV, the width \cite{PDG} $\Gamma_T=20.5$
MeV and $\Gamma_{ee}=0.272$ keV, the branching ratio $B_r(\Upsilon(4S)\to B\bar{B})=0.96$, the derived decay constant $f_\Upsilon=340.7$ MeV, and the electric
charge of the heavy quark $Q_b=-1/3$. The derived
coupling constant without FSI is $g_{\Upsilon B\bar{B}}=23.9$. We adopt a little larger
cutoff $\Lambda_p$=1.1 GeV for this case.

\begin{table}[htb]
\centering \caption{The S- wave $B\bar{B}$ scattering lengths in
unit of fm. NV (VC) indicates the contributions from
vector mesons are omitted (included). The number of * in the
table indicates that of the binding solutions. The binding
energies are given in Table \ref{EB}.}\label{SBBbar}
\tabcolsep=9pt
\begin{tabular}{c|c|cccccc}\hline
$m_\sigma$  & Vector meson  &\multicolumn{6}{c}{$\Lambda$ (GeV)}\\
(MeV)&exchange& 0.8 & 1.0 & 1.2 & 1.5 & 2.0 & $\infty$\\\hline
600   &NV & 0.026&0.058&0.083&0.11&0.13&0.17\\
400  & NV &0.19&0.26&0.30&0.34&0.38&0.44\\
600  & VC & 0.039& 2.74 & -1.32(*)&-0.46(*)&1.08(*)&-3.02(***)\\
400  & VC &0.21&7.47 &-1.22(*)&-0.38(*)&1.76(*)&-2.56(***)\\
\hline
\end{tabular}
\end{table}

\begin{figure}
\centering
\begin{tabular}{cc}
\scalebox{0.6}{\includegraphics{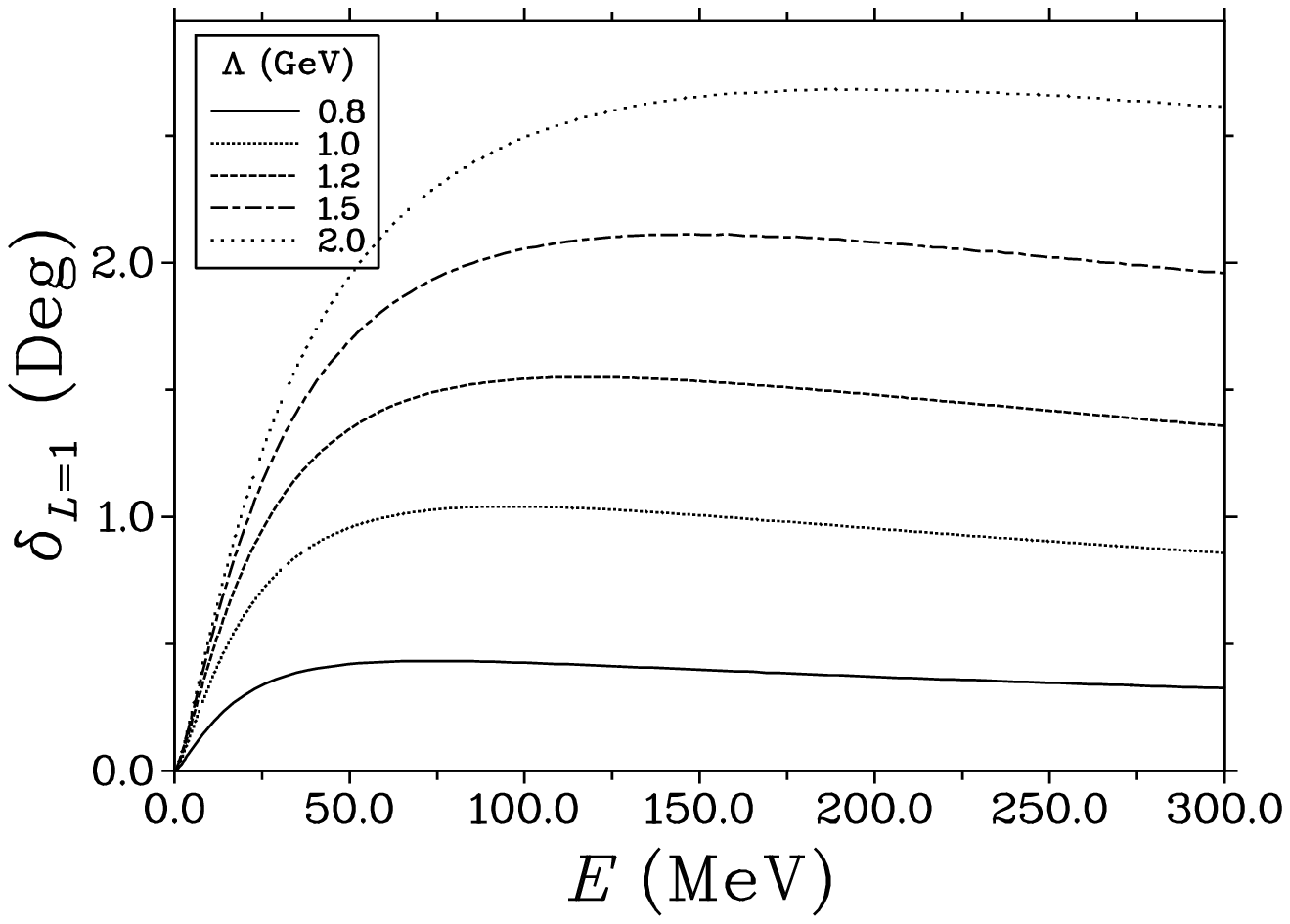}}&
\scalebox{0.6}{\includegraphics{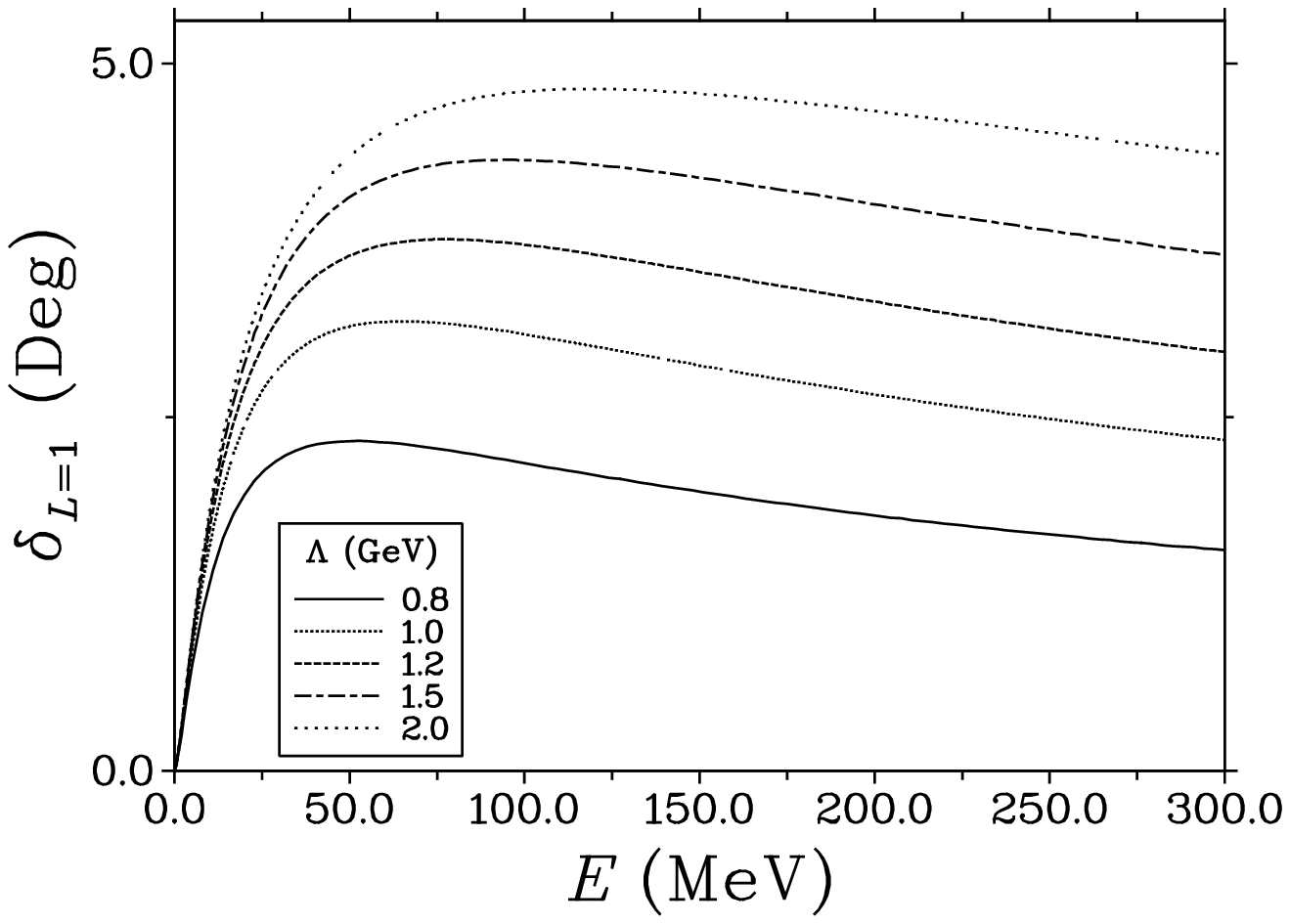}}\\
(a)&(b)\\
\scalebox{0.6}{\includegraphics{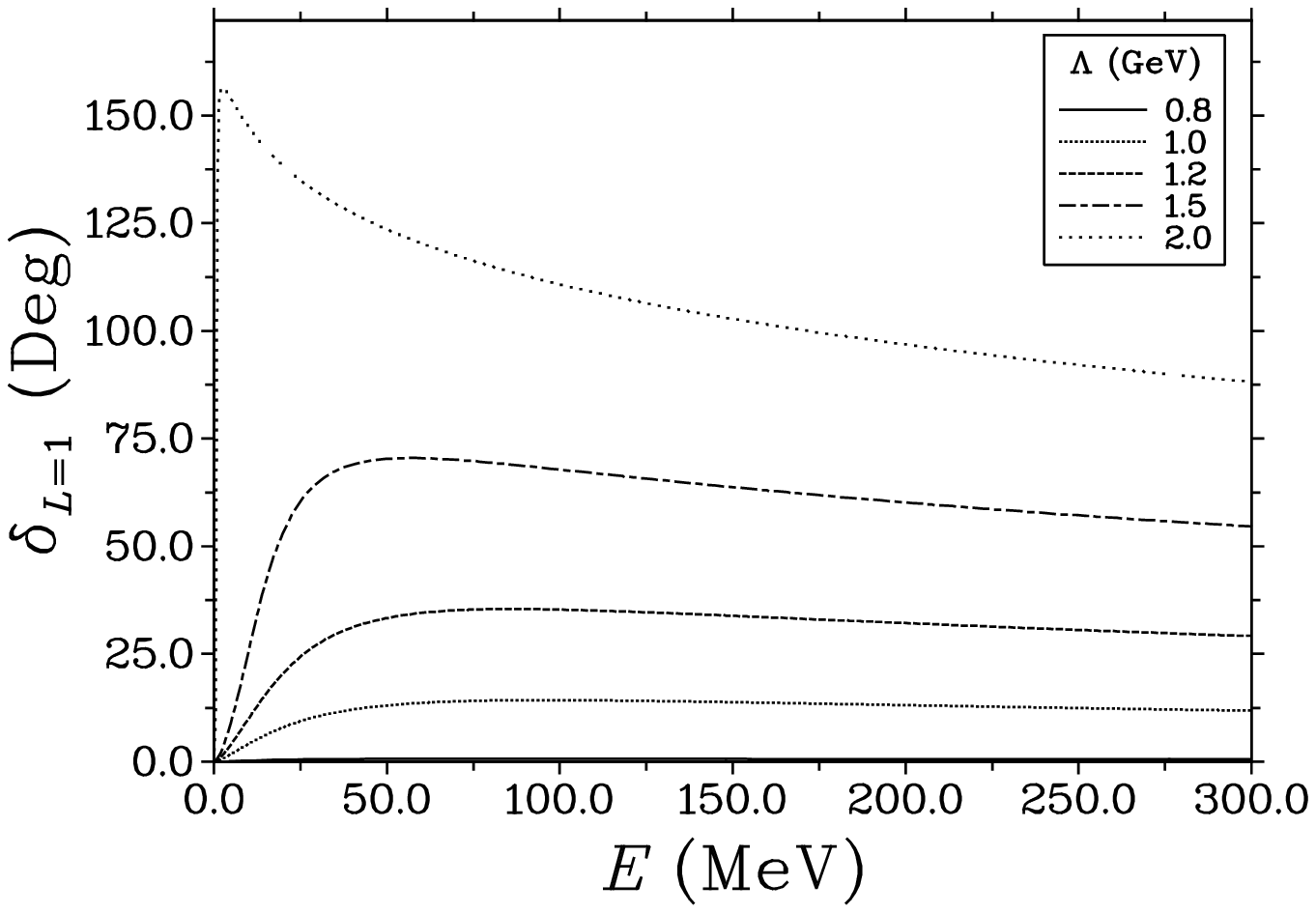}}&
\scalebox{0.6}{\includegraphics{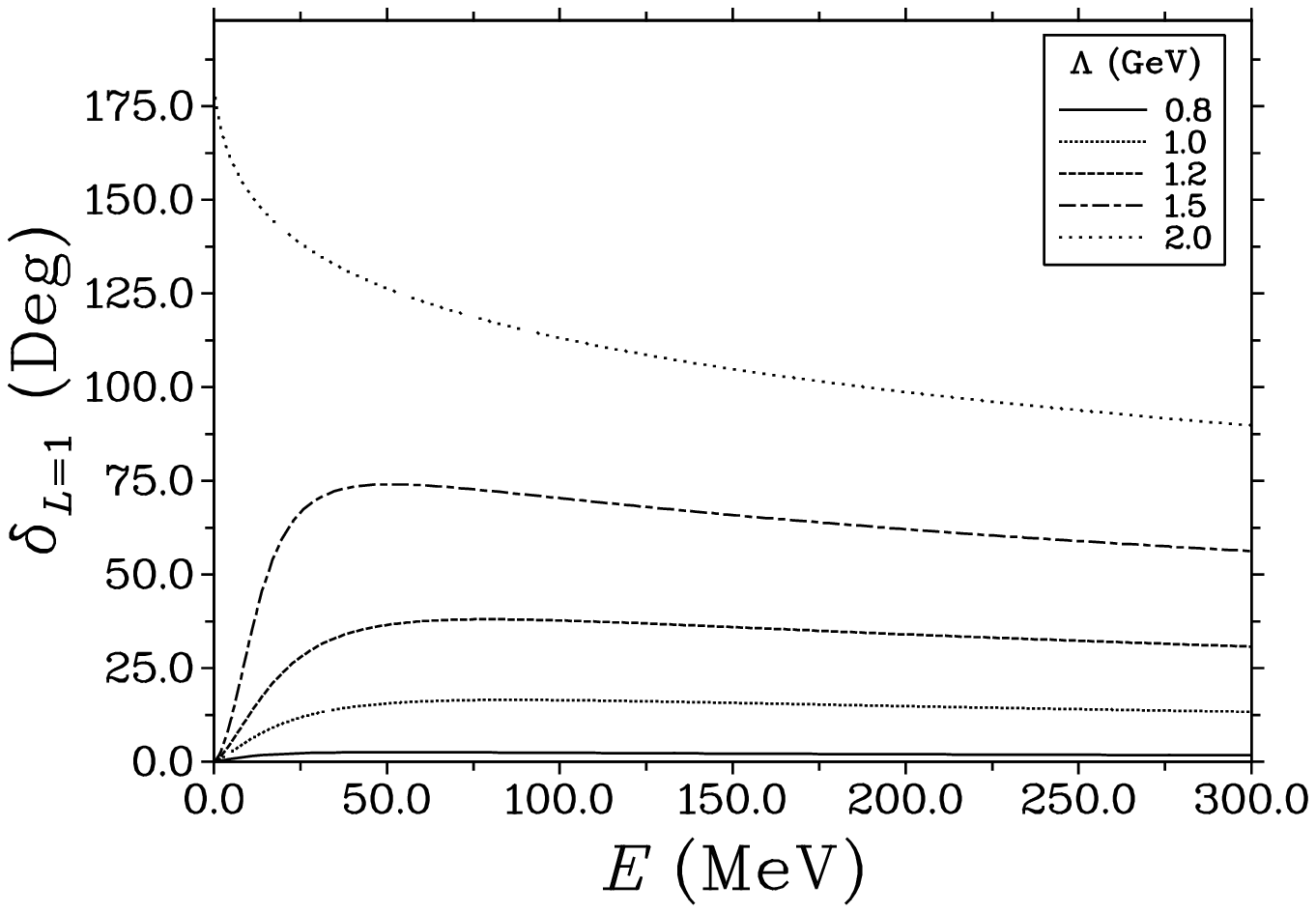}}\\
(c)&(d)
\end{tabular}
\caption{The phase shifts for the P- wave $B\bar{B}$ scattering
with various parameters. The upper (lower) two diagrams correspond
to the cases without (with) vector meson exchange contributions.
The left (right) two diagrams are obtained with $m_\sigma$=600
(400) MeV. The cutoff $\Lambda$ is in units of GeV.  }\label{ph-BBbar-P}
\end{figure}

The S- wave $B\bar{B}$ bound state is more likely to exist than the
$D\bar{D}$ one. Our calculated phase shifts do not exceed
16$^\circ$ when one considers only the scalar meson exchange
contribution, which indicates there is no bound state. After one
includes the vector meson contributions, the phase shift starts
from 180$^\circ$ with a cutoff $\Lambda$=1.2 GeV. It can go up to
210$^\circ$ from 180$^\circ$ with a stronger attraction
$\Lambda=2.0$ GeV. We present the derived scattering lengths in
Table \ref{SBBbar} and the binding solutions in Table \ref{EB}. In
the point particle limit, the observation that three solutions
exist and that the binding energies are large indicates this case
is not physical once again. A finite reasonable cutoff should be
slightly larger than that in the $D\bar{D}$ case because of the
heavy quark symmetry and the smaller size of the $B$ meson. As for $\Lambda$=2.0 GeV and VC, the scattering phase shift crosses $\delta_{L=0}=\pi$ at a finite $E$ and therefore the S-wave cross section becomes zero at that point.

\begin{table}[htb]
\centering \caption{The P- wave $B\bar{B}$ scattering volumes in
unit of fm$^3$. NV (VC) indicates the contributions
from vector mesons are omitted (included). The number of *
in the table indicates that of the binding solutions. The binding
energies are given in Table \ref{EB}.}\label{PBBbar}
\tabcolsep=9pt
\begin{tabular}{c|c|cccccc}\hline
$m_\sigma$  & Vector meson  &\multicolumn{6}{c}{$\Lambda$ (GeV)}\\
(MeV)&exchange& 0.8 & 1.0 & 1.2 & 1.5 & 2.0 & $\infty$\\\hline
600&NV & 0.0040&0.0067&0.0083&0.0092&0.0096&0.0099\\
400&NV & 0.042&0.047&0.048&0.050&0.050&0.050\\
600& VC & 0.0054&0.069&0.14&0.26&4.55&0.1$\sim$0.2(*)\\
400&VC & 0.043&0.11&0.19&0.34&-5.56(*)&0.2$\sim$0.3(*)\\
\hline
\end{tabular}
\end{table}

\begin{figure}
\centering
\begin{tabular}{cc}
\scalebox{0.6}{\includegraphics{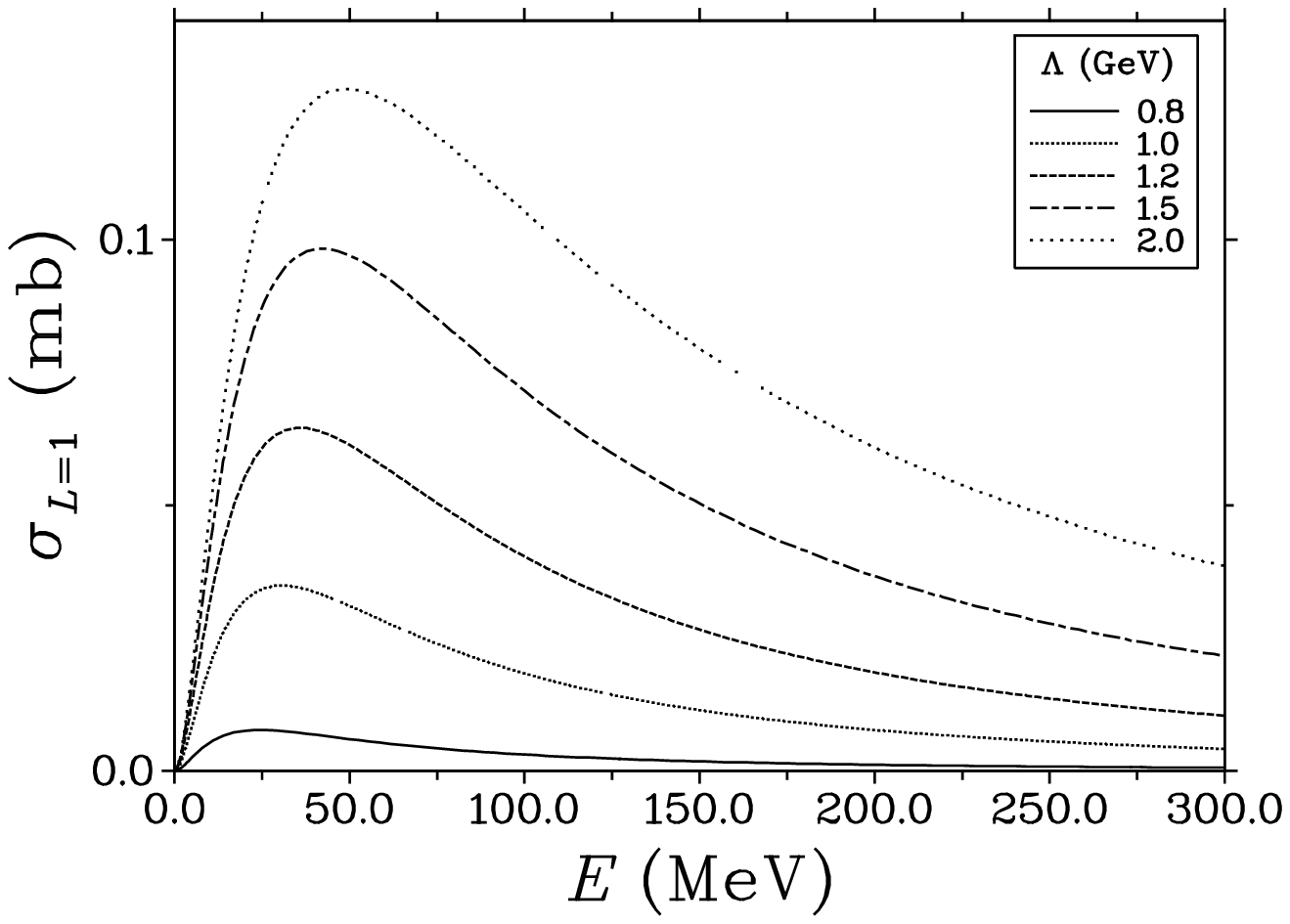}}&
\scalebox{0.6}{\includegraphics{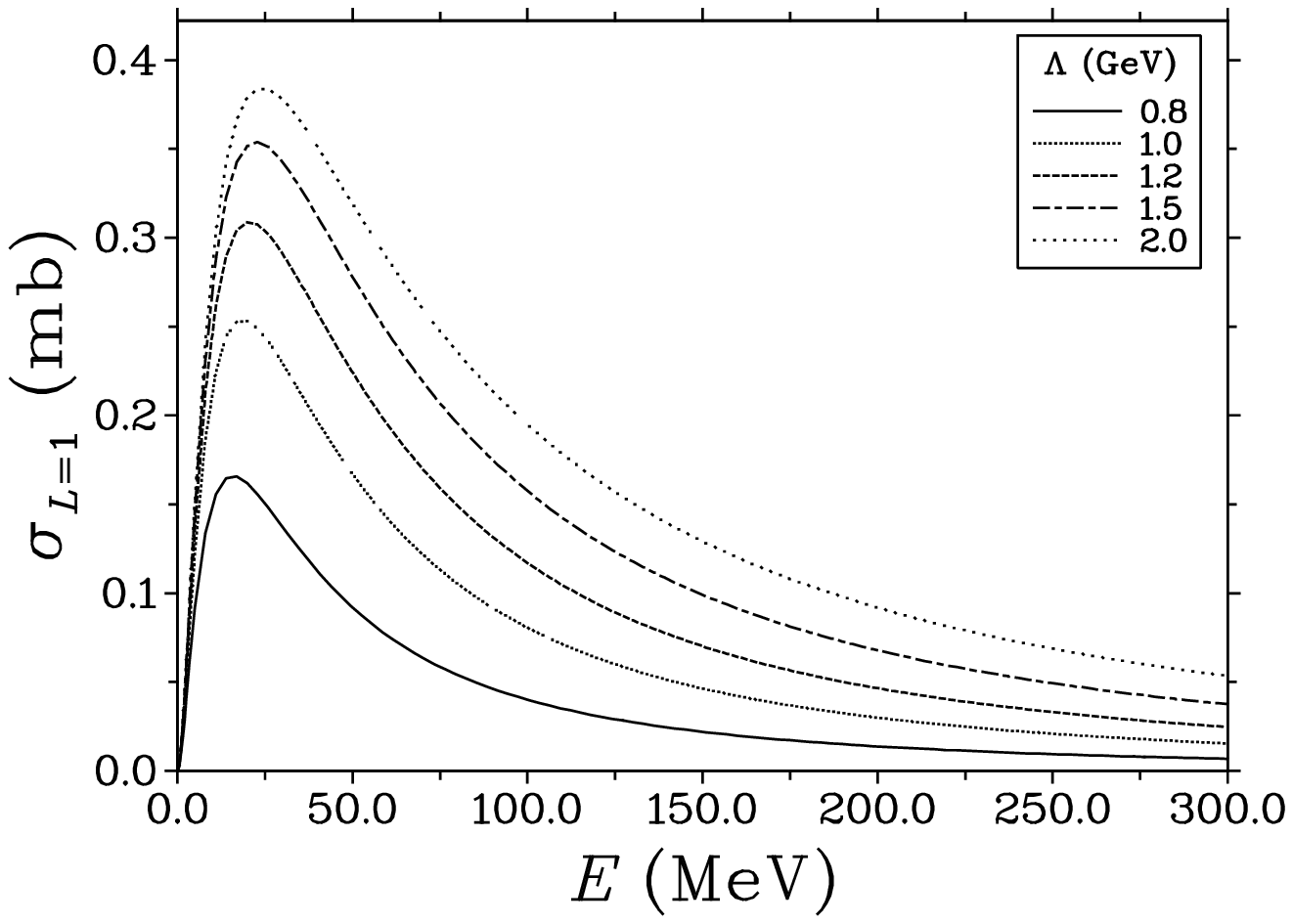}}\\
(a)&(b)\\
\scalebox{0.6}{\includegraphics{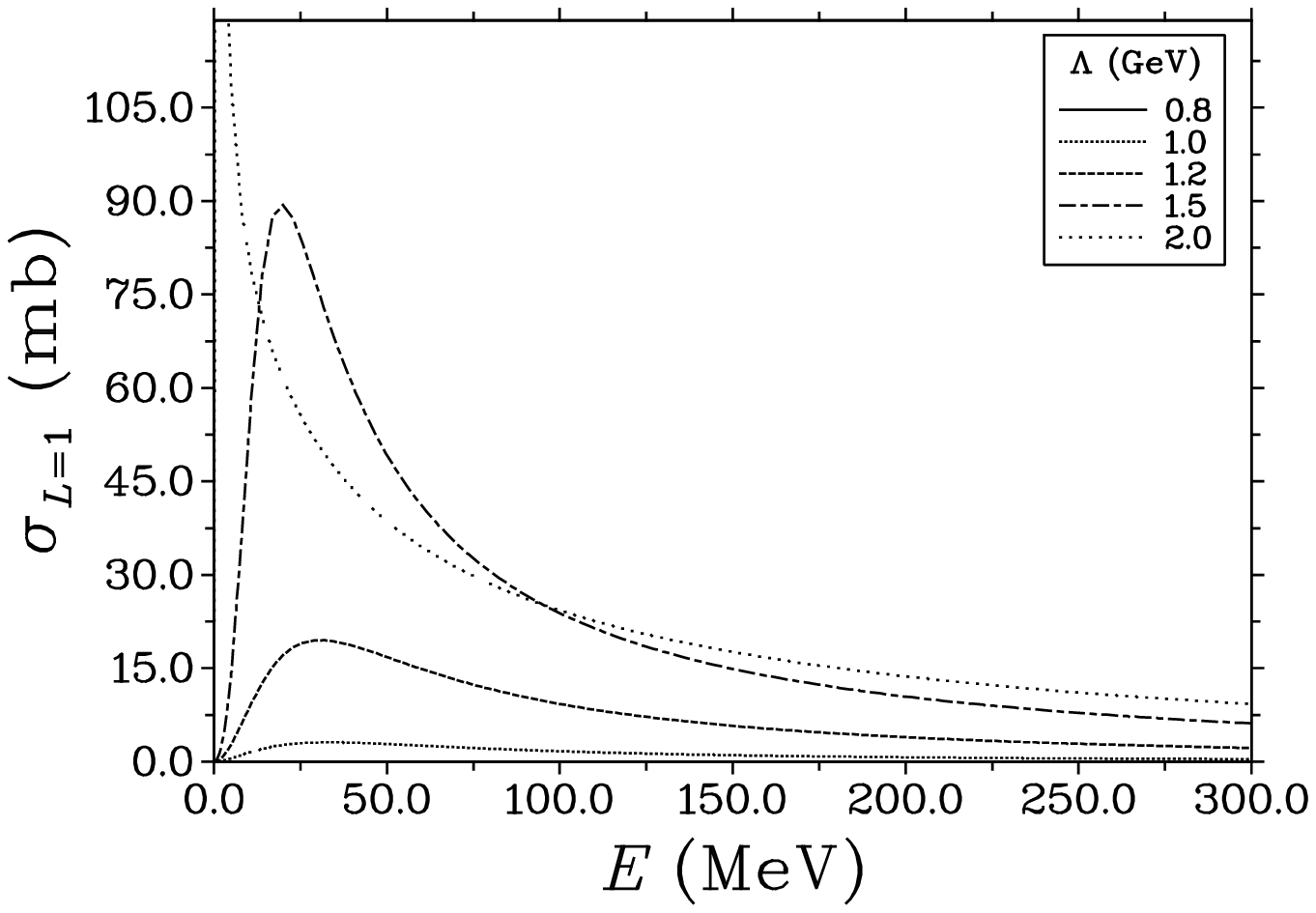}}&
\scalebox{0.6}{\includegraphics{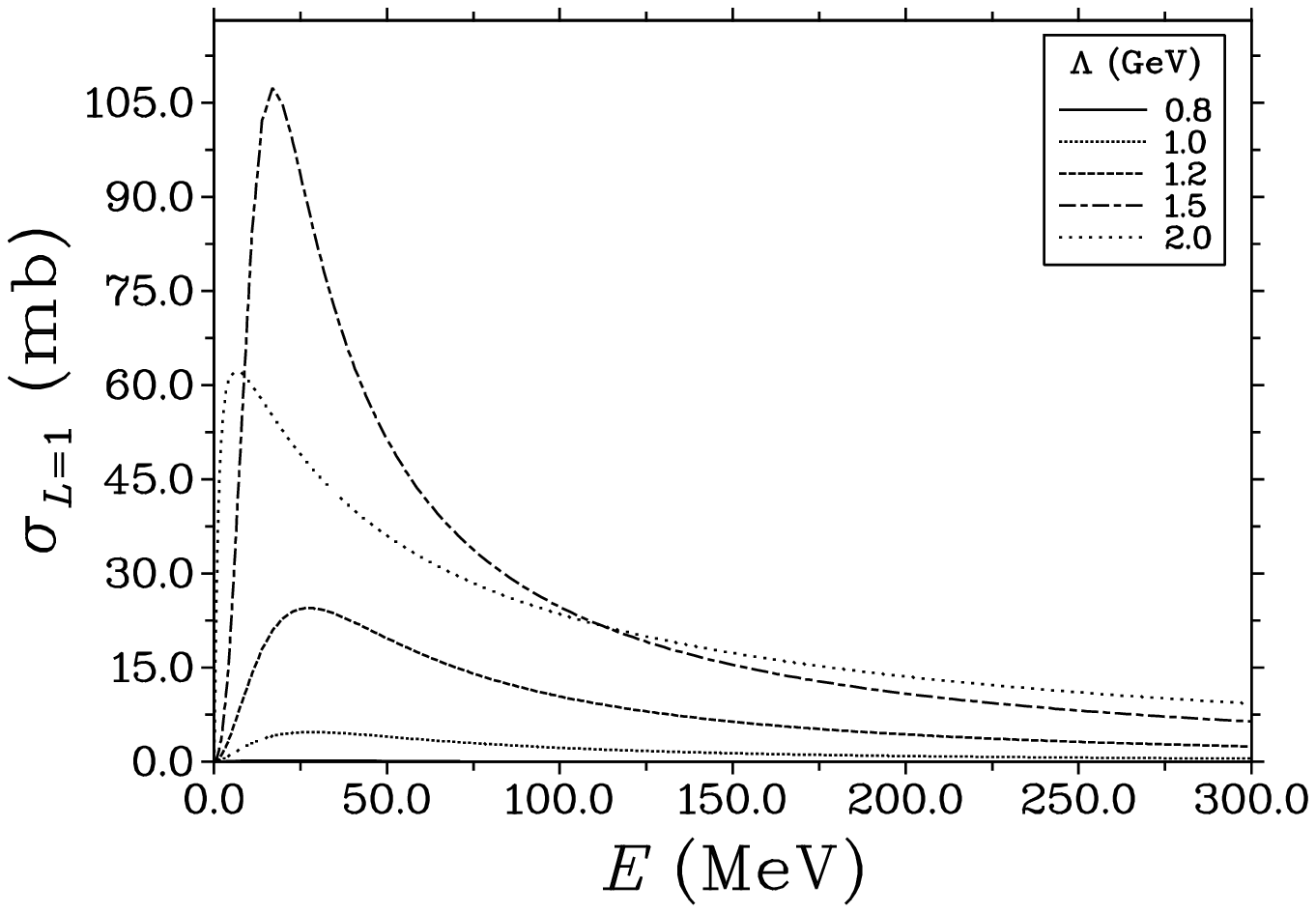}}\\
(c)&(d)
\end{tabular}
\caption{The P- wave total cross sections for the $B\bar{B}$ scattering
with various parameters. The upper (lower) two diagrams correspond
to the cases without (with) vector meson exchange contributions.
The left (right) two diagrams are obtained with $m_\sigma$=600
(400) MeV. The cutoff $\Lambda$ is in units of GeV.  }\label{cs-BBbar-P}
\end{figure}

For the P- wave interactions, we show the phase shifts with various
parameters in Fig. \ref{ph-BBbar-P}. It can reach 180$^\circ$ with
a strong attraction, e.g. $\Lambda=$2.0 GeV, $m_\sigma=400$ MeV
and VC. Note the line for $\Lambda=2.0$ GeV in the diagram in Fig. 9(c) is
a resonance while that in the diagram in Fig. 9(d) is a bound state. Therefore,
the results are more interesting than that in the $D\bar{D}$ case.
We give the derived scattering volumes in Table \ref{PBBbar} and
the binding solutions in Table \ref{EB}. In the point particle
limit, the positive scattering volume and the existence of one
binding solution come from the observation that the phase shift
goes up to $\sim 245^\circ$ from $180^\circ$. The large binding
energies in this limit again require a finite cutoff. If the
cutoff less than 1.5 GeV is reasonable, then no P- wave binding
solutions exist. We present the cross sections for different
parameters in Fig. \ref{cs-BBbar-P}. Comparing this figure with
Fig. \ref{cs-DDbar-P}, one observes the resonancelike structure
is more evident. It appears at a smaller energy of motion. Note
that the cross section for the case $\Lambda=2.0$ GeV
(c) also vanishes in the limit $E=0$, which is understood with Eq.
(\ref{cs0-P}) and the nonzero $1/a_1$.

\begin{figure}
\centering
\begin{tabular}{cc}
\scalebox{0.6}{\includegraphics{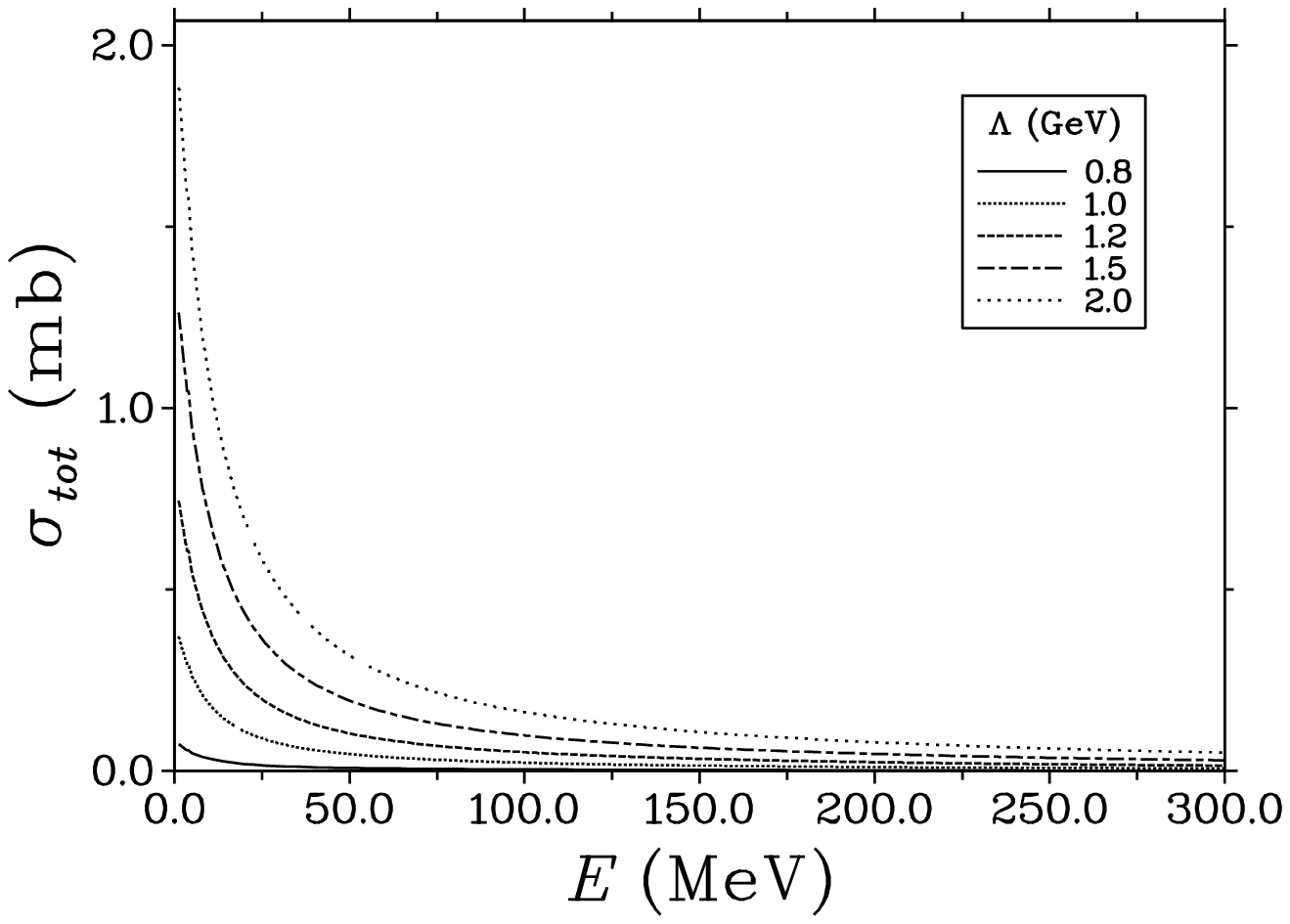}}&
\scalebox{0.6}{\includegraphics{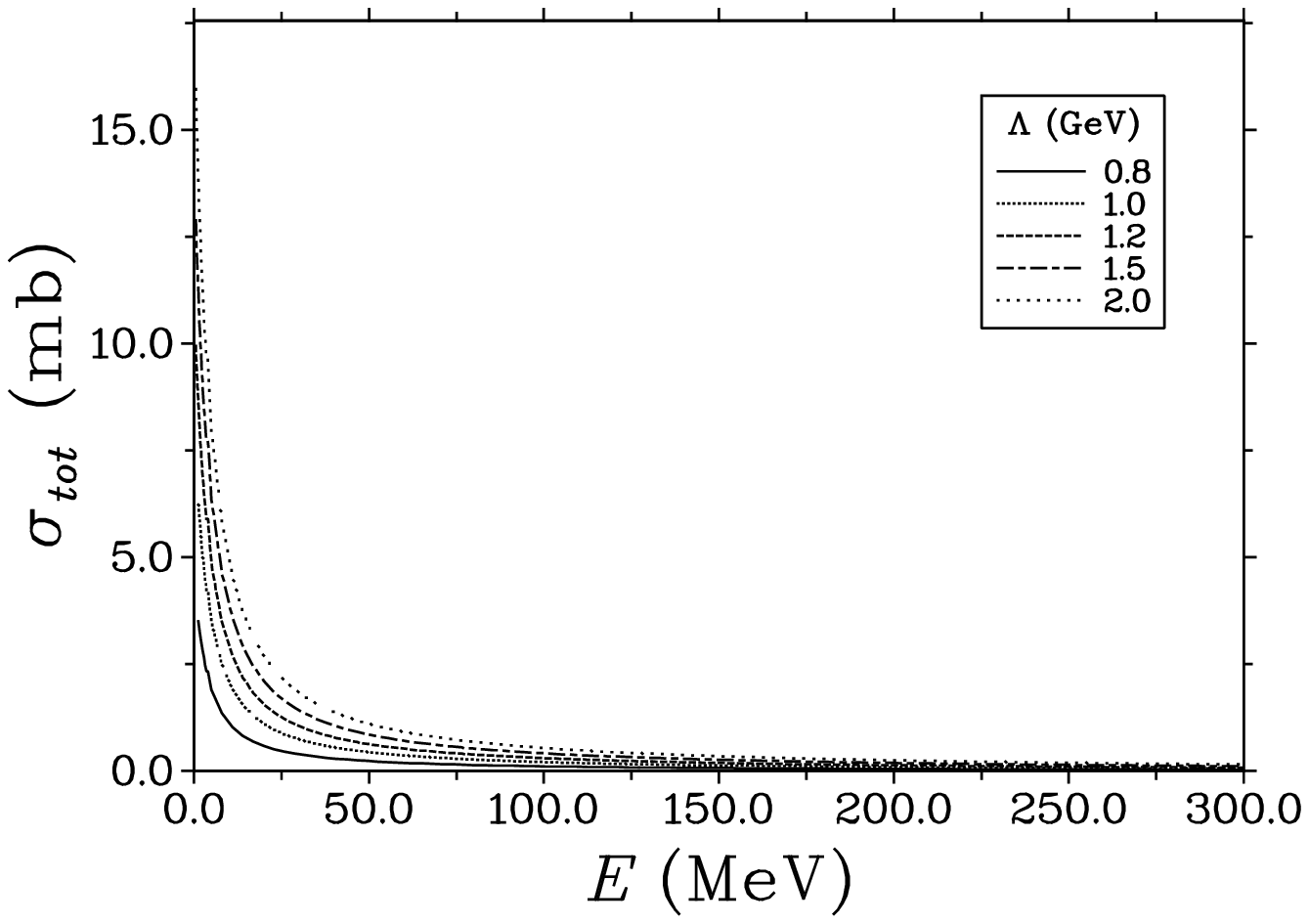}}\\
(a)&(b)\\
\scalebox{0.6}{\includegraphics{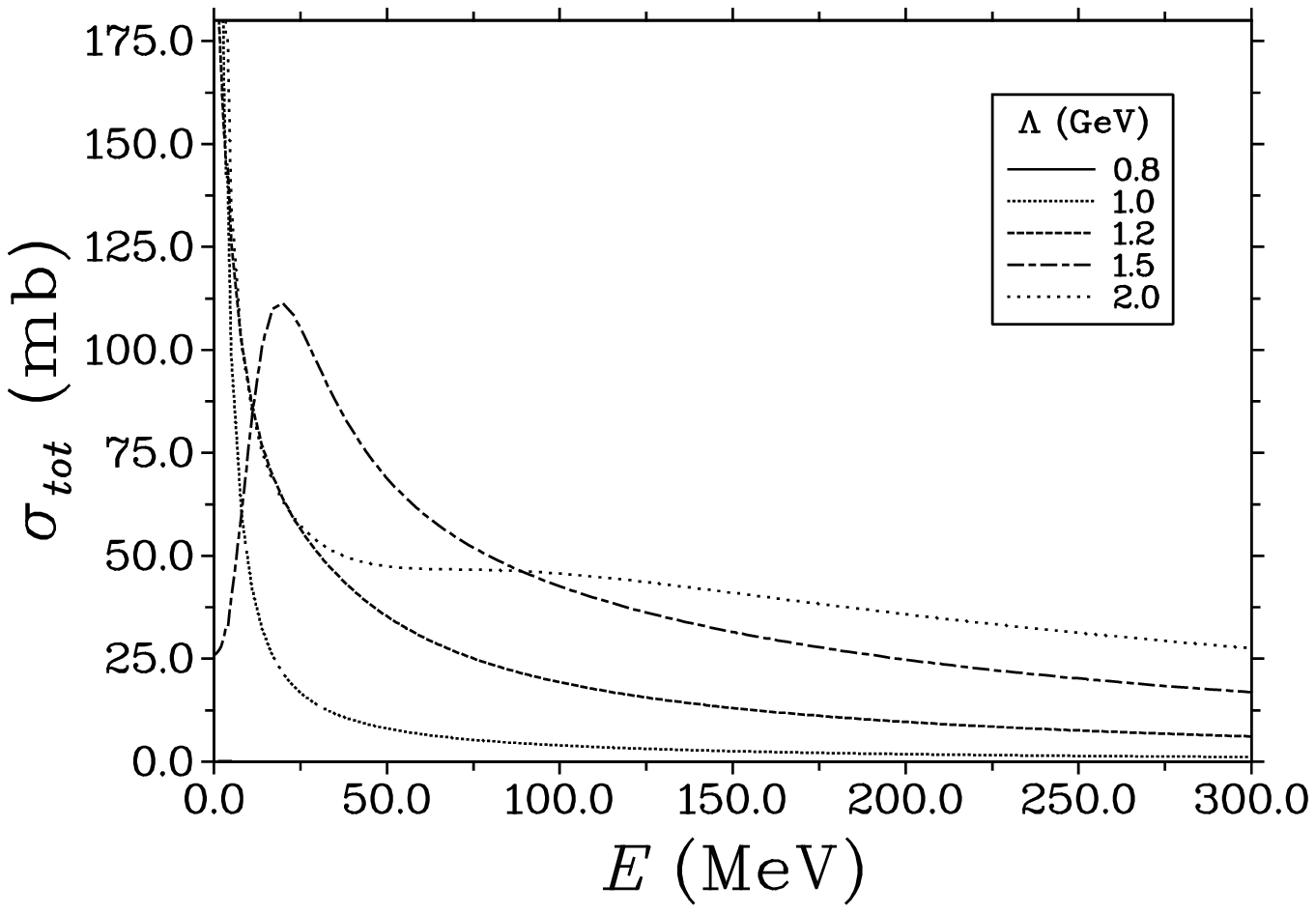}}&
\scalebox{0.6}{\includegraphics{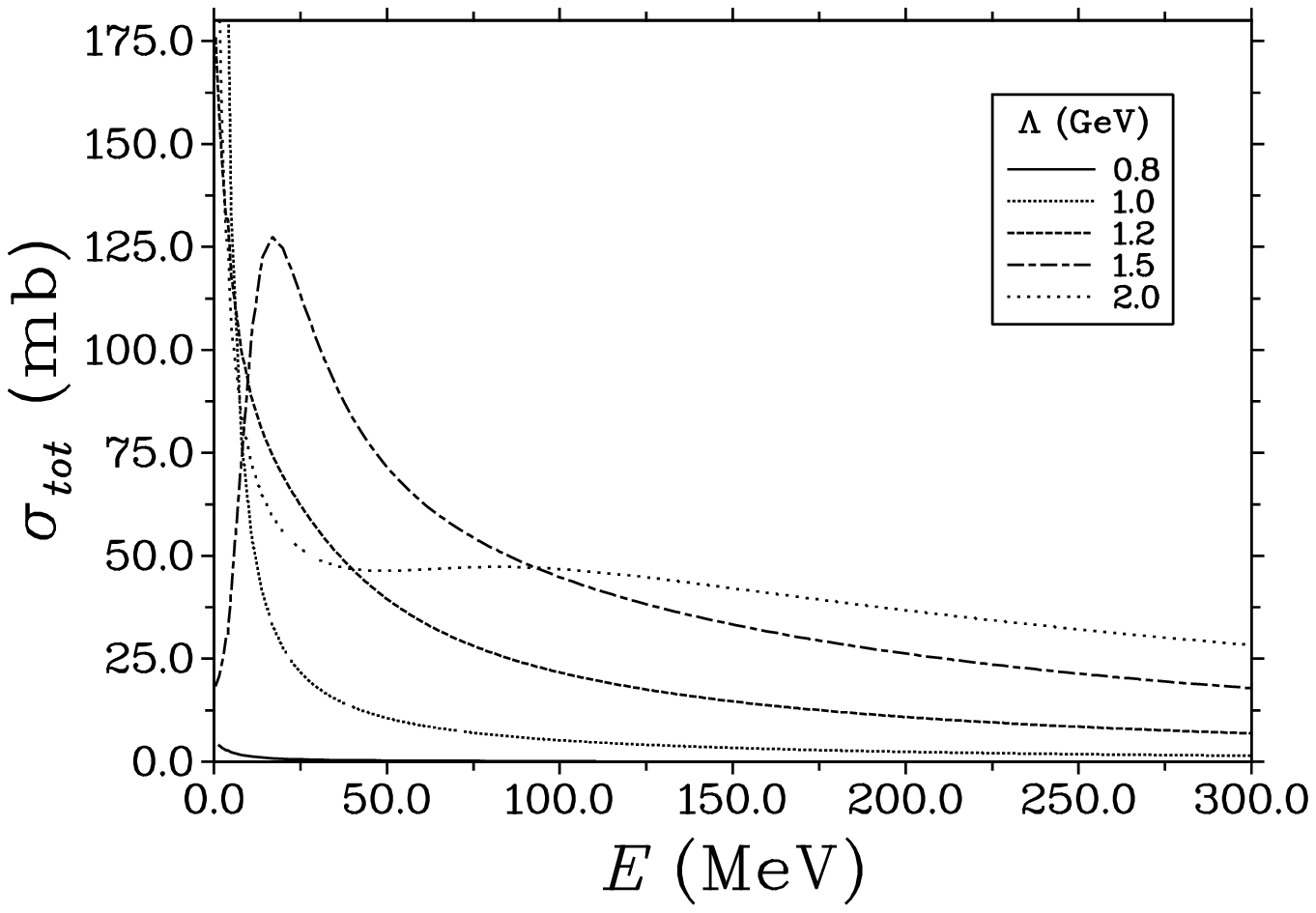}}\\
(c)&(d)
\end{tabular}
\caption{The total cross sections for the $B\bar{B}$ scattering
with various parameters. The upper (lower) two diagrams correspond
to the cases without (with) vector meson exchange contributions.
The left (right) two diagrams are obtained with $m_\sigma$=600
(400) MeV. The cutoff $\Lambda$ is in units of GeV.  }\label{totalcs-B}
\end{figure}

To get the total cross section, we have to sum up all the partial
wave contributions. In the $D\bar{D}$ case, the estimated maximum
partial wave was  $l_{max}\approx \frac{\sqrt{m_D
E}}{m_\sigma}\approx 1.87$, so we calculated up to P wave.
However, in the $B\bar{B}$ case, one should consider higher
partial wave contributions (up to F wave) since $l_{max}\approx
3.1$. For the D- wave scattering, the line shapes look like those in
the P- wave $D\bar{D}$ case, but the phase shift can go up to
100$^\circ$ with $m_\sigma$=400 MeV, $\Lambda\to\infty$, and
VC. For a finite cutoff $\Lambda\leq2.0$ GeV, a
resonancelike structure also appears in the cross sections,
although the phase shift does not exceed 90$^\circ$. The phase
shifts for the F- wave scattering do not exceed 40$^\circ$ even in
the point particle limit. After summing up these four partial wave
contributions, we get the total cross sections shown in Fig.
\ref{totalcs-B}. An interesting structure is there if one does not
ignore the vector meson exchange contributions. The bump structure
for the case $\Lambda=1.5$ GeV comes mainly from the P- wave
scattering.

Now we move on to the final state rescattering effects in the
process $e^+e^-\to B\bar{B}$. One may still choose the four cases
in the $D\bar{D}$ system. However, there are no strange line shapes
in the cases (a), (b), and (c). Since both a resonance and a bound
state are possible in the P- wave scattering with a finite cutoff,
here we consider the case (d): $m_\sigma=400$ MeV, $\Lambda=2.0$ GeV,
and VC; case (e): $m_\sigma=600$ MeV, $\Lambda=2.0$ GeV, and
VC; and an extreme case.

\begin{figure}
\centering
\begin{tabular}{cc}
\scalebox{0.6}{\includegraphics{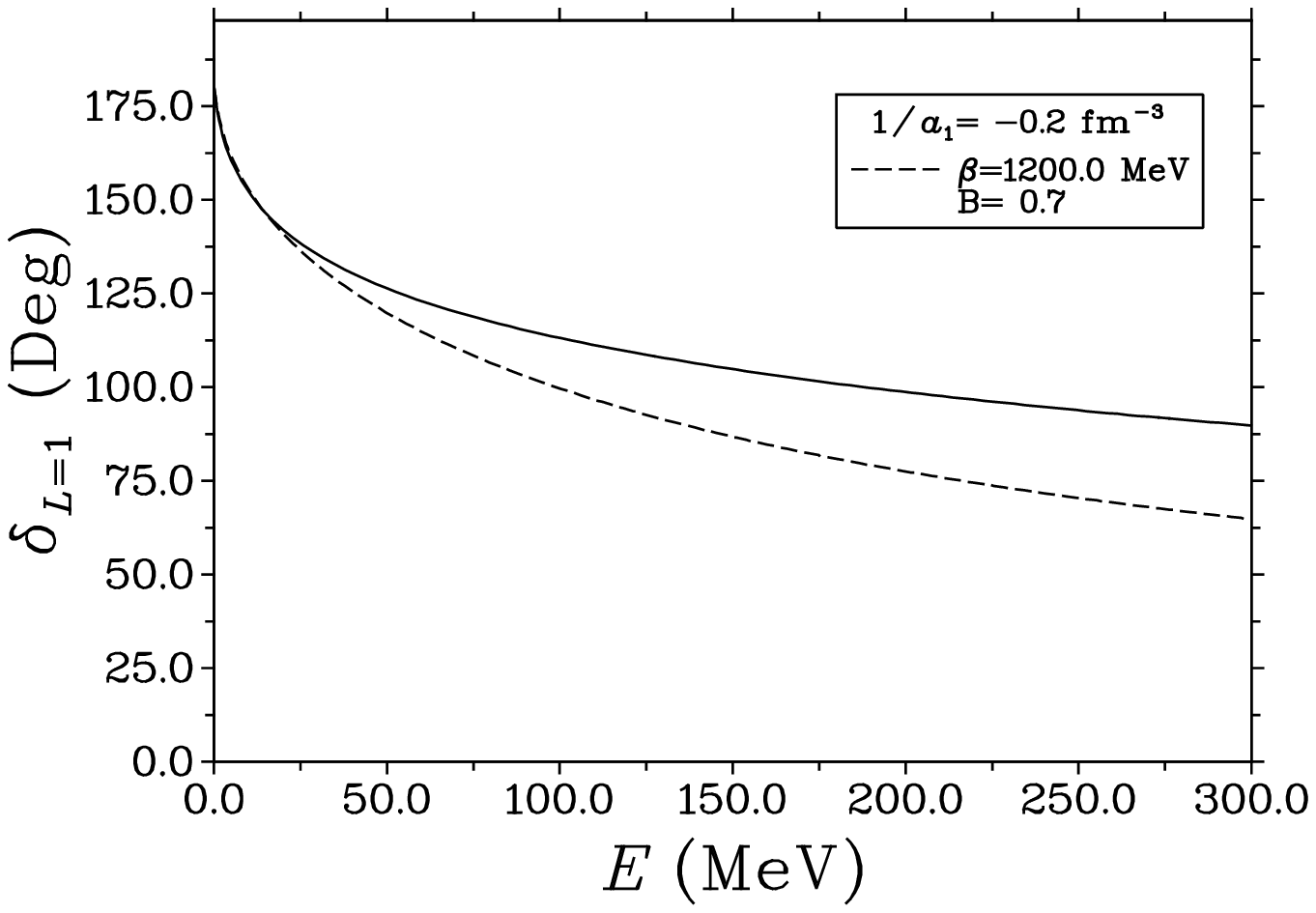}}&
\scalebox{0.6}{\includegraphics{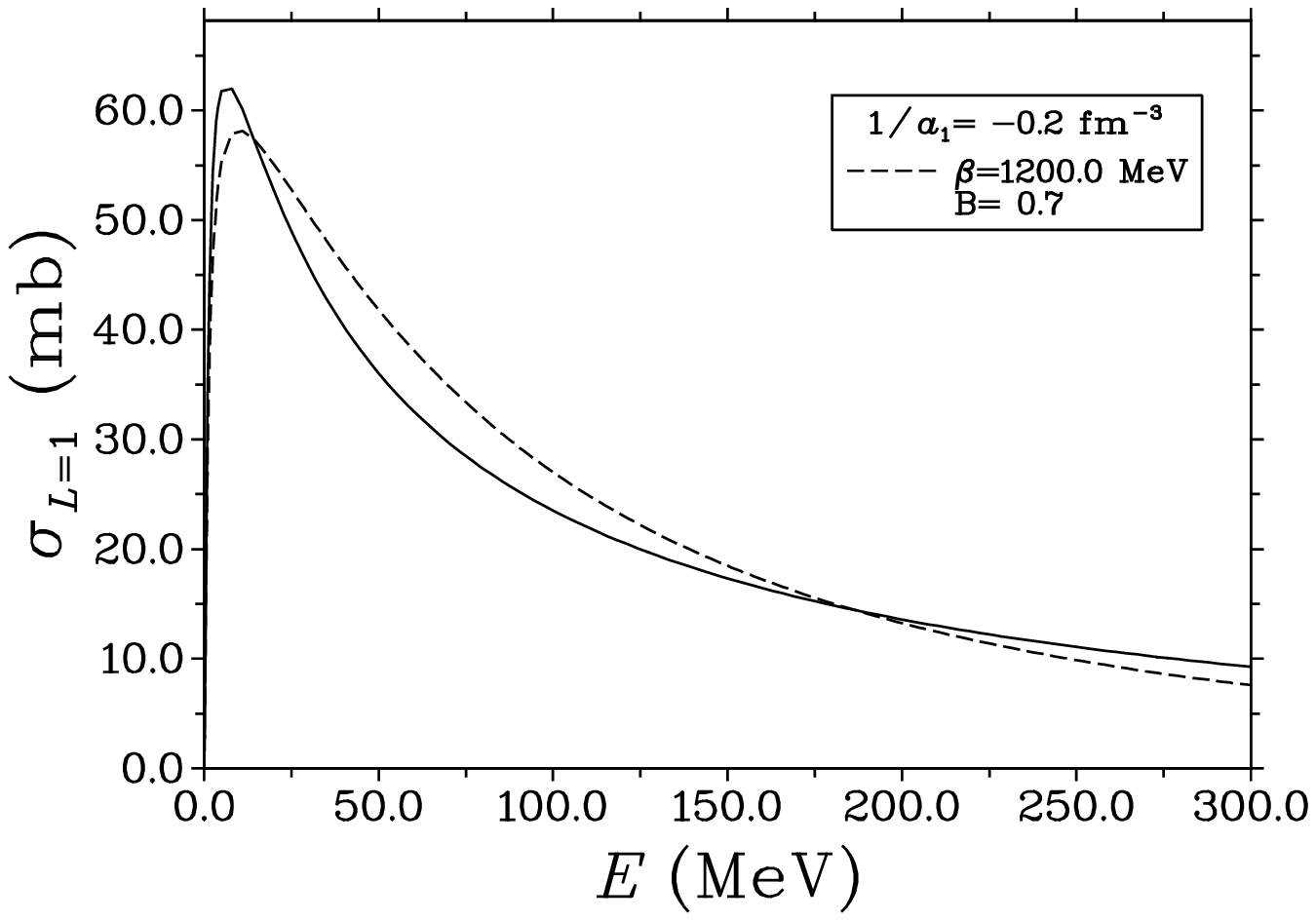}}\\
(d-1)&(d-2)\\
\scalebox{0.6}{\includegraphics{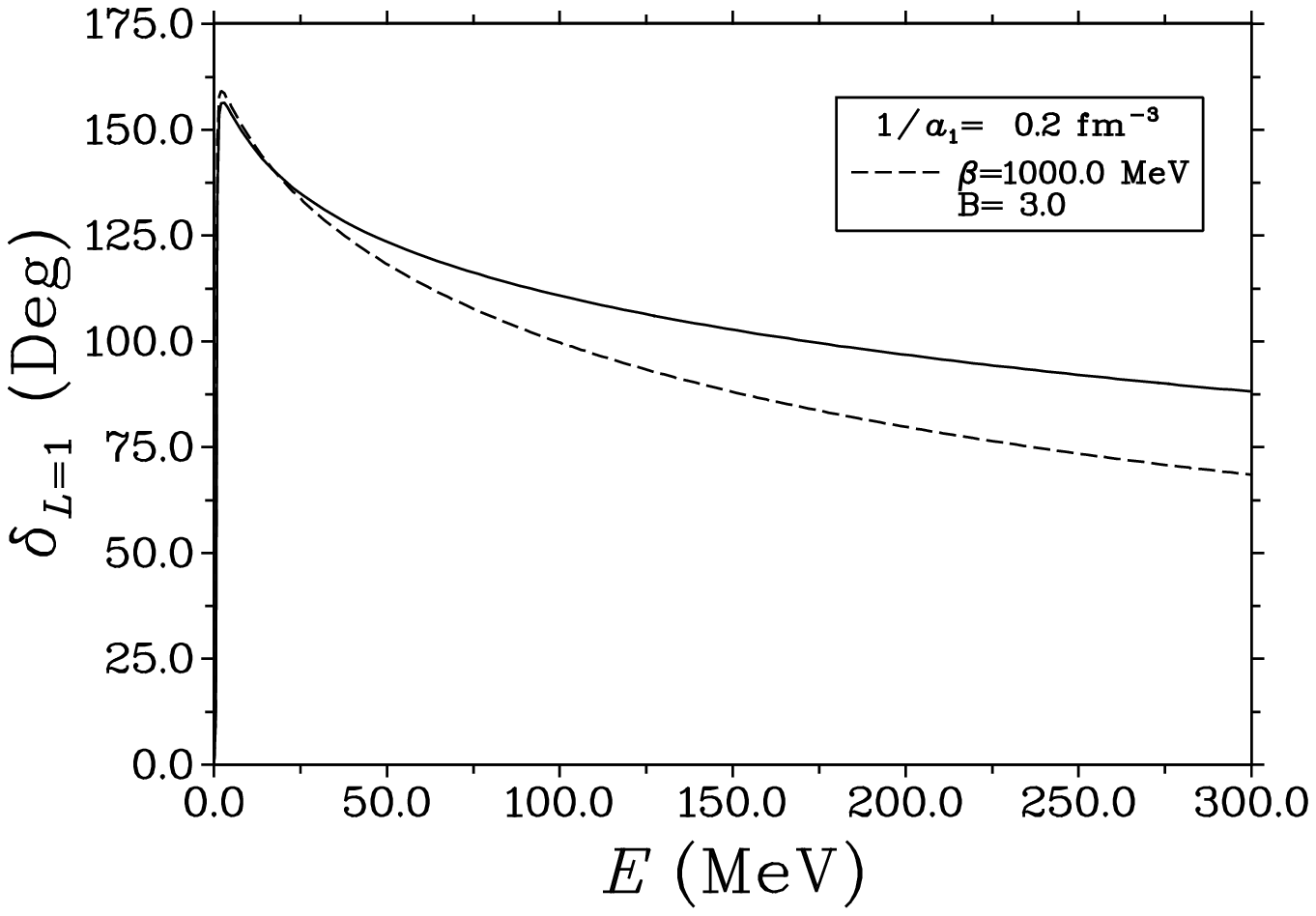}}&
\scalebox{0.6}{\includegraphics{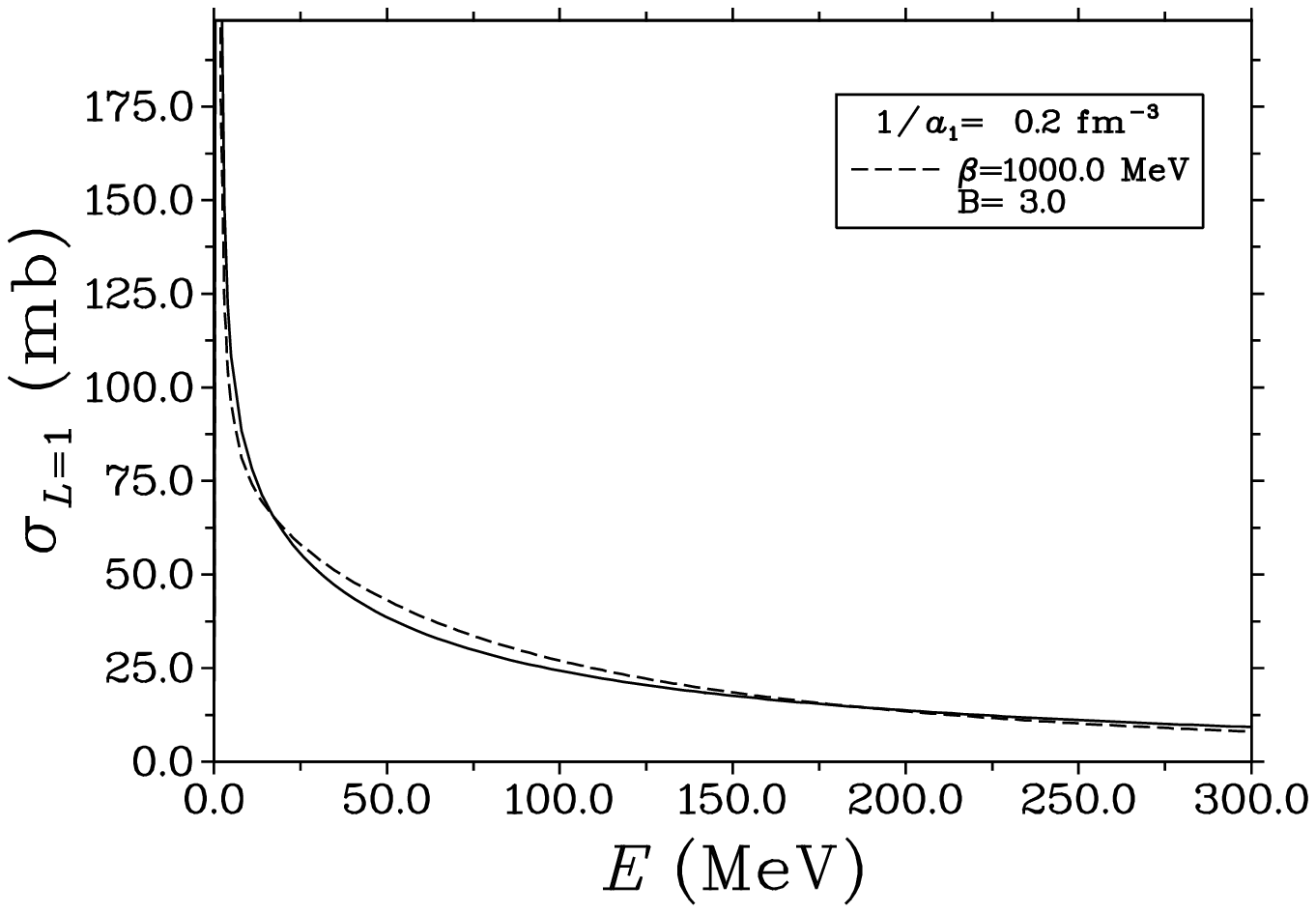}}\\
(e-1)&(e-2)
\end{tabular}
\caption{The first (last) two diagrams show the reproduced P- wave
$B\bar{B}$ phase shifts and scattering cross sections
corresponding to the case (d) [(e)] with $1/a_1$=-0.18 (0.22)
fm$^{-3}$, respectively.}\label{fit-ph-cs-B}
\end{figure}

\begin{figure}
\centering
\begin{tabular}{cc}
\scalebox{0.6}{\includegraphics{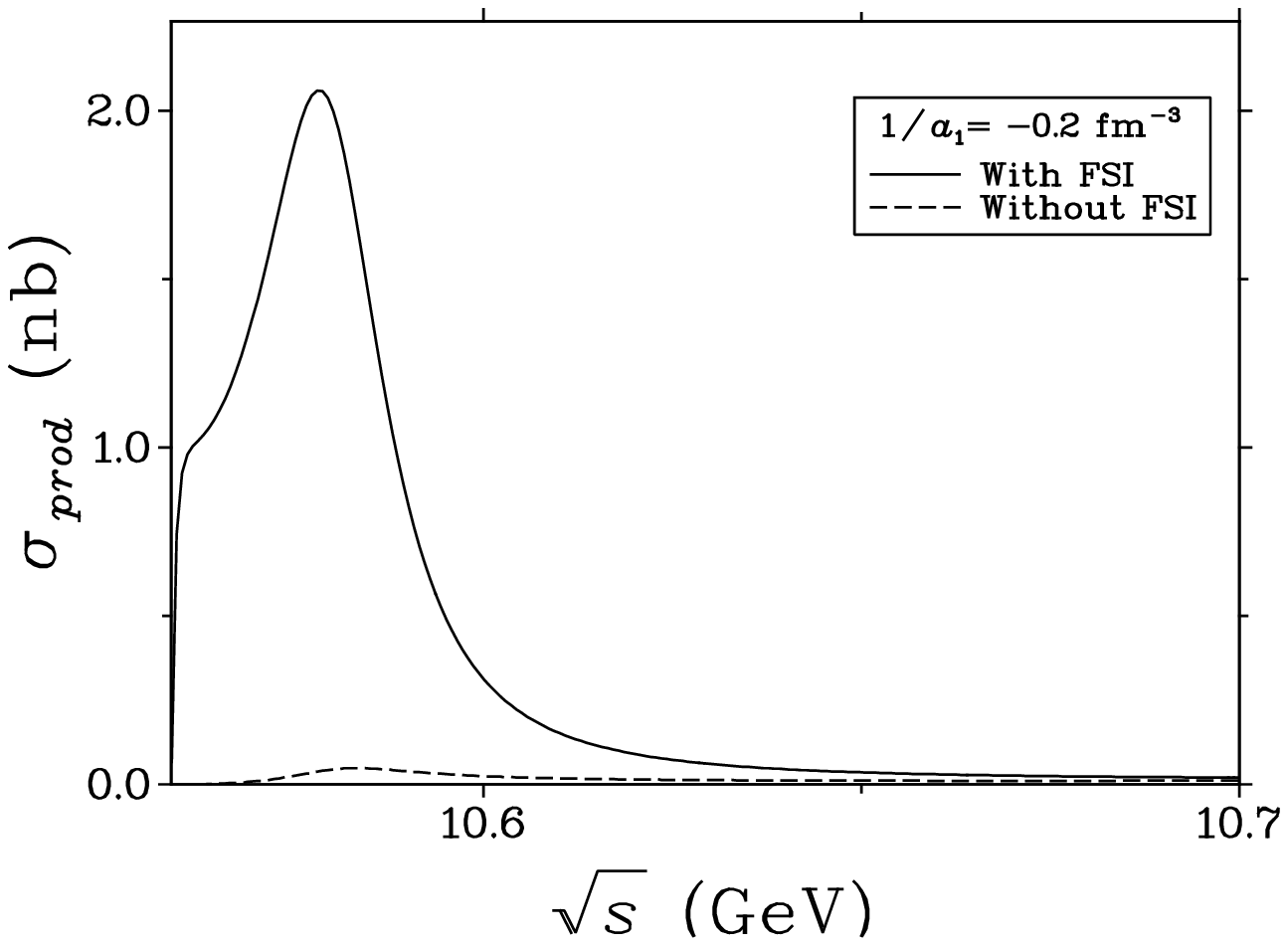}}&
\scalebox{0.6}{\includegraphics{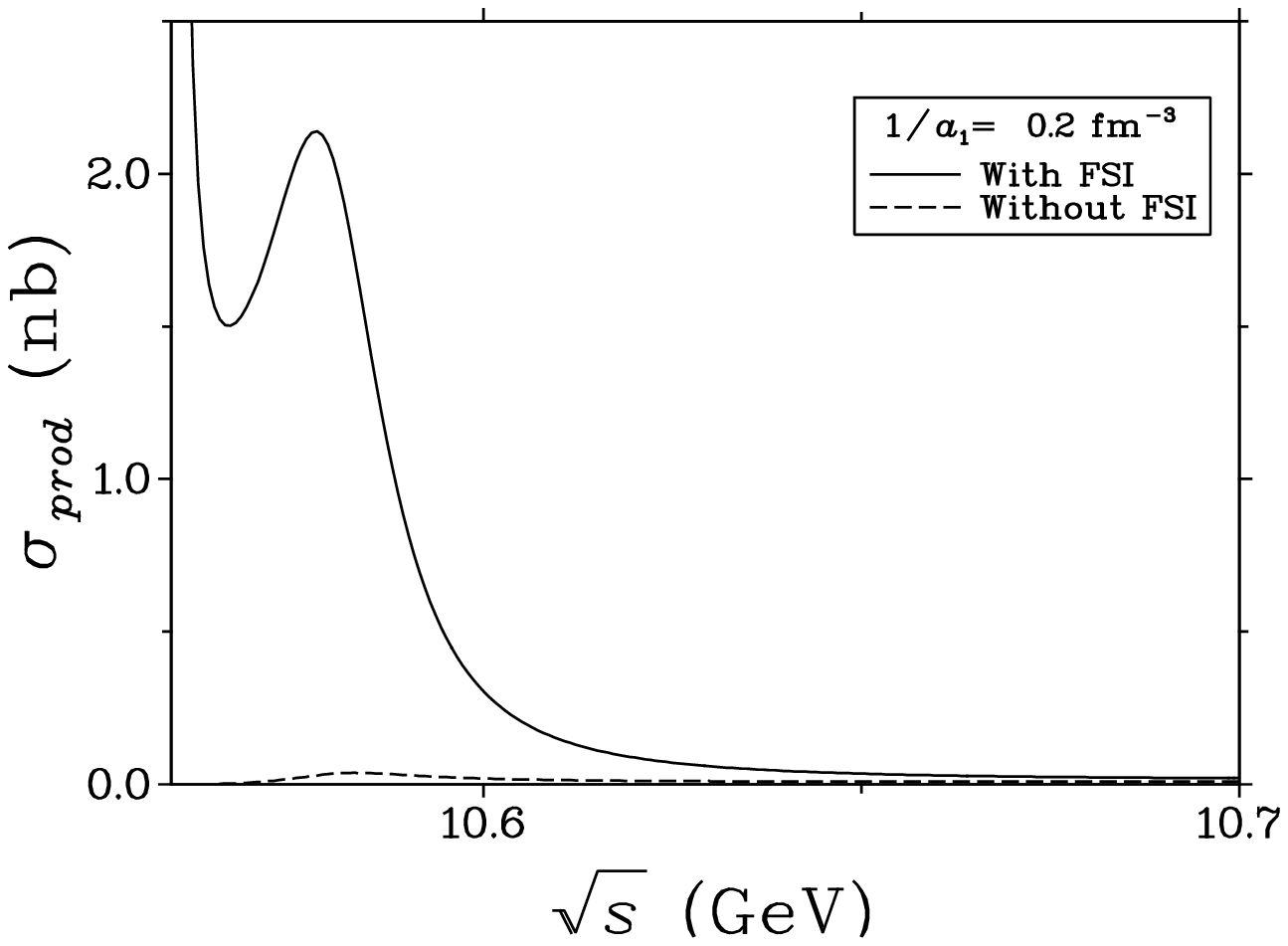}}\\
(d)&(e)
\end{tabular}
\caption{The obtained $B\bar{B}$ production cross sections
correspond to the case (d) $1/a_1$=-0.18 fm$^{-3}$ and (e)
$1/a_1$=0.22 fm$^{-3}$, respectively. We get the dashed lines by ignoring the rescattering part $\sigma_2$ of the production cross section $\sigma_{prod}=\sigma_1+\sigma_2$. }\label{pr-cs-B}
\end{figure}

\begin{figure}
\centering
\begin{tabular}{cc}
\scalebox{0.6}{\includegraphics{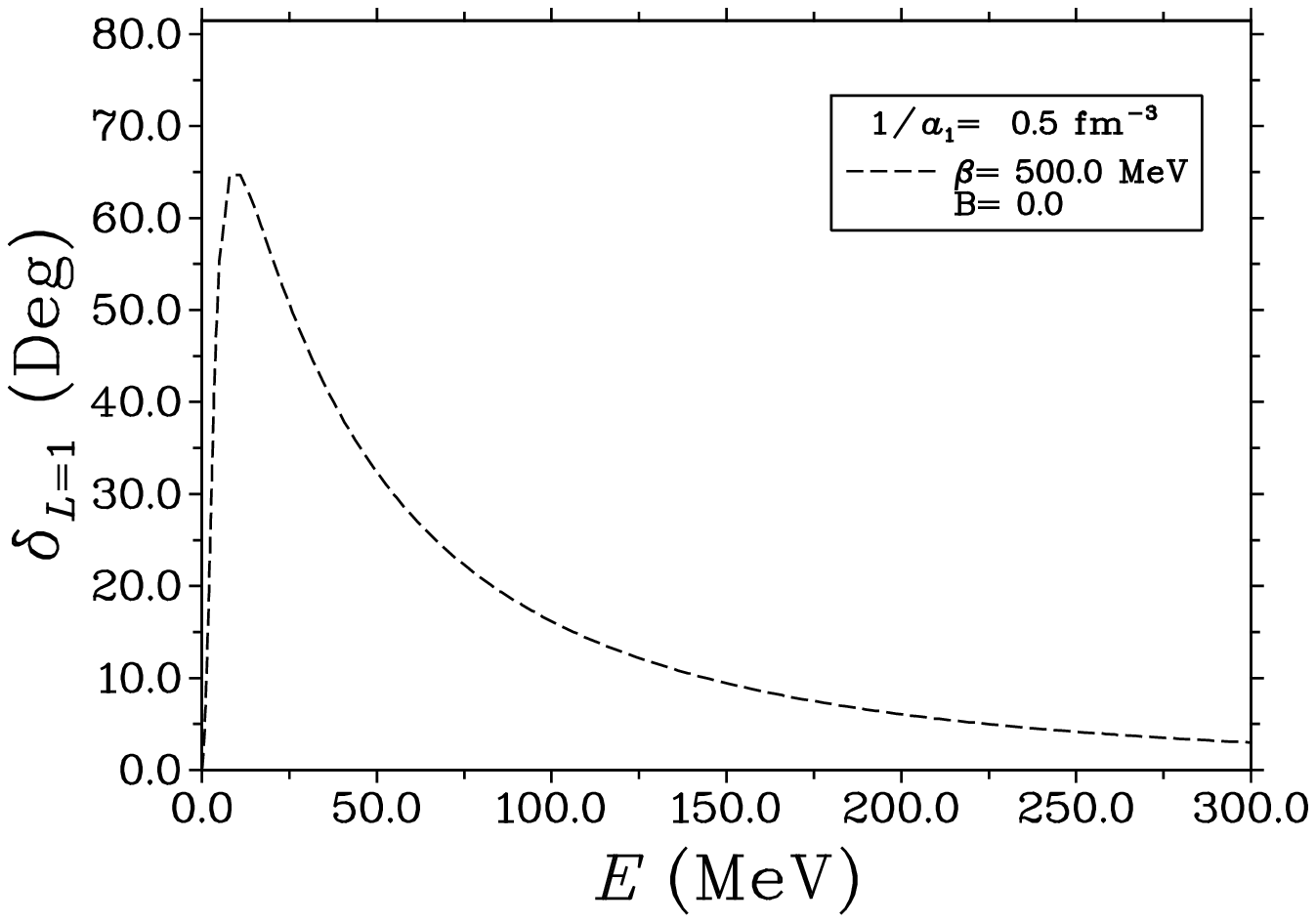}}&
\scalebox{0.6}{\includegraphics{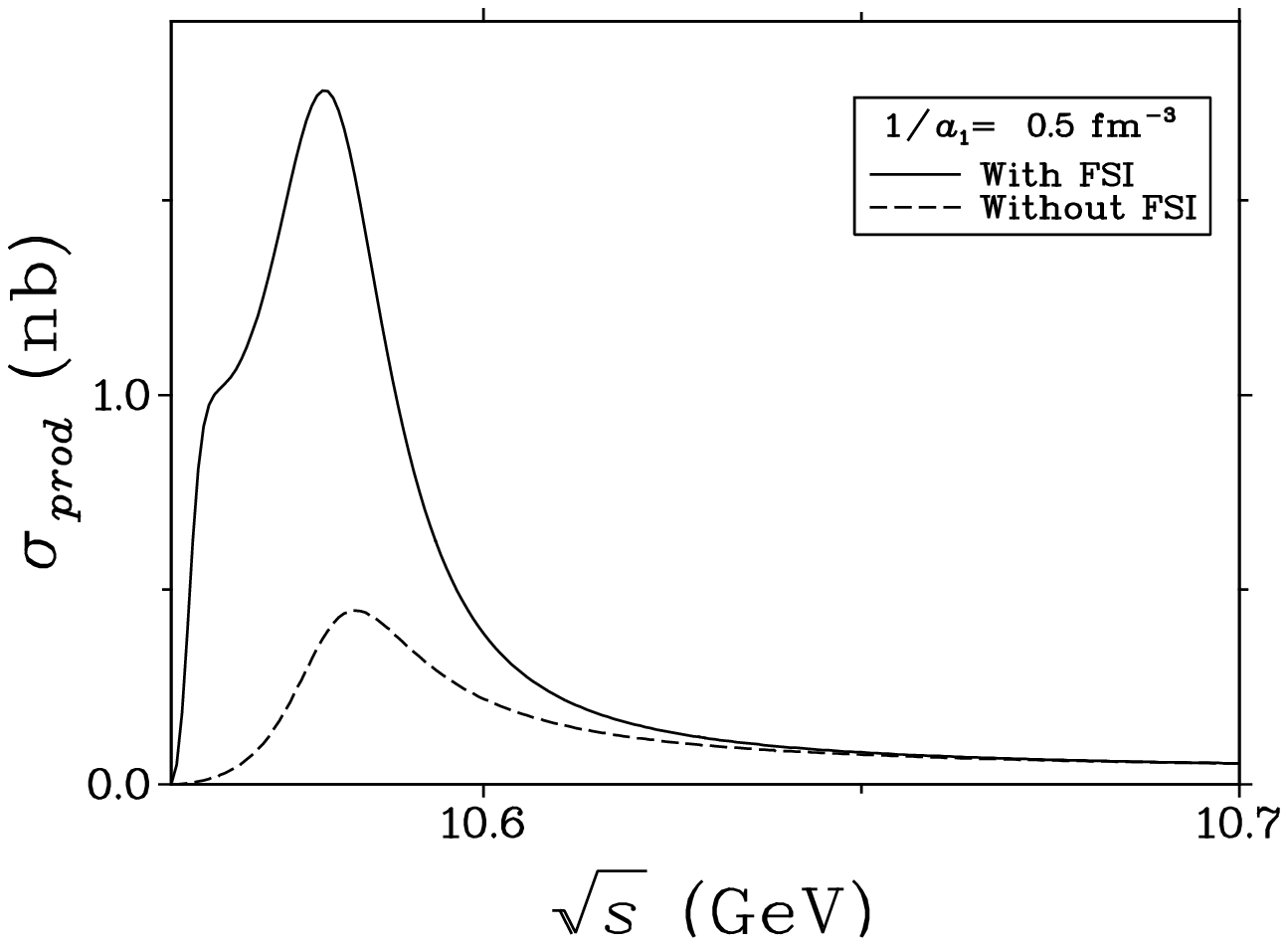}}
\end{tabular}
\caption{The obtained $B\bar{B}$ phase shifts and production cross
sections for the case $1/a_1$=0.5 fm$^{-3}$. We get the dash line in the right diagram by ignoring the rescattering part $\sigma_2$ of the production cross section $\sigma_{prod}=\sigma_1+\sigma_2$.}\label{assumed-B}
\end{figure}

For the case (d), we have $1/a_1=-0.18$ fm$^{-3}$. We illustrate
the results with $\beta=1200$ MeV and $B=0.7$. Figure
\ref{fit-ph-cs-B} shows the reproduced phase shifts and cross
sections while Fig. \ref{pr-cs-B} shows the calculated production
cross sections. We have adopted the coupling $g_{\Upsilon B\bar{B}}=3.5$ and the $D\bar{D}$ parameters $F_0=2.0$ and $\phi=\pi/2$. The anomalous line shape reflects the P- wave $B\bar{B}$ bound state. For the case (e), we plot, with $1/a_1=0.22$
fm$^{-3}$, $g_{\Upsilon B\bar{B}}=3.0$, $\beta=1000$ MeV, and $B=3.0$, the reproduced phase
shifts and scattering cross sections in Fig. \ref{fit-ph-cs-B} and
the corresponding production cross sections in Fig. \ref{pr-cs-B}.
One should note the production cross section vanishes at the
threshold. So we have two peaks in the second diagram of Fig.
\ref{pr-cs-B}. The sharp one near the threshold is due to a P- wave resonance. Similar to the $D\bar{D}$ case, we also consider an
extreme case: $1/a_1=0.5$ fm$^{-3}$, $\beta=500$ MeV and $B=0$.
Figure \ref{assumed-B} displays the derived phase shifts and the
obtained production cross sections where $g_{\Upsilon B\bar{B}}=11.3$ has been used. Now the anomalous structure is just a bump. The peaks of $\Upsilon(4S)$ in these cases are shifted to a lower position because of the strong final state interactions.

The big difference between the two diagrams in Fig. \ref{pr-cs-B}
results from the fact that the coupling constant in the separable
approximation is sensitive to the scattering volume. To see the
behavior of the line shape in the separable approximation, let us go
to Eq. (\ref{coup-sep}). From that equation, the coupling is stronger with a bigger value of the scattering volume. If the scattering volume is positive, this indicates that the stronger the $B\bar{B}$ interaction is, the more obvious the anomalous line shape is. On the other hand, if the scattering volume is negative, the weaker the $B\bar{B}$ interaction is, the more obvious the anomalous line shape is. Therefore, if there is a sharp P- wave resonance or a shallow P- wave bound state around the $B\bar{B}$ threshold, the structure may be observable in $e^+e^-\to B\bar{B}$. The sensitivity to the scattering volume explains why the two diagrams in Fig. \ref{pr-cs-B} look so different. Since the value is positive for a resonance, the peak is clearer. In one word, the scattering volume mainly affects the line shape of the production cross section. The other two parameters $\beta$ and $B$ mainly control the magnitude of the cross section. But then one may recover roughly the same cross section by adjusting $F_0$ and $\phi$.

In Ref. \cite{BBbar-babar}, the BaBar Collaboration measured $R_b(s)=\sigma_b(s)/\sigma_{\mu\mu}(s)$. One may derive a cross section around 1.2 nb near $\Upsilon(4S)$ while our result is around 2 nb. The reason may be that we considered only one resonance contribution. The interference between nearby resonances may reduce this number \cite{mao-2}. The normal line shape of the BaBar data tells us that the above three cases are not realistic, which means the P- wave $B\bar{B}$ interaction is not so strong. From this observation, we may get an upper limit of the cutoff, $\Lambda<2.0$ GeV. In fact, investigation in detail reduces the upper limit to 1.7 GeV where there is a shallow P- wave resonance.

\section{discussions}\label{sec7}

From the calculated partial wave phase shifts, we know that
neither the $D\bar{D}$ nor the $B\bar{B}$ bound state exists if
one does not consider the vector meson exchange interaction. One
gets the same conclusion from solving the bound state problem. Furthermore, in
this case, the interactions for the isoscalar and isovector
systems are the same, and therefore are not reasonable. The realistic
interactions in the meson exchange models should include the
vector meson contributions, and then the isoscalar interaction becomes more attractive. Our following discussions concentrate
on the case $I=0$ and VC.

First, we focus on the cutoff $\Lambda$ in our model study of
$D\bar{D}$ and $B\bar{B}$, which is to be almost universal for the bound
state problem, the scattering problem, and the production problem.
The sensitivity to the cutoff requires a reasonable range. This parameter for $B\bar{B}$ should be a little larger than that for $D\bar{D}$ and thus the upper limit from the former case applies to the latter case. Here we get an upper limit $\Lambda<1.7$ MeV from the $B\bar{B}$ production cross section and the P- wave interaction. We may constrain the cutoff from the P- wave $B\bar{B}$ process because the possible resonance or bound state in the rescattering mechanism changes the line shape of the cross section. Since neither a P- wave resonance nor a bound state exists, one cannot get the constraint from the $D\bar{D}$ production.

With the upper limit 1.7 GeV of the cutoff, we cannot exclude the possible $D\bar{D}$ S-wave bound state, but the binding energy is less than 10 MeV. There does not exist a P- wave $B\bar{B}$ bound state or a resonance, while the binding energy of the possible S- wave molecule should be below 100 MeV. We have used the vector meson coupling constants $g_V$ and $\beta_V$ derived from the vector meson dominance. The recent calculation with light cone QCD sum rule method gives a smaller $g_V\cdot\beta_V$ \cite{gDDV}. It will lead to shallower meson-meson bound states.

In our calculation, we have used a monopole type form factor in the meson exchange potential. One may alternatively choose a dipole type form factor. The potential in the latter case is weaker than that in the former one if the same cutoff is adopted. Provided one obtains the same scattering volume with $\Lambda_1$ in the monopole case and $\Lambda_2$ in the dipole case, $\Lambda_2$ is larger than $\Lambda_1$. The corresponding S- wave binding energies have the relation $|E_1|>|E_2|$, but the difference is small. From the normal line shape of the $B\bar{B}$ production cross section and the P- wave interaction, we find that the cutoff now should be smaller than 2.4 GeV. The corresponding binding energies for the possible S- wave molecules change a little, but we still have the results: $E_{D\bar{D}}< 10$ MeV and $E_{B\bar{B}}<100$ MeV. That is, the usage of a different form factor has small effect on the constrain for the S- wave binding energies.

If the $D\bar{D}$ S- wave bound state exists, its strong decay channel would be mainly $\eta_c\eta$ and $\chi_{c0}\pi\pi$. Other channels such as $\eta_c\pi\pi\pi$ and $J/\psi\pi\pi\pi$ are suppressed. This state may be produced in $B$ decay, $\gamma\gamma$ fusion and $p\bar{p}$ collision \cite{E835}. Future precision analysis of the process $e^+e^-\to J/\psi + scalar\, states$ may also offer a chance to find it. On the other hand, since the $B\bar{B}$ bound state is close to the thresholds of $\Upsilon(1S)\phi$ and $\chi_{b0}(2P)\pi\pi$, its dominant strong decay channels would be $\Upsilon(1S)\omega$ and $\chi_{b0}(1P)\pi\pi$. One has a chance to obtain this scalar state by $p\bar{p}$ collision.

In our approach, the S- and P- wave $D\bar{D}$ interactions depend on the same cutoff. Once the P- wave production including the final state interactions gives a constraint on its reasonable range, the S- wave interactions are better understood. This approach may be applied to new exotic resonances which are candidates of S- wave meson-antimeson molecules. For example, the P- wave $D^*\bar{D^*}$ production will be helpful to understand  $Z^+(4051)$ \cite{twocharged} and the relevant bound state problem \cite{RGG,vol-res}. The cutoff $\Lambda$ will not change significantly because of the heavy quark symmetry. Another
interesting example is the $D\bar{D}^*$ interaction. The
investigation on their P- wave production may deepen our knowledge about X(3872)
\cite{belle-DDstar,babar-DDstar}.

From the study of the P- wave $B\bar{B}$ production by including
the rescattering effects, we have seen the result in the separable
approximation is sensitive to both the amplitude and the sign of
the scattering volume $a_1$. If there is a shallow bound state or
a sharp resonance, the line shape of the production cross section
may reflect that structure, which is difficult to identify just
from the scattering cross sections. The S- wave production
processes should have a similar feature. Since the S- wave system
has stronger attraction, that process is more interesting. Such a
study is expected to be helpful to understand some of the newly
observed near-threshold structures.

In summary, we have explored the bound state and the scattering
problem for the isoscalar $D\bar{D}$ system, and the rescattering
effects in the $e^+e^-\to D\bar{D}$ process. We have also
considered the corresponding bottom cases. From the binding energies and the
phase shifts, the S- wave $D\bar{D}$ bound state would exist if the vector
meson exchange interaction plays a major role. From the line shape
of the calculated $D\bar{D}$ production cross section, it is
difficult to understand the BES observation by the $D\bar{D}$ rescattering
effect. From the line shape of the $B\bar{B}$ production cross section, we have estimated the upper limit of the
cutoff $\Lambda$ in the coupling form factor (monopole type) to be 1.7 GeV. Assuming this is the
case, we get an upper limit of the S- wave binding energy: 10 MeV for $D\bar{D}$ and 100 MeV for $B\bar{B}$. The future measurement of the S-
and P- wave phase shifts, scattering length or volume, and the
production cross sections may provide more information
about the near-threshold resonances.

\section*{Acknowledgments}

We thank Professors Z.Y. Zhang, P.N. Shen, X.Y. Shen, and Q. Zhao
for helpful discussions. This project was supported partly by the
Japan Society for the Promotion of Science under Contract No.
P09027; KAKENHI under Contract Nos. 17070002 (Priority area), 19540275, 20540281, 22105503, and 21$\cdot$09027; National Natural Science Foundation of China under Grant Nos.
10625521, 10675008, 10705001, 10775146, 10721063, and 10805048; the Foundation
for the Author of National Excellent Doctoral Dissertation of P.R.
China (FANEDD) under Contract No. 200924; the Doctoral Program
Foundation of Institutions of Higher Education of P.R. China under
Grant No. 20090211120029; and the Program for New Century Excellent
Talents in University (NCET) by the Ministry of Education of P.R.
China under Grant No. NCET-10-0442.

\appendix

\section{Production cross section}

Here, we illustrate the procedure to derive the production cross section after considering the rescattering effects. We use the PDG state normalization
$\langle \vec{p}|\vec{q}\rangle=(2\pi)^3\delta^3(\vec{p}-\vec{q})$
\cite{PDG}. In the center of mass (c.m.) frame, the $D\bar{D}$ system with the c.m.
momentum $\vec{p}$ also has this normalization. Therefore, one has
\begin{eqnarray}\label{iden}
1=\sum_f\int\frac{d^3\vec{p}_f}{(2\pi)^3}|\vec{p}_f\rangle\langle
\vec{p}_f|.
\end{eqnarray}
Our basic formula is the differential cross section for a $2\to2$
production process \cite{PDG}
\begin{eqnarray}
d\sigma=\frac{|M|^2}{4F_f}(2\pi)^4\delta^4(P-\sum
p_f)\frac{d^3\vec{p}_{f1}
d^3\vec{p}_{f2}}{(2\pi)^3(2E_{f1})(2\pi)^3(2E_{f2})},
\end{eqnarray}
where $M$ is the Lorentz-invariant scattering amplitude, $F_f$ is
the flux factor and $\vec{p}$ ($E_{f}$) is the 3-momentum (energy)
of the final state meson.

To consider FSI, we insert Eq. (\ref{iden}) into $|M|^2$:
\begin{eqnarray}
|M|^2&=&|\langle f|\hat{O}|i\rangle|^2=|\sum_m\int\frac{d^3\vec{p}_m}{(2\pi)^3}\langle f|\hat{O}_2|m\rangle\langle m|\hat{O}_1|i\rangle|^2\nonumber\\
&=&\sum_{m_1,m_2}\int\frac{d^3\vec{p}_{m_1}}{(2\pi)^3}\int\frac{d^3\vec{p}_{m_2}}{(2\pi)^3}\Big[\langle
i|\hat{O}_1^\dag|m_2\rangle \langle
m_1|\hat{O}_1|i\rangle\Big]\Big[ \langle
m_2|\hat{O}^\dag_2|f\rangle \langle f|\hat{O}_2|m_1\rangle\Big].
\end{eqnarray}
Here, the operator $\hat{O}_1$ describes the production of
$D\bar{D}$ and $\hat{O}_2$ describes the rescattering of the final
states $D\bar{D}$. $\vec{p}_{m_1}$ and $\vec{p}_{m_2}$ are the
momenta in the c.m. frame. They have different angles but the same
amplitude. One may get the total cross section
\begin{eqnarray}\label{sigma-cs}
\sigma&=&\frac{1}{4F_f}\int\frac{d^3\vec{p}_{m_1}}{(2\pi)^3}\int\frac{d^3\vec{p}_{m_2}}{(2\pi)^3}\frac{1}{(2E_{m_1})(2E_{m_2})}\Big[\langle i|\hat{O}_1^\dag|m_2\rangle \langle m_1|\hat{O}_1|i\rangle\Big]\int\frac{d^3\vec{p}_f}{(2\pi)^3}\Big[ \langle m_2|\hat{O}^\dag_2|f\rangle \langle f|\hat{O}_2|m_1\rangle\Big]\nonumber\\
&&\times(2\pi)\delta(E-2E_f)\nonumber\\
&=&\frac{1}{2s}\int\frac{d^3\vec{p}}{(2\pi)^3}\int\frac{d^3\vec{q}}{(2\pi)^3}\frac{1}{(2E_{p})(2E_{q})}\Big[\langle
i|\hat{O}_1^\dag|\vec{q}\rangle \langle
\vec{p}|\hat{O}_1|i\rangle\Big]S(E),
\end{eqnarray}
where we have ignored the mass of the electron and have used
$F_f=s/2$.

We calculate the first part as follows
\begin{eqnarray}\label{prodformu}
\langle i|\hat{O}_1^\dag|\vec{q}\rangle\langle
\vec{p}|\hat{O}_1|i\rangle=\frac{8e^4}{s^2}[-4\vec{k}\cdot\vec{p}\vec{k}\cdot\vec{q}+
\vec{p}\cdot \vec{q}s]\times |{\rm f.f.}|^2,
\end{eqnarray}
where $\vec{k}$ is the momentum of the initial electron in the c.m.
frame and f.f. has been given in Eq. (\ref{f.f.}).

For the second part in Eq. (\ref{sigma-cs}), we have
\begin{eqnarray}
S(E)&\equiv& \int\frac{d^3\vec{p}_f}{(2\pi)^3}\Big[ \langle \vec{q}|\hat{O}^\dag_2|f\rangle \langle f|\hat{O}_2|\vec{p}\rangle\Big](2\pi)\delta(E-2E_f)\nonumber\\
&=&-2 \,{\rm Im}\Big[ \langle \vec{q}| \hat{G}(E)
|\vec{p}\rangle\Big]
\end{eqnarray}
with $\hat{G}(E)=[E-\hat{H}+i\epsilon]^{-1}$. When there is no
FSI, $\hat{G}(E)=\hat{G}^0(E)=[E-\hat{H}^0+i\epsilon]^{-1}$,
$H^0=2(m_D+\frac{p^2}{2m_D})\approx2E_f$. In general,
\begin{eqnarray}
\hat{G}(E)&=&\hat{G}^0(E)+\hat{G}^0(E)V_{int}\hat{G}^0(E)+\cdots\nonumber\\
&=&\frac{1}{1-\hat{G}^0(E)V_{int}}\hat{G}^0(E),
\end{eqnarray}
where $V_{int}$ is the potential.

To get the analytical expression of the cross section, we adopt
the Yamaguchi separable approximation \cite{separable}. For the P-wave $D\bar{D}$ interaction, one
may write down as
\begin{eqnarray}
\langle
\vec{p}|V_{int}|\vec{q}\rangle=-\frac{\lambda}{m_D}g(\vec{p})\cdot
g(\vec{q}),\quad g(\vec{p})=\beta t(p)\vec{p}.
\end{eqnarray}
We determine
$\lambda$, $\beta$ and other parameters in the function $t(p)$ through
reproducing the calculated phase shifts and the scattering cross section. After some calculations, one
finally gets
\begin{eqnarray}\label{fsiformu}
S(E)=(2\pi)^4\delta(E-E_p-E_q)\delta^3(\vec{p}-\vec{q})+\frac{2\lambda}{m_D}\,{\rm
Im}\Big[\frac{\tilde{g}(\vec{p})\cdot
\tilde{g}(\vec{q})}{1+\frac{\lambda W}{3m_D}} \Big].
\end{eqnarray}
In this formula,
\begin{eqnarray}
\tilde{g}(\vec{k})&=&{g}(\vec{k})[E-2m_D-\frac{k^2}{m_D}+i\epsilon]^{-1},\\
W&\equiv&\int\frac{d^3\vec{k}}{(2\pi)^3} \frac{g(\vec{k})\cdot g(\vec{k})}{E-2m_D-k^2/m_D+i\epsilon}.
\end{eqnarray}
With the equation $\frac{1}{E-H+i\epsilon}=\frac{\cal P}{E-H}-i\pi\delta(E-H)$, it is easy to get the real part and the imaginary part
\begin{eqnarray}
ReW&=&-m_D\frac{\beta^2}{2\pi^2} {\cal P}\int_0^\infty dk \frac{k^4}{k^2-\alpha^2}[t(k)]^2,\\
ImW&=&-m_D\frac{\beta^2 \alpha^3}{4\pi}[t(\alpha)]^2,
\end{eqnarray}
where $\alpha^2\equiv m_D(E-2m_D)$.

By combining Eqs. (\ref{sigma-cs}), (\ref{prodformu}), and
(\ref{fsiformu}), we obtain the resultant cross section
$\sigma=\sigma_1+\sigma_2$,
\begin{eqnarray}
\sigma_1&=&\frac{\pi}{3}\alpha_e^2 \frac{(s-4m_D^2)^{3/2}}{s^{5/2}}\times |{\rm f.f.}|^2,\\
\sigma_2&=&\frac{64}{9}\frac{\pi^2\alpha_e^2\lambda}{s^2m_D}{\rm
Im}\left\{(1+\frac{\lambda W}{3m_D})^{-1}U^2 \right\}\times |{\rm
f.f.}|^2,
\end{eqnarray}
where $\alpha_e=1/137$ and
\begin{eqnarray}
U&\equiv&\beta\int\frac{d^3\vec{ p}}{(2\pi)^3}\frac{p^2}{(2E_p)} \frac{t(p)}{E-2m_D-\frac{p^2}{m_D}+i\epsilon}\nonumber\\
&\simeq&4\pi m_D\beta\int_0^\infty\frac{d p}{(2\pi)^3}\frac{p^4}{p^2+2m_D^2} \frac{t(p)}{E-2m_D-\frac{p^2}{m_D}+i\epsilon},\\
ReU&=&-m_D^2\frac{\beta}{2\pi^2} {\cal P}\int_0^\infty d k\frac{k^4}{(k^2+2M^2)(k^2-\alpha^2)}t(k),\\
ImU&\approx&- m_D\frac{\beta\alpha^3}{4\pi\sqrt{s}}
t(\alpha).
\end{eqnarray}
If one defines
$W^r=1+\frac{\lambda }{3m_D} ReW$, $W^i=\frac{\lambda }{3m_D}
ImW$, then
\begin{eqnarray}
\sigma_2&=&\frac{64}{9}\frac{\pi^2\alpha_e^2\lambda
}{s^2m_D}\left\{\frac{2(ReU\times
ImU)W^r-[(ReU)^2-(ImU)^2]W^i}{{(W^r)}^2+ {(W^i)}^2}\right\}\times
|{\rm f.f.}|^2.
\end{eqnarray}
When there is no FSI, $\lambda=0$ and $\sigma_2$ vanishes.

\section{The parameters}

Before the numerical evaluation, we have to choose the form of the
function $t(k)$ and determine the relevant parameters. In Ref.
\cite{sep-dirac}, the separable P- wave potential was given with
$t(k)=\frac{1}{(k^2+\beta^2)^2}$. However, we find the following
choice is better when reproducing the scattering cross sections:
\begin{eqnarray}\label{tk}
t(k)&=&\frac{1}{(k^2+\beta^2)^2}+\frac{B}{\beta^2(k^2+\beta^2)},
\end{eqnarray}
where $B$ is a dimensionless parameter. One notes the second term
of $g(\vec{k})$ does not have a good behavior at large $k$. This
is easy to see after the Fourier transformation. Since we consider
only low-energy interactions, this form is acceptable.

From the scattering amplitude, one has
\begin{eqnarray}
k^3\cot(\delta_1)&=&\frac{12\pi}{\lambda\beta^2 [t(k)]^2}+\frac{4\pi F(k)}{\beta^2 [t(k)]^2},\\
F(k)&=&\frac{\beta^2}{2\pi^2}{\cal P}\int_0^\infty d {q} \frac{q^4
}{k^2-q^2}[t(q)]^2,
\end{eqnarray}
where $\delta_1$ is the P- wave phase shift. According to the
definition of the scattering volume $a_1$,
\begin{eqnarray}
\frac{1}{a_1}&=&\frac{4\pi}{\beta^2}\frac{ F(0)}{
[t(0)]^2}+\frac{12\pi}{\lambda\beta^2 [t(0)]^2}.
\end{eqnarray}
So the coupling constant is
\begin{eqnarray}\label{coup-sep}
\lambda&=&\frac{12\pi}{\frac{1}{a_1}\beta^2[t(0)]^2-(4\pi)F(0)}.
\end{eqnarray}
If one does not explicitly use the coupling constant in expressing
the P- wave cross section, we have
\begin{eqnarray}
k^3\cot(\delta_1)&=&\frac{4\pi [F(k)-F(0)]}{\beta^2 [t(k)]^2}+\frac{1}{a_1}\frac{[t(0)]^2}{[t(k)]^2},\\
\sigma_{L=1}&=&\frac{12\pi
k^4}{k^6+[k^3\cot(\delta_1)]^2}.\label{cs0-P}
\end{eqnarray}
By inserting the function $t(k)$ into the above formulas, it is
easy to get the explicit expressions which we do not present here.
The scattering volume has been derived in Sec \ref{sec4}. Now the parameters we have to determine are $\beta$ and $B$. They are extracted by reproducing the $\delta_1$ and $\sigma_{L=1}$.

\end{document}